\pdfoutput=1  

\documentclass[aps,twocolumn,groupedaddress,floatfix,preprintnumbers,nofootinbib,eqsecnum]{revtex4}
\usepackage{amssymb,amsmath,graphicx,multirow,graphics}
\graphicspath{{TimeEvolPlots/}{ContourPlots/}}

\begin{document}

\title{Dynamical Dark Matter from Strongly-Coupled Dark Sectors}
\author{Keith R. Dienes$^{1,2}$\footnote{E-mail address:  {\tt dienes@email.arizona.edu}},
      Fei Huang$^{1}$\footnote{E-mail address:  {\tt huangfei@email.arizona.edu}},
      Shufang Su$^{1}$\footnote{E-mail address:  {\tt shufang@email.arizona.edu}},  
      Brooks Thomas$^{3}$\footnote{E-mail address:  {\tt thomasbd@lafayette.edu}}}
\affiliation{
     $^1$ Department of Physics, University of Arizona, Tucson, AZ  85721  USA\\
     $^2$ Department of Physics, University of Maryland, College Park, MD  20742  USA\\
     $^3$ Department of Physics, Lafayette College, Easton, PA  18042 USA}

\begin{abstract}
Dynamical Dark Matter (DDM) is 
an alternative 
framework for dark-matter physics 
in which the dark sector 
comprises a vast ensemble of particle species whose Standard-Model decay widths are balanced against
their cosmological abundances.
Previous studies of this framework have focused on a
particular class of DDM ensembles --- motivated primarily by Kaluza-Klein 
towers in theories with extra dimensions --- in which the density of dark states 
scales roughly as a polynomial of the mass.
In this paper, by contrast, we study the properties of a different class
of DDM ensembles in which the density of dark states grows {\it exponentially}\/ with
mass.  Ensembles with this Hagedorn-like property arise naturally as
the ``hadronic'' resonances associated with the confining phase of
a strongly-coupled dark sector;  
they also arise naturally as the gauge-neutral bulk states of Type~I string theories.
We study the dynamical properties of such ensembles, 
and demonstrate that an appropriate DDM-like balancing
between decay widths and abundances can emerge naturally --- even with an exponentially
rising density of states.
We also study the effective equations of state for such ensembles, and 
investigate some of the model-independent observational constraints on such 
ensembles that follow directly from these equations of state. 
In general, we find that such constraints tend to introduce correlations
between various properties of these DDM ensembles such as their associated
mass scales, lifetimes, and abundance distributions.
For example,
we find that
these constraints allow DDM ensembles 
with energy scales ranging from the GeV scale all the way to the Planck scale,
but that the total present-day cosmological abundance of the dark sector must be spread across an increasing number of
different states in the ensemble as these energy scales 
are dialed from the Planck scale down to the GeV scale.
Numerous other correlations and constraints are also discussed.
\end{abstract}

\maketitle

\newcommand{\newc}{\newcommand}
\newc{\gsim}{\lower.7ex\hbox{$\;\stackrel{\textstyle>}{\sim}\;$}}
\newc{\lsim}{\lower.7ex\hbox{$\;\stackrel{\textstyle<}{\sim}\;$}}
\makeatletter
\newcommand{\biggg}{\bBigg@{3}}
\newcommand{\Biggg}{\bBigg@{4}}
\makeatother

\def\vac#1{{\bf \{{#1}\}}}

\def\beq{\begin{equation}}
\def\eeq{\end{equation}}
\def\beqn{\begin{eqnarray}}
\def\eeqn{\end{eqnarray}}
\def\calM{{\cal M}}
\def\calV{{\cal V}}
\def\calF{{\cal F}}
\def\half{{\textstyle{1\over 2}}}
\def\quarter{{\textstyle{1\over 4}}}
\def\ie{{\it i.e.}\/}
\def\eg{{\it e.g.}\/}
\def\etc{{\it etc}.\/}


\def\inbar{\,\vrule height1.5ex width.4pt depth0pt}
\def\IR{\relax{\rm I\kern-.18em R}}
 \font\cmss=cmss10 \font\cmsss=cmss10 at 7pt
\def\IQ{\relax{\rm I\kern-.18em Q}}
\def\IZ{\relax\ifmmode\mathchoice
 {\hbox{\cmss Z\kern-.4em Z}}{\hbox{\cmss Z\kern-.4em Z}}
 {\lower.9pt\hbox{\cmsss Z\kern-.4em Z}}
 {\lower1.2pt\hbox{\cmsss Z\kern-.4em Z}}\else{\cmss Z\kern-.4em Z}\fi}
\def\TBBN{T_{\mathrm{BBN}}}
\def\OmegaCDM{\Omega_{\mathrm{CDM}}}
\def\OmegaDM{\Omega_{\mathrm{CDM}}}
\def\Omegatot{\Omega_{\mathrm{tot}}}
\def\rhotot{\rho_{\mathrm{tot}}}
\def\rhocrit{\rho_{\mathrm{crit}}}
\def\tnow{t_{\mathrm{now}}}
\def\tMRE{t_{\mathrm{MRE}}}
\def\TMRE{T_{\mathrm{MRE}}}
\def\arcsinh{\mbox{arcsinh}}
\def\Omegatotnow{\Omega_{\mathrm{tot}}^\ast}
\def\mij{m_{jj}}
\def\mijmin{m_{jj}^{(\mathrm{min})}}
\def\mijmax{m_{jj}^{(\mathrm{max})}}
\def\mmax{m_{\mathrm{max}}}
\def\epsig{\epsilon_{\mathrm{sig}}}
\def\Lint{\mathcal{L}_{\mathrm{int}}}
\def\MT2{M_{T2}}
\def\MTtwomax{M_{T2}^{\mathrm{max}}}
\def\BR{\mathrm{BR}}
\def\weff{w_{\mathrm{eff}}}
\def\mstring{M_{s}}
\def\tLB{t_{\mathrm{LB}}}
\newcommand{\Dsle}[1]{\hskip 0.09 cm \slash\hskip -0.23 cm #1}
\newcommand{\Dirsl}[1]{\hskip 0.09 cm \slash\hskip -0.20 cm #1}
\newcommand{\met}{{\Dsle E_T}}


\input epsf





\section{Introduction\label{sec:Intro}}


Dynamical Dark Matter (DDM)~\cite{DDM1,DDM2} is an alternative framework for
dark-matter physics in which dark-matter stability is not required.
Instead, 
the dark sector within the DDM framework comprises a vast ensemble of 
individual constituent particles exhibiting a variety of different masses, lifetimes, and
cosmological abundances.  The phenomenological viability of such a dark sector
is then ensured through a non-trivial {\it balancing}\/ between 
cosmological abundances and Standard-Model (SM) decay widths across the ensemble. 
Indeed, under this balancing, 
those ensemble constituents with shorter lifetimes must have smaller 
cosmological abundances, while states with longer lifetimes may
have larger cosmological abundances.
As a result, the dark sector in such a scenario is {\it dynamic}\/:
states in the dark sector are continually decaying into visible-sector states
throughout the evolution of the universe ---  not just in previous epochs but even
at the present time and into the future.  Quantities such as the total energy density
$\Omega_{\rm CDM}$ and the effective equation-of-state parameter $w_{\rm eff}$ are thus 
time-dependent quantities, 
and it is only an accident
that these quantities happen to take particular values at the present time.
Many methods have been developed for testing this framework, 
spanning from collider signatures~\cite{DDMcolliders,DDMcolliders2} 
to signatures in direct-detection~\cite{DDMdirect} 
and indirect-detection~\cite{DDMindirect,DDMboxes1,DDMboxes2} experiments.  

Of course, many of the constraints on such DDM ensembles depend on model-specific
details associated with the ensemble in question, such as the specific particle nature of the 
individual dark constituent fields and the precise form of their decays into SM states.
By contrast, other phenomenological properties of (and constraints on) 
these DDM ensembles depend simply on the manner in which the lifetimes and abundances
of ensemble constituents
scale with respect to each other, and thus have a greater degree of model-independence.  For example, 
the effective equations of state for these ensembles
 are governed in large part solely by these scaling relations.
As a result, all phenomenological/observational constraints on 
the equations of state of the dark sector
are essentially constraints on 
the types of balancing relations that DDM ensembles may exhibit.
These are thus model-independent constraints which can be placed
on such ensembles simply as a result of their inherent scaling relations.

One general class of DDM ensembles consisting of  
large numbers of dark particle species exhibiting suitable scaling relations between 
lifetimes and cosmological abundances 
are those whose constituents are the 
Kaluza-Klein (KK) modes of a gauge-neutral bulk field in a 
theory with extra spacetime dimensions in which 
cosmological abundances are established through misalignment production~\cite{DDM1}.  
Indeed, explicit realizations of DDM ensembles of this type have been 
constructed~\cite{DDM2,DDMAxion}.  
Although many aspects of these ensembles depend 
on the details of the particular fields under study,
certain general properties are common across all such ensembles in this class.
One of these is that the 
cosmological abundance of each component scales as a power  of the 
lifetime of that component.
Likewise, the density of states within such ensembles is either insensitive 
to mass or scales 
roughly as a polynomial function of mass across the ensemble.   
For these reasons, 
most phenomenological studies of the DDM framework have focused on ensembles
exhibiting polynomial scaling relationships. 

Polynomial scaling relations also emerge in other (purely four-dimensional) contexts as well.
For example, under certain circumstances, thermal freeze-out mechanisms for abundance 
generation can also lead to appropriate polynomial inverse scaling relations between lifetimes and
abundances~\cite{DesigningDDM}.
In fact, such inverse scaling relations can even emerge {\it statistically}\/ in contexts 
in which the dynamics underlying the dark sector is essentially random~\cite{RandomDDM}.

There are, however, other well-motivated theoretical constructions which 
do not give rise to dark sectors with polynomial scaling relations.
One example is a dark sector 
consisting of a set of fermions (dark ``quarks'') charged under a non-Abelian gauge 
group $G$ which becomes confining below some critical temperature $T_c$.  At 
temperatures $T \lesssim T_c$, when the theory is in the confining phase, 
the physical degrees of freedom are composite states (dark ``hadrons'').  
Another well-motivated type of DDM ensemble consists of the bulk (\ie, closed-string) states in 
Type~I string theories.   Such bulk states are typically neutral with respect to
all brane gauge symmetries, and interact with those brane states only gravitationally.
As such, from the perspective of brane-localized observers, these bulk states too are dark matter.

At first glance, these two latter types of ensembles may seem to have little in common with each other.
Indeed, many aspects of the detailed phenomenologies  associated with these ensembles will 
be completely different.
However, they nevertheless exhibit certain underlying model-independent commonalities  
which are relevant for their viability as DDM ensembles.    
Indeed, these features are identical to those 
which characterize the ``visible'' sector of ordinary hadrons, namely 
\begin{itemize}
\item   mass distributions which follow linear Regge trajectories (\ie, $\alpha' M^2_n\sim {n}$
where $\alpha'$ is a corresponding Regge slope), and 
\item exponentially growing (``Hagedorn-like'') degeneracies of states (\ie, $g_n\sim e^{\sqrt{n}} \sim
e^{\sqrt{\alpha'} M_n}$).
\end{itemize}
These features --- especially the appearance of an {\it exponential}\/ scaling of the state degeneracies
with mass --- represent a behavior which is markedly different from that 
exhibited by DDM ensembles with polynomial scaling relations.  
For example, as a result of their exponentially growing densities of states,
such ensembles have a critical temperature~\cite{Hagedorn} beyond which their partition functions diverge.

In this paper, we shall study the generic properties of DDM ensembles which exhibit the two 
features itemized above.  We shall calculate the effective equations of state $w_{\rm eff}(t)$ for such 
ensembles, and subject these ensembles to those  
immediate model-independent observational constraints 
that follow directly from these equations of state.
We shall therefore be able to place zeroth-order model-independent bounds on some of 
the quantities that parametrize these features, such as the 
effective Regge slope as well as the rate of exponential growth in the state degeneracies.
Our primary motivation is to understand the 
phenomenology that might apply to strongly-coupled dark sectors in their confined (``hadronic'') phase,
imagining nothing more than that our DDM ensemble resembles the visible hadronic sector in the two respects
itemized above.
However, the results of such analyses might also be useful in constraining the 
bulk sector of various classes of string theories, since these bulk sectors also give rise
to ensembles of dark-matter states which
share these two grossest features.
We shall therefore aim to keep our discussion as model-independent as possible,
subject to our assumption of the above two properties itemized above.
In this way, our analysis and the constraints we obtain
can serve as useful phenomenological guides 
in eventually building realistic dark-matter models of this type.

This paper is organized as follows.  
In Sect.~\ref{sec:DensityOfStates}, we begin by reviewing the properties that we shall assume
for the mass spectrum and density of states of our DDM dark ``hadron'' ensemble.
We shall also discuss the physical interpretations of these properties in terms of a variety of
underlying flux-tube models and string theories.
This section will also serve to establish our conventions and notation.
Then, in Sect.~\ref{sec:Balancing}, we discuss how the required balancing between lifetimes and 
abundances naturally arises for such DDM ensembles.
In particular, we examine the mechanism through which primordial abundances for 
these hadron resonances are generated, and we determine how these abundances scale across
the ensemble as a function of the hadron mass.  We also discuss the
scaling behavior of the decay widths that characterize the decays of the hadronic 
ensemble constituents to SM states,
as well as the assumptions that enter into such calculations.
In Sect.~\ref{sec:OmegaEtaWeff}, we then derive
expressions for the total abundance $\Omegatot(t)$, the tower fraction $\eta(t)$, 
and the effective equation-of-state parameter $\weff(t)$ for these DDM ensemble as functions
of time.  As discussed in Refs.~\cite{DDM1,DDM2} and reviewed in Sect.~\ref{sec:OmegaEtaWeff}, 
these three functions characterize the time-evolution of DDM ensembles
and allow us to place a variety of general, model-independent
constraints on such ensembles.
In Sect.~\ref{sec:Results}, we then present the results of our analysis of the 
phenomenological viability of such DDM ensembles, identifying those regions of the corresponding
parameter space which lead to the most promising ensembles and uncovering generic phenomenological
behaviors and correlations across this space.
One of our key findings is that these DDM ensembles can satisfy our constraints across a broad range
of energy scales
ranging from the GeV scale all the way to the Planck scale,
but that the present-day cosmological abundance of the dark sector must be distributed across an increasing number of
different states in the ensemble as the fundamental mass scales associated with the ensemble
are dialed from the Planck scale down to the GeV scale.
Finally, in Sect.~\ref{sec:Conclusion}, we summarize our results and
discuss possible avenues for future work.


\section{DDM ensembles of dark hadrons: Fundamental assumptions  \label{sec:DensityOfStates}}

 
As discussed in the Introduction,
in this paper we are primarily concerned with the properties of DDM ensembles 
whose constituents are the ``hadronic'' composite states or resonances of a strongly-coupled 
dark sector.  As has been well known since the 1960's, many of the attributes of 
such an ensemble can be successfully modeled by strings.  
These attributes include linear Regge trajectories,
linear confinement, 
an exponential rise in hadron-state degeneracies,
and $s$- and $t$-channel duality.
It is not a complete surprise that there is a deep connection between hadronic
spectroscopy and the spectra of string theory.
Hadronic resonances (particularly mesons) can be viewed 
as configurations of dark ``quarks'' linked together by flux tubes.  The spectrum of 
excitations in such a theory therefore corresponds to the spectrum of fluctuations 
of these flux tubes.
However, it is well known that these flux tubes can be modeled as non-critical strings.
Thus string theory
can provide insight into the properties of such collections of composite states.

In what follows, we shall use this analogy between hadronic physics and string theory
to motivate our parametrization for the mass spectrum and for the density of 
states of our dark-``hadronic'' DDM ensembles.  
We shall also make recourse to modern string technology, when needed, for refinements
of our basic picture.
Throughout, however, we shall attempt to keep our parametrizations as general as possible
so that they might apply to the widest possible set 
of DDM ensembles sharing these properties.   As discussed in the Introduction, this will allow
our analysis and eventual constraints
to serve as useful guides in future attempts to build realistic models exhibiting these features.

\subsection{The mass spectrum:~   Regge trajectories}

The first feature that we shall assume of our hadronic dark sector is a mass spectrum consistent
with the existence of 
Regge trajectories.  The existence of such trajectories follows directly from 
nothing more than 
our assumption that our dark-sector bound states can be modeled by dark quarks 
connected by the confining flux tube associated with a strong, attractive, dark-sector interaction.
Taking meson-like configurations as our guide 
and temporarily assuming massless quarks, 
it can easily be shown that the mass $M_n$ associated with a relativistic rotating flux tube 
scales with the corresponding total angular momentum $n$ 
as \mbox{$n\sim \alpha' M_n^2$},   where $\alpha'$ is the so-called Regge slope.
In the {\it visible}\/ sector, 
this successfully describes the so-called leading Regge trajectory of the observed mesons,
with \mbox{$\alpha'\sim 1~{\rm (GeV)}^{-2}$} appropriate for QCD.~
Moreover, there also 
exist subleading (parallel) Regge trajectories of observed mesons which have the 
same Regge slope but different intercepts:   $n\sim \alpha' M_n^2 +\alpha_0$.

Regge trajectories of this form, both leading and subleading, also emerge in string theory.
For example, the perturbative states of a quantized open bosonic string 
have masses $M$ and spins $J=0,1,...,J_{\rm max}$ 
which satisfy $J_{\rm max}= \alpha' M^2  + 1$ where $\alpha'$ is
now the Regge slope associated with string theory [typically assumed to be $\sim (M_{\rm Planck})^{-2}$].
The states with $J=J_{\rm max}$ thus sit along the leading Regge trajectory, while those with
smaller values of $J$ sit along the subleading Regge trajectories.
Similar results also hold for superstrings and heterotic strings.

Given these observations, in this paper 
we shall assume that the states of our dark ``hadronic'' DDM ensemble
have discrete positive masses $M_n$ of the general form
\begin{equation}
  M_n^2 ~=~ n \mstring^2  + M_0^2~.
\label{eq:MassSpectrum}
\end{equation}
where $n$ is an index labeling our states in order of increasing mass.
Here $M_s\equiv 1/\sqrt{\alpha'}$ is the corresponding ``string scale'',  
while $M_0$ represents the mass of the lightest ``hadronic'' constituent in the 
DDM ensemble.
Indeed, since we do not expect to have any tachyonic states in our DDM ensemble, we shall
assume throughout this paper that $M_0^2\geq 0$.  
We shall avoid making any further assumptions about 
the nature of the dark sector by treating both $M_s$ and $M_0$ as free parameters
to be eventually constrained by cosmological data.

Our choice of sign for $M_0^2$ perhaps deserves further comment.   
For the visible sector, most hadrons lie along Regge trajectories with $M_0^2 \geq 0$.
While there do exist Regge trajectories with $M_0^2 <0$, the lowest states in such
trajectories are of course absent.    In string theory, by contrast, all Regge trajectories
have $M_0^2 <0$.   However, just as in the hadronic case, all tachyonic
states which might result for small $n$ are ultimately removed from the string spectrum
by certain ``projections'' which are ultimately required for the self-consistency of
the string.    In other words, for Regge trajectories with $M_0^2 <0$, 
one could equivalently relabel our remaining states by shifting $n\to n-1$ and thereby obtain
an ``effective'' $M_0^2 \geq 0$.   This is not normally done in string theory
because in string theory the index $n$ is correlated with other physical quantities such as the
spin of the state.   However we are making no such assumption for the
states of our dark sector, and are treating the index $n$ as a mere labelling
parameter.  Our assumption of a tachyon-free dark sector then leads us to
take $M_0^2 \geq 0$.   

There is also another motivation for taking $M_0^2 \geq 0$.
All of the above results which treat $n$ as an angular momentum assume
massless quarks at the endpoints of the flux tube.   However, while
such an approximation holds well for the lightest states in the visible
sector, we do not wish to make such an approximation for our unknown
dark sector.     We shall therefore assume $M_0^2 \geq 0$ in 
what follows, recognizing that this parameter may in principle also implicitly
include the positive contributions from dark quark masses as well.

\subsection{Degeneracy of states:~   Exponential behavior}

The second generic feature associated with hadronic spectroscopy
is the well-known exponential rise in the degeneracies of hadrons as
a function of mass:   \mbox{$g_n\sim e^{\sqrt{n}}$}.
This behavior was first predicted and observed for hadrons (both mesons and baryons) in Ref.~\cite{Hagedorn},
and also holds as a generic feature for both bosonic and fermionic states in string theory~\cite{StringReviews}.

In general, we can understand this behavior as follows.
If we model our hadrons as quarks connected by flux tubes, 
the degeneracy $g_n$ of hadronic states at any mass level $n$ can be written as the product of
two contributions:   
one factor $\kappa$ representing a multiplicity of states due to the degrees of freedom associated with the quarks
(such as the different possible configurations of quantities like spin and flavor),
and a second factor $\hat g_n$ representing the multiplicity of states due to the degrees of freedom 
associated with the flux tube.
We thus have
\begin{equation}
  g_n ~\approx~ \kappa \, \hat g_n~.
\label{eq:gnInTermsOffn}
\end{equation}
While $\kappa$ is a constant which is independent of the particular mass level $n$,
the remaining degeneracy factor $\hat g_n$ counts the 
rapidly increasing number of ways
in which a state of given total energy $n$ can be realized as a combination of the vibrational, rotational,
and internal excitations of the different harmonic oscillators which together comprise a quantized string.
It is this quantity which grows exponentially with mass,
and in string theory the leading behavior of $\hat g_n$ for large $n$ 
generally takes the form~\cite{StringReviews}
\beq
          \hat g_n ~\approx~  A n^{-B} e^{C\sqrt{n}}~~~~~ {\rm as}~ n\to \infty~,
\label{eq:fnForLargen}
\eeq
where $A,B,C$ are all positive quantities which depend on the particular type of string model under study.
Indeed, for any $B$ and $C$, it turns out that the proper normalization for $\hat g_n$ in string theory
is given by 
\beq
        A ~=~ {1\over \sqrt{2}} \left( {C\over 4\pi}\right)^{2B-1}~.    
\eeq
Thus our asymptotic degeneracy of states is parametrized by two independent quantities $B$ and $C$,
and we shall assume that this continues to be true in our dark sector as well.

The most salient property of the expression in Eq.~(\ref{eq:fnForLargen}) is that it rises exponentially 
with $\sqrt{n}$, or equivalently with the mass $M_n$ of the corresponding state.  
This represents a crucial difference relative to the
KK-inspired DDM ensembles previously considered in 
Refs.~\cite{DDM1,DDM2,DDMAxion} (or even the purely four-dimensional
DDM ensembles considered in Refs.~\cite{DesigningDDM, RandomDDM}).
For example, the KK states corresponding to a single flat extra spacetime dimension
have degeneracies $\hat g_n$ which are constant, or which become so above the $n=0$ level.
The key difference here is that the degrees of freedom associated with our flux tube 
consist of not only KK excitations (if the flux tube happens to be situated within a spacetime with
a compactified dimension), but also so-called {\it oscillator}\/ excitations representing the internal 
fluctuations of the flux tube itself.
It is these oscillator excitations which give rise to the exponentially growing degeneracies
and which are a direct consequence of the non-zero spatial extent of the flux tube.
As such, they are intrinsically stringy and would not arise in theories involving fundamental
point particles.

Unfortunately, the asymptotic form in Eq.~(\ref{eq:fnForLargen}) is not sufficient for our purposes.
Although we are interested in the 
behavior of all states across the DDM ensemble,
it is the lighter states 
rather than the heavier states
which are most likely to have longer lifetimes and therefore greater cosmological abundances.
Thus, even though we want to keep track of all of the states in our ensemble,
we need to be particularly sensitive 
to the degeneracies of the lighter states, \ie, the states with smaller values of $n$.
This poses a problem because 
the asymptotic expression in Eq.~(\ref{eq:fnForLargen})
is fairly accurate in the large-$n$ limit
but is not especially accurate in the small-$n$ limit.
 
Fortunately, for values of $B$ and $C$ which correspond to self-consistent 
strings (to be discussed below),
the tools of 
modern string technology (specifically conformal field theory and modular invariance)
furnish us with a more precise approximation for $\hat g_n$ 
which remains accurate even for very small values of $n$.
This expression is given by~\cite{Cudell,HR,KV,missusy}
\beq
  \hat g_n ~\approx~ 2\pi
    \left(\frac{16\pi^2 n}{C^2} - 1\right)^{\frac{1}{4} - B} 
    I_{|2B - \frac{1}{2}|}\left(C\sqrt{n - \frac{C^2}{16\pi^2}}\right)~,
\label{eq:fnBetterApprox}
\eeq
where $I_\nu (z)$ denotes the modified Bessel function of the first kind
of order $\nu$.  
Use of the approximation $I_\nu(z)\approx e^z /\sqrt{2\pi z}$ for $z\gg1$
then reproduces the result in Eq.~(\ref{eq:fnForLargen}).
However, the expression in Eq.~(\ref{eq:fnBetterApprox}) remains valid to within only 
a few percent all the way down to $n=1$,
assuming $C\leq 4\pi$ (so that the argument of the Bessel function
remains real even for $n=1$).

In what follows, we therefore shall adopt the expression in  
Eq.~(\ref{eq:fnBetterApprox}) 
as our general parametrization for the degeneracy of states $\hat g_n$ for
arbitrary values of $B$ and $C\leq 4\pi$ and for all $n\geq 1$.
For values of $B$ and $C$ corresponding to bona-fide string theories,
this expression yields results for the state degeneracies which, though not necessarily
integral, are highly accurate
for all values of $n\geq 1$.
An explicit example of this will be provided below.
More generally, however, this expression is smooth and well-behaved for all values of
the $B$ and $C$ parameters,
and in all cases exhibits the exponential Hagedorn-like behavior
whose primary effects we seek to analyze in this paper.
For $n=0$, by contrast, we shall define $\hat g_0\equiv 1$, 
representing the unique ground state of our flux tube.

\subsection{Physical interpretation of ensemble parameters}

Thus far we have introduced four parameters to describe our dark ``hadron'' DDM ensemble:
$M_s$, $M_0$, $B$, and $C$.~
The first two parameters have immediate interpretations:  $M_0$ is the mass of the lightest
state in the DDM ensemble, while $M_s$ parametrizes the splitting between the states.
We would now like to develop analogous physical interpretations of $B$ and $C$.~

Clearly $B$ and $C$ describe the dynamics of the flux tube.
However, in the case of the 
ordinary strong interaction,
many possible theories governing this dynamics have been proposed. 
These range from early examples such as 
the scalar (Nambu) string~\cite{Nambu},
the Ramond string~\cite{Ramond}, 
and the Neveu-Schwarz~(NS) string~\cite{NS} 
to more modern examples such as Polyakov's ``rigid string''~\cite{Polyakov},
Green's ``Dirichlet string''~\cite{Green}, 
and the Polchinski-Strominger ``effective string''~\cite{PolchinskiStrominger}.
Many other possibilities and variants have also been proposed.

All of these theories begin by imagining a one-dimensional line of flux energy (\ie, a string) which sweeps 
out a two-dimensional flux-sheet (or worldsheet) as it sweeps through an external $D$-dimensional spacetime.    
Here $D$ is the number of 
spacetime dimensions which are effectively uncompactified with respect to the fundamental energy 
scale $M_s$ associated with the flux tube.
As such, as it propagates,
our string/flux tube is free to fluctuate into any of the \mbox{$D_\perp\equiv D-2$} spatial dimensions  
transverse to the string.
We can describe such fluctuations 
by specifying $D_\perp$ embedding functions $X^i(\sigma_1,\sigma_2)$, \mbox{$i=1,.., D_\perp$},
which are nothing but the transverse spacetime locations 
of any point on the flux-tube worldsheet with coordinates $(\sigma_1,\sigma_2)$.
As such, these embedding functions may be regarded as fields on the two-dimensional flux-tube worldsheet.
The dynamics of this system is then governed by the Polyakov action
\beq
       S ~\sim ~  M_s^2 \int d^2\sigma \,
     \sum_{i=1}^{D_\perp}
            \left( {\partial\over \partial \sigma^\alpha} X^i\right)
            \left( {\partial\over \partial \sigma_\alpha} X^i\right) ~.
\label{scalarstring}
\eeq
Minimizing this action is classically equivalent to minimizing the area of the flux-tube worldsheet.

By itself,
the expression in Eq.~(\ref{scalarstring}) describes the action of the 
so-called $D_\perp$-dimensional ``scalar'' string.
In some sense this theory provides the simplest possible description of a strongly-interacting flux tube,
with the term in Eq.~(\ref{scalarstring})
representing the bare minimum that must always be present for any flux-tube description.
The various possible refinements of this basic theory then differ in the extra terms that might be 
added to this action.
Some of these theories mentioned above introduce extra terms which correspond to additional, purely internal degrees 
of freedom [\eg, additional fields analogous to $X^i(\sigma_1,\sigma_2)$ but without interpretations as the coordinates of uncompactified spacetime dimensions]
on the flux-tube worldsheet.
By contrast, other theories introduce extra interaction terms for the $X^i$-fields which 
alter their short-distance behavior.

The action in Eq.~(\ref{scalarstring}) can be interpreted as that of a two-dimensional (2D) field theory
(where the two dimensions are those of the flux-tube worldsheet),
and we immediately see that it is endowed with a 2D conformal symmetry.
There are good reasons
to expect that the long-distance limit of any self-consistent flux-tube theory should exhibit such a symmetry,
since we expect the physics of this system to be invariant under reparametrizations of our 
flux-tube worldsheet coordinates.
As a result, those flux-tube theories that augment the scalar string by introducing extra  
purely internal degrees of freedom on the flux-tube worldsheet must not break this conformal symmetry;
this requirement constrains what kinds of terms can be added.
By contrast, the theories that introduce extra interaction terms for the $X^i$ fields
do break this conformal symmetry,
but they do so only in the short-distance limit.
The 2D conformal symmetry of the long-distance limit is then preserved as an effective symmetry.

In any 2D conformal field theory, either exact or effective,
the total number of degrees of freedom 
is encoded within the so-called {\it central charge}\/ $c$.
Each $X^i$ field contributes a central charge $c=1$, and thus the minimal scalar-string action 
in Eq.~(\ref{scalarstring}) describes a theory with central charge $c=D_\perp$.
However the introduction of additional degrees of freedom on the flux-tube worldsheet will 
necessarily increase the central charge, producing a theory with $c>D_\perp$.

Given a particular action for our flux-tube dynamics, it is straightforward to quantize 
the fields in question.   In this way, we can determine the corresponding spectrum of the
theory at all mass levels.    
These calculations are standard in string theory (see, \eg, Ref.~\cite{StringReviews}), and ultimately
one obtains~\cite{HR,KV,missusy} asymptotic state degeneracies $\hat g_n$ 
of the forms given in Eq.~(\ref{eq:fnForLargen}) or 
Eq.~(\ref{eq:fnBetterApprox}).
Remarkably, one finds a relatively straightforward connection between
the parameters $(B,C)$ appearing in our state degeneracies and the parameters $(D_\perp,c)$ of
our underlying flux-tube theory~\cite{Cudell,HR,KV,missusy}:
\beq
    \begin{cases}
       B = {1\over 4}(3+D_\perp) \cr
       C = \pi \sqrt{2 c/3}~.\cr
     \end{cases}
\label{stringvars}
\eeq
Indeed, for any value of $B$ and $C$, we may regard the total central charge $c$ as having two contributions:
one contribution $c_{\rm fluc}= D_\perp$ associated with the degrees of freedom associated with the
transverse uncompactified spacetime fluctuations of the flux tube,
and a remaining contribution 
\beq
          c_{\rm int} ~\equiv~  c - D_\perp 
            ~\equiv~ {3 C^2 \over 2\pi^2} - 4B + 3  
\label{stringvars2}
\eeq
associated with those additional, purely internal degrees 
of freedom which might also exist within the full flux-tube theory
(including those associated with any {\it compactified}\/ spacetime dimensions which may also exist).

At first glance, it might seem that our dark sector must have $D_\perp=2$, just as
does our visible sector.  
This would certainly be true if our dark-sector flux tube 
were to experience the same spacetime geometry as does the visible sector.
However, we emphasize that in a string-theoretic or ``braneworld'' context, 
the dark sector could correspond to physics in the ``bulk'' --- \ie, physics 
perpendicular to the brane on which  the visible-sector resides.
The degrees of freedom in the bulk would then be able to interact with those on the brane 
at most gravitationally, 
and would thus constitute dark matter by construction.  
However, the geometric properties of the bulk will generally differ from those of the brane --- the bulk
might contain not only extra spacetime dimensions which are effectively large (\ie, uncompactified)
with respect to the fundamental string scale, 
but also extra spacetime dimensions which are small (\ie, compactified).
The bulk may also be populated by additional fields with no spacetime interpretations at all.
It is for this reason that we make no assumptions about the values of $c$ or $D_\perp$ associated
with the dark sector.

Once our flux-tube theory is specified and the corresponding values of $B$ and $C$ determined, 
we may calculate the corresponding effective static-quark potential $V(R)$
between two quarks a distance $R$ apart.
We find~\cite{Cudell}
\beqn
         V(R) &=& \left({M_s\over 2\pi}\right) \sqrt{ (M_s R)^2 - (C/2)^2}\nonumber\\
           &\approx& 
                {M_s^2 R\over 2\pi} - {C^2\over 16\pi} {1\over R} ~+~...~~~~ {\rm for}~ R\gg M_s^{-1}~.~~~~~~
\eeqn
The first term in the final expression indicates a linear confinement potential, as expected;
this is nothing but the classical energy in the flux tube.
By contrast, the second term resembles a Coulomb term but is actually 
an attractive universal quantum correction (or Casimir energy) which arises due to
the transverse zero-point vibrations of the flux tube.
  
For visible-sector hadrons, it is natural to take $D=4$.
As a result, the $D_\perp=2$ scalar string with $c_{\rm int}=0$
(corresponding to $B=5/4$ and $C=2\pi/\sqrt{3}\approx 3.63$)
is the ``minimal'' string that we expect to underlie all descriptions of the actual visible-sector QCD flux tube.
In fact, 
it has been shown in Ref.~\cite{Cudell} that this minimal $D_\perp=2$ scalar string
with $\kappa=36$ provides an excellent fit to hadronic data, both for low energies (which are sensitive to the Casimir
energy within the confinement potential) as well as high energies (which are governed by the 
asymptotic degeneracy of hadronic states and the corresponding Hagedorn temperature).
As discussed in Ref.~\cite{Cudell}, this success --- coupled with
the appearance of the same quantity $C$ in both places ---
provides a highly non-trivial test of the classical
conformal invariance of the QCD string.

In this paper, we shall imagine that our DDM ensemble of dark-sector hadrons mimics that of the visible-sector
hadrons to the extent that it corresponds to a set of masses $M_n$ and state degeneracies $\hat g_n$ 
parametrized by the functional forms given in Eqs.~(\ref{eq:MassSpectrum})
and (\ref{eq:fnBetterApprox}).
However, we shall not insist on an actual string interpretation governing our dark-sector confinement dynamics, 
and as discussed above we shall therefore regard
$B$ and $C$ as free parameters which may be adjusted at will (subject to certain constraints to be 
discussed below).  Nevertheless it is only when $B$ and $C$ correspond
to appropriate values of $D_\perp$ and $c$ via the relations in Eq.~(\ref{stringvars})
that we may describe our resulting spectrum as 
corresponding to that of a classically self-consistent string moving in a specific geometry. 
Moreover, motivated by our experience with visible-sector hadrons, we shall continue to regard the special 
scalar-string case with \mbox{$B=5/4$} 
and \mbox{$C= 2\pi/\sqrt{3}$} as our ``minimal'' theory, corresponding to the action in Eq.~(\ref{scalarstring}) 
with \mbox{$D_\perp=2$}.
Adjusting the value of $B$ above or below $5/4$ can then be interpreted as changing the effective number of 
uncompactified spacetime dimensions felt by our dark-sector flux tube
(\ie, the number of uncompactified spacetime dimensions into which it can experience fluctuations), 
while increasing the value 
of $C$ beyond $2\pi/\sqrt{3}$ corresponds
to introducing additional purely internal degrees of freedom with central charge $c_{\rm int}$ 
into our flux-tube theory.

Note, in this regard, that the degrees of freedom associated with fluctuations
into extra {\it compactified}\/ spacetime dimensions count towards $c_{\rm int}$ rather than $D_\perp$.
Thus, in terms of its effects on the dark sector, 
the act of compactifying a spacetime dimension to a radius below the associated string scale
preserves the central charge $c$ (and thus the 
coefficient $C$)
and merely shifts 
the associated degrees of freedom from $D_\perp$ to $c_{\rm int}$.
The resulting change in the asymptotic state degeneracies $\hat g_n$
due to the change in $B$ then reflects the appearance of new Kaluza-Klein resonances
in the total flux-tube spectrum.

\subsection{Constraints on parameters}

Even though $M_s$, $M_0$, $B$, and $C$ are henceforth to be viewed as unrestricted quantities
parametrizing our hadron-like DDM ensemble, 
they are nevertheless
subject to certain self-consistency constraints.

First, we note that while the asymptotic form for $\hat g_n$ in 
Eq.~(\ref{eq:fnBetterApprox}) is remarkably accurate within those regions of $(B,C)$ parameter
space for which actual string realizations exist,
there are other regions of $(B,C)$ parameter space within which this approximation provides unphysical
results.   
For example,
given that the expression for $\hat g_n$ in Eq.~(\ref{eq:fnBetterApprox}) 
multiplies a growing Bessel function against a falling monomial,
for any given value of $B$
it is in principle possible for there to
exist a critical value of $C$ 
below which $\hat g_n$ is not always monotonically increasing for all $n\geq 0$.
Such a situation is clearly unphysical, 
implying that the number of accessible flux-tube states fails to grow
with the total energy in the flux tube. 
We therefore demand that 
\beq
    \hat g_{n+1} ~>~ \hat g_{n} ~~~~ {\rm for~ all}~ n\geq 0~.
\label{strongconstraint}
\eeq
Given that we have taken $\hat g_0=1$,
it turns out throughout the parameter range of interest that this 
requirement is tantamount to demanding 
\beq
        \hat g_1~ > ~1~.
\label{weakconstraint}
\eeq

If we further wish to demand that our ensemble of dark ``hadrons'' admit a string-theoretic 
description, then certain additional consistency conditions on the parameters $B$ and $C$ 
must be satisfied as well.  For example, since $D_\perp \in \IZ > 0$ in any self-consistent
string construction, we must have 
\begin{equation}
  B ~\in~ \IZ/4 ~>~ 3/4~.
\label{eq:BStringConsistCond}
\end{equation}
Likewise, as discussed above, any self-consistent string theory 
will also have $c \geq D_\perp$ (or $c_{\rm int}\geq 0$), which in turn
implies 
\begin{equation}
  C^2 ~\geq~ \frac{2\pi^2}{3}(4B - 3)~.
  \label{eq:CStringConsistCond}
\end{equation}
There are, of course, further string-derived constraints that might be imposed.  For example,
the allowed set of worldsheet central charges $c$
that can be realized in such non-critical string theories  
depends crucially on the types of string models under study and the types of
conformal field theories used in their constructions.
However, the constraints in Eqs.~(\ref{eq:BStringConsistCond}) and (\ref{eq:CStringConsistCond})
can be taken as a minimal model-independent set of constraints
that must be satisfied as a prerequisite to any possible string interpretation.

\begin{figure}[t]
\centering
  \includegraphics[width=0.45\textwidth]{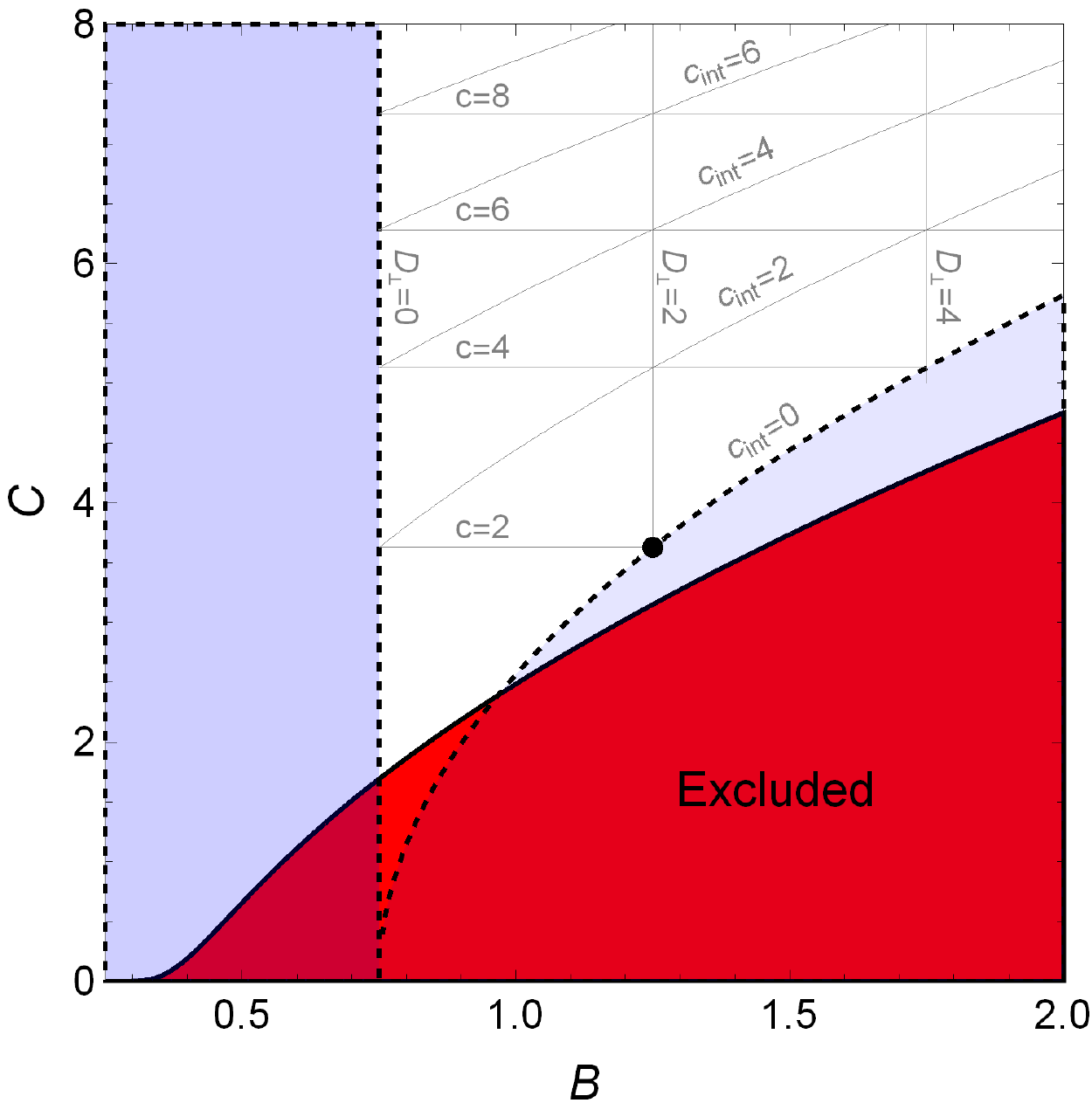}
\caption{The region of $(B,C)$ parameter space of interest for a DDM ensemble of
  dark ``hadrons.''  The red shaded region is excluded by the theoretical
  self-consistency condition $\hat g_1 \geq 1$.  By contrast, the blue shaded regions
   are excluded by the constraint $B>3/4$ 
   as well as by the constraint in 
       Eq.~(\ref{eq:CStringConsistCond}),
   and thus correspond to regions in which
  it would not be possible to interpret the ensemble constituents 
  as the states of a quantized string.
   Note that locations for which $B\not\in \IZ/4$ would also suffer
   from this difficulty.
    Within the (unshaded) string-allowed region, we have indicated 
     contours of $D_\perp$, $c$, and $c_{\rm int}$, 
    as defined in Eqs.~(\ref{stringvars}) and (\ref{stringvars2}).
     The black dot indicates the point in 
  parameter space corresponding to the minimal \mbox{$D_\perp=2$} scalar string
    with $c_{\rm int}=0$.
   As demonstrated in Ref.~\protect\cite{Cudell}, 
  this model provides the best fit to the visible hadron spectrum.}
\label{fig:g1BCExclusion}
\end{figure}

In Fig.~\ref{fig:g1BCExclusion}, we indicate the region of $(B,C)$ parameter space which 
is consistent with the constraints in Eqs.~(\ref{weakconstraint}),
(\ref{eq:BStringConsistCond}), and
(\ref{eq:CStringConsistCond}).
We emphasize that the first of these constraints
must always be satisfied as a matter of internal self-consistency.  
By contrast, as discussed above, the latter two conditions 
need to be satisfied only if one imposes the additional stipulation that our ensemble of 
dark ``hadrons'' admit a string-theory description.  
We observe in this connection that the first constraint 
is always weaker than the remaining string-motivated
constraints.  In other words,  a string-based description with $B\in \IZ/4\geq 1$ is always 
guaranteed to have monotonically growing degeneracies $\hat g_n$.
In Fig.~\ref{fig:g1BCExclusion} 
we also highlight 
the point $(B,C) = (5/4, 2\pi/\sqrt{3})$
corresponding to the ``minimal'' \mbox{$D_\perp=2$} scalar string.
While this theory need not necessarily
provide the best-fit description for our dark hadrons 
(as it does for the visible hadrons),
its minimality nevertheless provides a useful benchmark for exploring the parameter 
space of our DDM model.
Finally, we observe from
Fig.~\ref{fig:g1BCExclusion}
that our combined constraints imply that
\beq
           C ~\gsim~ 1.693~.
\label{Cmin}
\eeq
Indeed, this is the allowed range in $C$ for which $\hat g_1 > 1$ when $B=3/4$.

\begin{figure}[t!]
\centering
  \includegraphics[width=0.425\textwidth]{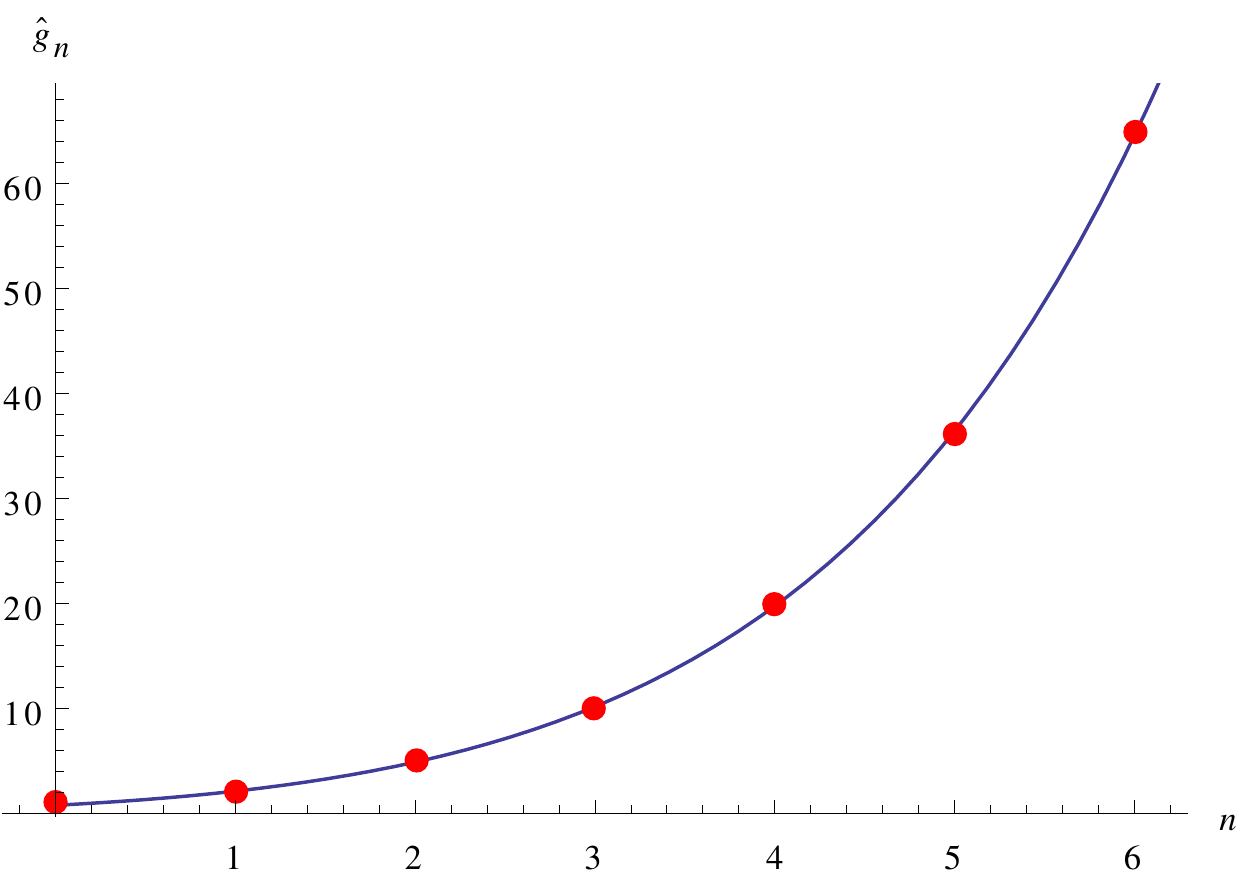}
\caption{State degeneracies $\hat g_n$ for the \mbox{$D_\perp=2$} 
           scalar-string flux-tube model of Eq.~(\ref{scalarstring}) (red circles),
       with the asymptotic functional form in Eq.~(\ref{eq:fnBetterApprox}) superimposed (blue line).
     It is clear that our asymptotic functional form succeeds in modelling the state degeneracies extremely
       accurately all the way down to the ground state, as we shall require for our analysis.}
\label{comparison}
\end{figure}

As an illustration of the results of this section, let us focus
further on this  ``minimal'' \mbox{$D_\perp=2$} scalar string. 
As noted above, the action for this string is given in Eq.~(\ref{scalarstring}).
Quantizing this theory then gives rise to a discrete spectrum of states
whose exact degeneracies are\footnote{These degeneracies $\hat g_n$ may be
   extracted as the coefficients of $q^n$ in a small-$q$ power-series expansion of
   the infinite product \mbox{$\prod_n (1-q^n)^{-2}$}.   With only minor modifications
    and a proper physical definition for $q$,
   this infinite product turns out to be the partition function of the \mbox{$D_\perp=2$} scalar string
   theory in Eq.~(\ref{scalarstring}).}
\mbox{$\hat g_{n\geq 0} =\lbrace 1,2,5,10,20,36,65, 110, 185, ...\rbrace$}.
Indeed it is only because of the existence of a quantized string formulation that we are even
able to calculate the degeneracies of the corresponding ensemble from first principles.
However, as we have asserted, these
degeneracies are extremely well approximated by 
the expression in  
Eq.~(\ref{eq:fnBetterApprox}) with \mbox{$(B,C)=(5/4, 2\pi/\sqrt{3})$}.
This is shown in Fig.~\ref{comparison}, where we plot both the discrete exact degeneracies $\hat g_n$
and the approximate functional form in Eq.~(\ref{eq:fnBetterApprox}).
As evident from Fig.~\ref{comparison}, our functional form matches these discrete 
values of $\hat g_n$ extremely well for all values of $n\geq 0$ --- even though the degeneracies
$\hat g_n$ are necessarily integers and even though our functional form was 
originally designed to be accurate only in the asymptotic $n\to\infty$ limit! 
Indeed, as claimed above, this functional form 
is accurate to within two percent over the entire range of $n$.
This demonstrates the power of the functional form we have adopted, as 
well as the utility of an underlying string formulation for our flux tube.

\section{Lifetimes and cosmological abundances for hadronic DDM ensembles}
\label{sec:Balancing}


In the previous section, we discussed the spectra of our dark ``hadronic'' DDM ensembles.
Our next step, then, is to consider the lifetimes and cosmological abundances of the individual states within
these ensembles.

\subsection{Cosmological abundances}

As we have seen, the degeneracy of states $g_n$ for our ensemble of dark ``hadrons'' 
grows exponentially with the mass of the state, with asymptotic behavior
$g_n\sim e^{\sqrt{n}} \sim e^{M_n/M_s}$.   
This exponential rise in the state degeneracies 
places severe constraints on the possible, physically consistent 
cosmological production mechanisms 
by which the corresponding abundances $\Omega_n$ 
might be established.
Indeed, unless the corresponding abundances $\Omega_n$ fall sufficiently rapidly with $n$,
our ensemble is likely to encounter severe phenomenological difficulties.

Fortunately, our interpretation of the individual components 
of such an ensemble as dark hadrons 
suggests a natural mechanism
through which the corresponding abundances $\Omega_n$ 
are generated with an exponential
suppression factor capable of overcoming this exponential rise in $g_n$.
As we have discussed, we have been imagining that these dark ``hadrons'' 
emerge as the result of a dark-sector confining phase transition
triggered by the strong interactions of some dark-sector gauge group $G$.
This phase transition occurs when the temperature $T$ 
in the dark sector drops below the critical temperature $T_c$ associated 
with this phase transition.  
This event marks the time $t_c$ at which the primordial abundances of our
individual hadrons are established.
Moreover, it is reasonable to assume that residual $G$
interactions establish thermal equilibrium among these hadrons at
$T \sim T_c$. 
Thus, the primordial abundances $\Omega_n$ of our hadrons can be  
assumed to follow a Boltzmann distribution at $t=t_c$:
\beq
 \Omega_n(t_c) \equiv {\rho_n(t_c)\over \rho_{\rm crit}(t_c)} =
   {1\over 3 \widetilde M_P^2 H(t_c)^2}\,
   \int {d^3 {\bf p} \over (2\pi)^3} \, 
     E_{\bf p} \, e^{-E_{\bf p} /T_c}~ 
\label{eq:OmeganPrimordial}
\eeq
where $E_{\bf p}\equiv \sqrt{ {\bf p}\cdot{\bf p} + M_n^2 }$
and $\rho_{\rm crit}(t) \equiv 3 \widetilde M_P^2 H(t)^2$ 
where $\widetilde M_P\equiv M_P/\sqrt{8\pi}=1/\sqrt{8\pi G_N}$ 
is the reduced Planck mass and $H(t)$ the Hubble parameter.
Indeed, we may equivalently regard these abundances as emerging from
an infinitely rapid succession of thermal freeze-outs.
Evaluating Eq.~(\ref{eq:OmeganPrimordial}) explicitly,
we find
\beqn
  \Omega_n(t_c) &=&  X \bigg\lbrace (M_n T_c)^2 \, K_2(M_n/T_c) \nonumber\\
  &&  + \half \, M_n^3 T_c\,  \bigg\lbrack
           K_1(M_n/T_c) + K_3(M_n/T_c) \bigg\rbrack\bigg\rbrace~~~~~~~~
\label{Besselexact}
\eeqn
where
$K_\nu(z)$ are modified Bessel functions of the second kind and where
$X\equiv [6\pi^2 \widetilde M_P^2 H(t_c)^2]^{-1}$
is a common overall multiplicative factor.

In general, a given state with mass $M$ produced at temperature $T_c$ will
be non-relativistic (behaving like massive matter) 
if $T_c\lsim M$ and relativistic (behaving like radiation) otherwise.
In such limiting cases, the abundances in Eqs.~(\ref{eq:OmeganPrimordial})
and (\ref{Besselexact})
take the simplified forms
\beq
   \Omega_n(t_c)  \approx
     \begin{cases}
     \sqrt{\pi/2}  \,X M_n  (M_nT_c)^{3/2}\, e^{-M_n/T_c} &   \mbox{non-rel} \cr
       6 X T_c^4    &   \mbox{rel} 	~.\cr
       \end{cases}
\label{reducedforms}
\eeq

At first glance, it may seem
that any value for $T_c$ might be phenomenologically permissible.
However, this production mechanism can only be self-consistent if it injects
a finite total energy density into our system.
In other words, as a bare minimum, we must require that
\beq
 \Omega_{\rm tot}(t_c) ~\equiv~ \sum_{n=0}^\infty g_n \Omega_n(t_c)~ < ~ \infty~.
\label{finitetotal}
\eeq
However, this condition is sensitive to the behavior of the abundances 
$\Omega_n(t_c)$ for extremely
large $n$, corresponding to states which are non-relativistic.
For such states, we see from Eq.~(\ref{reducedforms}) that $\Omega_n(t_c)\sim e^{-M_n/T_c}$.
With $g_n\sim n^{-B} e^{C\sqrt{n}}$ as $n\to\infty$,
we find using Eqs.~(\ref{eq:MassSpectrum})
and (\ref{eq:OmeganPrimordial})
that Eq.~(\ref{finitetotal}) can only hold if
\beq
          { T_c\over M_s} ~\leq ~ {1\over C}~.
\label{Tclimit}
\eeq
This then becomes a hard bound on the allowed values of $T_c$, one which
ensures that the Boltzmann exponential suppression factor in 
Eq.~(\ref{eq:OmeganPrimordial}) ultimately  overcomes the exponential rise in
the degeneracy of states $g_n$.
Indeed, Eq.~(\ref{Tclimit}) reflects nothing more than  
the statement that $T_c\leq T_H$, where $T_H\equiv M_s/C$ is the Hagedorn temperature
of our dark ensemble.
For the {\it visible}\/ hadronic sector, one often assumes that $T_c$ and $T_H$ are related
to each other parametrically, with $T_c$ either directly identified as $T_H$ 
or positioned not too far below $T_H$.  We shall implicitly make
the same assumption for the dynamics of our dark sector as well.

The next question is to determine which of our ensemble components are produced
relativistically or non-relativistically at $T=T_c$.
To do this, we shall henceforth assume that
$T_c, M_s, M_0 > T_{\mathrm{MRE}}$ where
$\tMRE$ and $\TMRE$ are the time and temperature associated with matter-radiation equality.
This assumption, which parallels what occurs for the hadrons of the visible sector,
ensures that our abundances $\Omega_n(t)$ are
established during the radiation-dominated era prior to matter-radiation equality
and that all ensemble constituents have become
effectively non-relativistic by $\tMRE$.
Note that the assumption that $T_c>\TMRE$ follows from our expectation that our dark degrees of freedom
prior to $t_c$ (\ie, prior to ``hadronization'' in the dark sector) are likely
to be relativistic, thereby 
reinforcing the radiation-dominated nature of the era prior to $T_{\rm MRE}$
and making matter-radiation equality impossible to achieve
using only visible-sector matter, as would have been required had we taken $T_c<\TMRE$.
Similarly, the assertion that $M_s>T_c$ follows directly from our assumption that
$T_c>\TMRE$, given the constraints in Eqs.~(\ref{Cmin}) and (\ref{Tclimit}).
Finally, although it is not impossible to imagine self-consistent scenarios
in which $M_0<\TMRE$,
taking $M_0>\TMRE$ also helps to preserve $\tMRE$ at its standard cosmological value.
We shall nevertheless make no assertion regarding the relative sizes of $M_0$ and $T_c$.

The above assumptions 
enable us to determine 
which of the components
of our ensemble are relativistic or non-relativistic
at $T=T_c$.
To do this, we simply compare $T_c$ against the ensemble masses $M_n$
given in Eq.~(\ref{eq:MassSpectrum}).
Given the constraint in Eq.~(\ref{Tclimit}),
it is straightforward to demonstrate that
\beq
    T_c ~\leq~ {M_s\over C} ~\leq~ { M_1\over C}~.
\eeq
Since $C>1$ [as follows from Eq.~(\ref{Cmin})],
we conclude that {\it all
of our ensemble
components
with $n\geq 1$ are necessarily non-relativistic at $t=t_c$.
By contrast, the $n=0$ component
will be relativistic at $t=t_c$ if $T_c \gsim M_0$, and non-relativistic otherwise.}

Eq.~(\ref{eq:OmeganPrimordial}) describes 
the abundances of our dark-sector hadrons at the time $t_c$ when these hadrons come
into existence as the result of a dark-sector confining transition.
However, 
once established, 
these abundances then 
evolve non-trivially with time as a result of two effects.  The first of these is
Hubble expansion; the second is particle decay.  
We shall treat each of these effects separately.

In order to evaluate
the effect of Hubble expansion on the abundances $\Omega_n(t)$, we
shall assume a standard cosmological history in which the universe remains 
radiation-dominated (RD) from 
very early times up to the time $\tMRE$ of matter-radiation equality.  We shall
also approximate the universe as matter-dominated (MD) throughout the subsequent epoch.  
In general,
we recall that 
the abundance $\Omega(t)$ of non-relativistic matter scales as $t^{1/2}$ during 
an RD epoch but remains constant in an MD epoch;
by contrast, the abundance of relativistic matter remains constant during 
an RD epoch but scales as $t^{-2/3}$ during an MD epoch.
Likewise, we recall that 
the temperature $T$ of the universe scales as $T\sim t^{-1/2}$
during RD but $T\sim t^{-2/3}$ during MD.
Thus any ensemble component of mass $M$ which is ``born'' relativistic at $T=T_c\gg M$ will eventually
transition to non-relativistic behavior as the temperature ultimately drops below $T\sim M$.

Collecting these observations, 
we then find that
the net effect of Hubble expansion 
is to rescale the original abundance of given state of mass $M$ 
by a factor which depends on whether that state was non-relativistic or relativistic 
at the time $t_c$ of its production:
\beq
       \Omega(t) ~=~ \Omega(t_c) \times
     \begin{cases}
      \sqrt{\tMRE/t_c} &  \mbox{non-rel}~\cr
      \sqrt{\tMRE/t_M} &  \mbox{rel}\cr
       \end{cases}
\label{magnifications}
\eeq
where $t_M$ denotes the time at which $T=M$.
Note that this result is valid for any time $t\geq \tMRE$.
Since it follows from our assumptions that $t_c,t_M < \tMRE$, we  
see that the abundances of all of our ensemble states are necessarily {\it enhanced}\/
before reaching the current MD era.
However, as evident from Eq.~(\ref{magnifications}),
these abundances are not enhanced 
equally:  the abundance of a non-relativistic component is enhanced more greatly than
that of any relativistic component of mass $M$ a factor $\sqrt{t_M/t_c}$. 

We have already seen that the states with $n\geq 1$ are all non-relativistic,
while the $n=0$ ground state is either relativistic or non-relativistic depending
on the value of $M_0/T_c$.
Thus, 
putting all of the pieces together, 
we find for all $n\geq 1$ that 
\beqn
  \Omega_{n}(t)  &=& 
           \sqrt{{\pi\over 2}} \, X\,  M_n (M_n T_c)^{3/2} e^{-M_n/T_c} \sqrt{ {t_{\rm MRE}\over t_c}}\nonumber\\
        &=&  \sqrt{\pi\over 2} \, X \left( {g_c\over g_{\rm MRE}}\right)^{1/4}
                             {(M_n T_c)^{5/2} \over T_{\rm MRE}} \, e^{-M_n/T_c} \nonumber\\  
        &=&  \sqrt{\pi\over 72}\, {1\over g_c^{3/4} g_{\rm MRE}^{1/4}}\, {M_n^{5/2} \over T_c^{3/2} \, T_{\rm MRE}} 
             \, e^{-M_n/T_c}~.\nonumber\\
\label{abundances_n}
\eeqn
Note that in passing to the second line we have exploited the standard time/temperature relationship 
suitable for an RD epoch, specifically\footnote{
   Note that the factor of $\sqrt{\pi/32}$ in Eq.~(\ref{timetemp}) is  
   consistent with our adoption of Boltzmann statistics in Eq.~(\ref{eq:OmeganPrimordial});
   for Bose-Einstein statistics this would instead become $\sqrt{45/16\pi^3}$.}
\beq
  t~=~ \sqrt{\pi\over 32}\, g_\ast(T)^{-1/2}\, {M_P\over T^2}~,
\label{timetemp}
\eeq 
where $g_\ast(T)$ tallies the number of effectively relativistic degrees of freedom  
driving the Hubble expansion at any temperature $T$, with 
$g_\alpha\equiv g_\ast(T_\alpha)$.
Likewise, in passing to the final line of Eq.~(\ref{abundances_n}) we have recognized 
that $H=1/(2t)$ for an RD epoch,
from which it follows that $X=1/(6 g_c T_c^4)$.

For $n=0$, however, the corresponding cosmological abundance   
is given by
\beq
  \Omega_0(t) =
\begin{cases}
         \displaystyle{\sqrt{\pi\over 72}\, {1\over g_c^{3/4} g_{\rm MRE}^{1/4}}\, 
               {M_0^{5/2} \over T_c^{3/2} T_{\rm MRE}}\,
             e^{-M_0/T_c}}      &   T_c\lsim M_0 \cr
           & \cr
         \displaystyle{{1\over g_c} \left( {g_{M_0}\over g_{\rm MRE}}\right)^{1/4} \left({M_0\over T_{\rm MRE}}\right)}
     & T_c\gsim M_0~.~~~~
\end{cases}
\label{abundances_0}
\eeq
    
As expected, the cosmological abundances in Eqs.~(\ref{abundances_n}) and (\ref{abundances_0}) depend non-trivially
on the three mass scales which parametrize our dark-hadron mass spectrum, namely
$M_0$, $T_c$, and $M_s$ (the latter appearing implicitly through $M_n$).
They also depend on the fixed mass scale $T_{\rm MRE}$.
However, if we disregard the numerical $g$-factors which appear in these results
and which only serve to parametrize the external time/temperature relationship,
we see that the {\it ratios}\/ between these abundances depend only on the {\it ratios}\/
between our input mass scales.
In particular, 
such abundance ratios are no longer anchored to a fixed external mass scale such as $T_{\rm MRE}$.
To make this point explicit, let us define the dimensionless quantities 
\beq
             r~\equiv~ {M_0\over M_s} ~~~~{\rm and}~~~~
             s~\equiv~ {T_c\over M_s}~
\eeq
and imagine that $g_\ast(T)^{1/4}$ does not change significantly between $T_c$ and $M_0$. 
(Note, indeed, that $g^{1/4}$ varies much more slowly than $g$.) 
We then find from Eqs.~(\ref{abundances_n}) and (\ref{abundances_0}) that
\beq
  {\Omega_{n\geq 1}(t)\over \Omega_0(t)} ~=~
 \begin{cases}
  \displaystyle{{(n+r^2)^{5/4} \over r^{5/2}} \, e^{-(\sqrt{n+r^2}-r)/s}} &   s\lsim r \cr
        & \cr
   \displaystyle{
    \sqrt{\pi\over 72}\, { (n+r^2)^{5/4}\over r\, s^{3/2} }\, e^{-\sqrt{n+r^2}/s}}~ & s\gsim r~.~~~~
\end{cases}
\eeq
Thus, up to an overall rescaling factor $\Omega_0$,
we see that all of our abundances $\Omega_n$ depend purely on the dimensionless ratios $r$ and $s$. 
It then follows that the cosmological abundance of each state in our dark-hadron ensemble is determined once
$\Omega_0$ is anchored to a particular numerical value
and specific values of $r$ and $s$ are chosen.
This observation will be important in what follows.

\subsection{Lifetimes and decays}

As indicated above, our derivation of the dark-sector cosmological abundances $\Omega_n(t)$ 
has thus far disregarded the effects of particle decays.
In other words, we have implicitly assumed that each ensemble component is 
absolutely stable once produced at $T_c$.  
As our final step, we shall therefore now incorporate the effects of such decays into 
our analysis.
In doing so,
we shall make several simplifying assumptions.
First, we shall assume that the net injection of energy density in the form of radiation from these 
decays has a negligible effect on the total radiation-energy density of the universe. 
Hence, this effect decouples from the effect of Hubble expansion.  Second, we
shall further assume that the contribution to the total decay width 
$\Gamma_n$ of each ensemble constituent from intra-ensemble decays is 
negligible.  In other words, we shall assume that $\Gamma_n$ is dominated by decays 
to visible-sector final states which do not include lighter ensemble constituents.
We shall discuss the consequences of relaxing this assumption in 
Sect.~\ref{sec:Conclusion}.  
Third, we shall assume that all states at a given mass level $n$ share a  
common decay width $\Gamma_n$, and that this 
width scales with $n$ across our dark-hadron ensemble according to
\begin{equation}
  \Gamma_n ~=~ \Gamma_0 \left( \frac{M_n}{M_0} \right)^{\xi}~
  \label{eq:Widths}
\end{equation}
where $M_n$ are the dark-hadron masses in Eq.~(\ref{eq:MassSpectrum})
and where $\Gamma_0$ (or, equivalently, the corresponding lifetime $\tau_0$) and 
the scaling exponent $\xi > 0$ are taken to be additional free parameters of our model.  
Thus each state in our dark-sector ensemble has a lifetime 
$\tau_n\equiv 1/\Gamma_n$ given by
\begin{equation}
  \tau_n ~=~ \tau_0 \left( \frac{n}{r^2} + 1 \right)^{-\xi/2}~.
  \label{eq:Lifetimes}
\end{equation}
Finally, for simplicity, we shall imagine that all states 
with lifetimes $\tau_n$ indeed actually decay at $t=\tau_n$.

Under these assumptions, the abundance $\Omega_n(t)$ of any
ensemble constituent at any time $t \geq t_c$ is
given by the expressions quoted above, but now multiplied by an additional decay factor
\beq
            e^{- (t-t_c)/\tau_n}~\approx~ 
            e^{- (\sqrt{n+r^2 }/r)^\xi t/\tau_0}~ 
\label{expfactor}
\eeq
where we have approximated $t \gg t_c$.   
For $s\lsim r$, we thus have
\beq
 \Omega_{n\geq 1}(t)  ~=~
  \Omega^{\rm (NR)}_0(t) \, {(n+r^2)^{5/4} \over r^{5/2}} \, {\cal E}_n^{\rm (NR)}(t)~
\label{omeganNR}
\eeq
where
\beq
{\cal E}_n^{\rm (NR)} (t)~\equiv~
   e^{-(\sqrt{n+r^2}-r)/s - [(\sqrt{n+r^2 }/r)^\xi-1] t/\tau_0}
\eeq
and where 
\beq
  \Omega_0^{\rm (NR)}(t)~=~ 
         \sqrt{\pi\over 72} {1\over g_c^{3/4} g_{\rm MRE}^{1/4}}
               \left({r\over s}\right)^{3/2}  \left( {M_0 \over T_{\rm MRE}}\right)
             e^{-r/s-t/\tau_0}~.
\label{omega0NR}
\eeq
By contrast, for $s\gsim r$, we have
\beq
   \Omega_{n\geq 1}(t) ~=~ 
    \sqrt{\pi\over 72}\, \Omega_0^{\rm (R)}(t)\, { (n+r^2)^{5/4}\over r\, s^{3/2} }
       \, {\cal E}_n^{\rm (R)} (t)
\label{omeganR}
\eeq
where
\beq
       {\cal E}_n^{\rm (R)} (t)~\equiv~
       e^{-\sqrt{n+r^2}/s - [(\sqrt{n+r^2 }/r)^\xi-1] t/\tau_0}~~
\eeq
and where
\beq
  \Omega_0^{\rm (R)}(t) ~=~ 
         {1\over g_c} \left( {g_{M_0}\over g_{\rm MRE}}\right)^{1/4} 
                      \left({M_0\over T_{\rm MRE}}\right)\, e^{-t/\tau_0}~.
\label{omega0R}
\eeq


\section{Cosmological constraints on the dark-hadron ensemble }
\label{sec:OmegaEtaWeff}


Having determined the abundances and lifetimes of each of
the individual components of our dark-hadron DDM ensemble,
we now proceed to study
the overall properties of our ensemble and its behavior as a function of time.
However, as we shall see, many of the phenomenological properties and constraints 
that apply to such an ensemble do not rest upon the properties of the individual 
ensemble components {\it per se}\/, but rather upon various aggregate 
quantities that collectively describe the ensemble as a whole.
Accordingly, in this section we shall begin by describing three aggregate quantities which
ultimately play the most important roles in characterizing and constraining such dark-hadron DDM ensembles. 
We shall then discuss of some of the most immediate cosmological constraints that 
can be placed upon these quantities.

\subsection{Total abundance, tower fraction, and effective equation of state
\label{quantities}}

Perhaps not surprisingly, the first aggregate property of a given dark-hadron DDM ensemble 
that shall concern us is its total abundance 
\beq
\Omegatot(t)~\equiv~ \sum_{n=0}^\infty  g_n \, \Omega_n(t)~=~ \kappa \sum_{n=0}^\infty \hat g_n \Omega_n(t)~.
\label{omegatot}
\eeq
Given our results in Eqs.~(\ref{omeganNR}) and (\ref{omeganR}), this total abundance takes the form
\beqn
&& \hskip -0.15 truein \Omegatot(t) ~=~ \nonumber\\
&& \hskip -0.15 truein \begin{cases}
         \displaystyle{\kappa \Omega^{\rm (NR)}_0(t)
                \left\lbrack
           1 + \sum_{n=1}^\infty   \hat g_n\, { (n+r^2)^{5/4}\over r^{5/2}}\,
                   {\cal E}_n^{\rm (NR)}(t)\right\rbrack} &   s\lsim r \cr
         \displaystyle{
         \kappa \Omega^{\rm (R)}_0(t) \left\lbrack
           1 + \sqrt{\pi\over 72} \,\sum_{n=1}^\infty   \hat g_n\, { (n+r^2)^{5/4}\over r\, s^{3/2}} \,
                   {\cal E}_n^{\rm (R)}(t)\right\rbrack} &   s\gsim r \cr
         \end{cases}\nonumber\\
\label{Omegatot}
\eeqn
where $\Omega_0^{\rm (NR,R)}(t)$ are given in Eqs.~(\ref{omega0NR}) and (\ref{omega0R}).
Indeed, we further note from Eqs.~(\ref{omega0NR}) and (\ref{omega0R}) that
\beq
       \Omega^{\rm (NR,R)}_0(t) ~=~ e^{-(t-\tnow)/\tau_0} \, \Omega^{\rm (NR,R)}_0(\tnow)~
\label{omega0timenow}
\eeq
where $\tnow\approx 4\times 10^{17}$~s denotes the current age of the universe.
We thus see from Eqs.~(\ref{Omegatot}) and (\ref{omega0timenow}) that the 
overall magnitude of $\Omega_{\rm tot}^{\rm (NR,R)}(t)$ can be viewed as being set by the
single number $\Omega_0^{\rm (NR,R)}(\tnow)$.

In characterizing the properties of our DDM ensemble and how they evolve with time, 
we are certainly interested in tracking 
$\Omegatot(t)$. 
However, we are also interested in tracking the
{\it distribution}
of this total abundance among the individual ensemble 
constituents.  One quantity of particular interest that provides essential information about
this distribution is the so-called
``tower fraction'' $0\leq \eta(t)\leq 1$ originally introduced in Ref.~\cite{DDM1}.  
This quantity is
typically defined in the DDM literature as the fraction of the abundance carried by 
all ensemble components {\it other}\/ than the dominant component,
where the dominant component is the one
making the largest 
individual contribution to $\Omegatot(t)$.  
As such, the quantity $\eta$ tracks the degree to which a single component carries
the bulk of the total abundance.   
When $\eta$ is close to zero, our ensemble effectively resembles a traditional single-component dark-matter
setup.
By contrast, when $\eta$ differs significantly from zero, 
our ensemble is more truly ``DDM-like'', with many of the ensemble constituents playing a non-trivial
role in together shaping the 
properties of the dark sector.

Such a definition for $\eta$ is appropriate in
cases in which each ensemble constituent has a unique mass and lifetime. 
Indeed, this has often been the case for the types of DDM ensembles previously studied.
However, for the dark-hadron DDM ensembles on which we are focusing here,
the states at a given Regge level $n$ have been assumed to have essentially equal masses and 
lifetimes.  Thus, in this paper, we shall adopt a modified definition for $\eta(t)$ in 
which the comparison is made between the  aggregate abundance contributions 
that accrue {\it level by level}\/ rather than state by state.
Specifically, we define
\begin{equation}
  \widehat{\Omega}_n(t) ~\equiv~ g_n\Omega_n(t) 
\end{equation}
as the aggregate cosmological abundance arising from all states at a particular oscillator level $n$.  
In terms of these aggregate abundances,
we then define 
\begin{equation}
  \eta(t) ~\equiv~ 1 - \frac{\max_n\{\widehat{\Omega}_n(t)\}}{\Omegatot(t)}~.
\end{equation}   
Thus we continue to have $0\leq \eta(t)\leq 1$, with $\eta\approx 0$ signifying a dark
sector resembling traditional single-component dark matter and
$\eta >0$ indicating (and quantifying) a DDM-like departure from this traditional scenario.
 
At first glance,
one might assume that the $n=0$ ground state(s) must always yield
the largest aggregate abundance $\widehat \Omega_n(t)$ because the primordial abundances 
$\Omega_n(t)$ for the states at all higher levels $n>0$ are exponentially suppressed by the 
corresponding Boltzmann factor in Eq.~(\ref{eq:OmeganPrimordial}).  However, for 
the DDM ensembles of dark hadrons studied here, 
it often turns out that the Hagedorn-like exponential 
growth of the degeneracies $g_n$ as a function of $n$ 
can more than compensate for the Boltzmann suppression for small values of $n$.  Indeed, 
this is true even for combinations of the ensemble parameters $B$, $C$, $r$, and $s$ which satisfy 
the consistency conditions discussed in Sect.~\ref{sec:DensityOfStates} and which yield 
a finite value of $\Omegatot(t_c)$.  
As a result of this
net balancing between these two competing exponential effects,
the level carrying the greatest aggregate  
cosmological abundance $\widehat \Omega_n(t)$  need not always be the $n=0$ ground state.
It need not even be fixed as a function of time.
This possibility must therefore be taken into account 
when evaluating $\eta(t)$.

Finally, another important quantity 
which can be taken to characterize our dark sector 
is the so-called equation-of-state parameter $w$.
For a single-component dark sector, this quantity is
nothing but the ratio between the pressure $p$ and energy density $\rho$ of the dark component:
$p=w\rho$.
However, we are dealing here with a multi-component dark sector in which each
component has its own individual lifetime and abundance.
As a result, the total 
energy density and pressure associated with our dark sector will generally experience a rather 
non-trivial time dependence
which causes our ensemble as a whole
to behave collectively as if it had a non-trivial $w$ --- even if
each individual component is taken to be pure matter with $w=0$.

To describe these collective effects, we therefore define~\cite{DDM1}
an {\it effective}\/ equation-of-state parameter $\weff(t)$ 
which describes the behavior of our ensemble as a single collective entity:
\begin{equation}
  \weff(t) ~\equiv~ -\left(\frac{1}{3H}\frac{d\log\rhotot}{dt}+1 \right)~.
\label{weffdef}
\end{equation}
Here $H$ is the Hubble parameter and $\rhotot = 3\widetilde M_P H^2\Omegatot$ is the total 
energy density of the ensemble.   
Note that the definition in Eq.~(\ref{weffdef}) is nothing but the usual definition of
$w$ prior to any assumptions of dark-sector minimality. 
As discussed above, we are primarily concerned with the evolution of the ensemble 
during the present matter-dominated epoch, within which $H(t)\approx 2/(3t)$.  Thus, 
the effective equation-of-state parameter for our DDM ensemble within this epoch 
is given by   
\begin{equation}
  \weff(t) ~=~ -\frac{t}{2\Omegatot}\frac{d\Omegatot(t)}{dt}~.
  \label{eq:wdef}
\end{equation}
As discussed in Sect.~\ref{sec:Balancing}, the only explicit dependence of 
$\Omegatot(t)$ on $t$ within a matter-dominated epoch is due to the exponential decay factor 
(\ref{expfactor}) within each individual abundance $\Omega_n(t)$.  We thus find that
\begin{equation}
  \weff(t) ~=~ \frac{t}{2\tau_0 \Omegatot(t)}\sum_{n=0}^\infty 
    g_n 
\left(\frac{\sqrt{n + r^2}}{r}\right)^\xi \Omega_n(t)~.~~
  \label{eq:weffExplicit}
\end{equation}
Note that even though each of the individual components of our ensemble has been taken to be
matter-like (with $w=0$), the collective equation-of-state parameter $w_{\rm eff}(t)$ for
our ensemble as a whole is {\it positive}\/, reflecting the fact that the ensemble 
as a whole
is continually losing abundance as its individual components decay.
Indeed, it is only in the $\tau_0\to\infty$ limit that $w_{\rm eff}(t)\to 0$.
As we shall see in Sect.~\ref{sec:Constraints}, $\weff(t)$ plays an important
role in constraining the parameter space of these DDM ensembles.


\subsection{Cosmological constraints \label{sec:Constraints}}


Given our time-dependent aggregate quantities $\Omegatot(t)$, $\eta(t)$, and $w_{\rm eff}(t)$,
we now turn to the cosmological constraints that bound these functions.
In this way, we shall ultimately be placing non-trivial 
constraints on the parameter space underlying these hadronic DDM ensembles.

In this connection,
we again stress that our aim in this paper is not to perform a detailed analysis of 
the astrophysical and/or cosmological constraints on this parameter space.  Such a detailed analysis
would clearly be an important but extensive task which is beyond the scope of this paper.
Moreover, such an analysis would require 
a host of further assumptions concerning the particular nature of our ensemble,
the specific decay modes of its constituents into SM states, and so forth. 
Rather, in this paper, our goal is to simply to obtain a rough initial sense
of those regions of parameter space
in which a DDM ensemble of dark ``hadrons'' might have
at least the {\it potential}\/ of phenomenological viability.
Accordingly, in what follows, we shall put
forth a set of requirements which directly constrain
the fundamental quantities $\Omegatot(t)$,  $\eta(t)$, and $\weff(t)$
we have defined above,
but which do not require any further information concerning these hadronic ensembles
beyond those properties already discussed.
In some sense, then, these might be viewed as the immediate ``zeroth-order'' model-independent
constraints that any DDM ensemble of this sort must satisfy.

Our first constraint is an obvious one:   despite the presence of an infinite tower of dark-hadronic
resonances, each with its own cosmological abundance and lifetime,
we shall demand that
\beq
 \Omegatot(t_{\rm now}) ~=~ \OmegaCDM ~\approx~ 0.26~.
\label{constraint1}
\eeq
This requirement is clearly predicated on the assumption that our dark-hadronic ensemble 
represents the totality of the dark sector;
for other cases we would simply require that
 $\Omegatot(t_{\rm now}) \lsim 0.26$.
As we shall see, in either situation this is
a severe and unavoidable constraint which 
ultimately ``anchors'' our entire construction in terms of actual numbers and mass scales.

Second, we may also consider the {\it time-variation}\/ of $\Omegatot(t)$. 
The time-variation of this total abundance is constrained by experimental
probes which yield information about the dark-matter abundance during
different cosmological epochs.  For example, CMB 
data~\cite{Planck} provides information about the dark-matter abundance around 
the time of last scattering --- \ie, at a redshift $z \approx 1100$, or equivalently
a time of roughly $2.7 \times 10^{-5} \tnow$.  On the other 
hand, observational data on baryon acoustic oscillations~\cite{BOSS} and the
relationship between luminosity and redshift for Type~Ia supernovae~\cite{SCP}
provide information about $H(t)$ and the dark-energy abundance $\Omega_\Lambda$ at
subsequent times, down to redshifts of around $z\approx 0.5$.  Within the context 
of the $\Lambda$CDM cosmology, the agreement between these different measurements 
implies that the dark-matter abundance has not changed dramatically since the time 
of last scattering.  

In order be consistent with this result, we shall therefore
demand that the total abundance of our DDM ensemble 
not vary by more than 5\% between an early ``look-back'' time $t_{\rm LB}$
and today:
\begin{equation}
  \frac{\Omega(t) - \Omega(\tnow)}{\Omega(\tnow)} ~\leq ~ 0.05~~~~~
       {\rm for~all}~~  t_{\rm LB}\leq t \leq \tnow~.
\label{constraint2}
\end{equation}
In what follows, we shall choose a look-back time $\tLB = 10^{-6}\tnow$,
which lies comfortably before the recombination epoch.  

In addition to these constraints on the time-variation of the dark-matter abundance, 
there are further considerations which constrain the decays 
of the DDM-ensemble constituents more directly.
These constraints depend on the decay properties of the dark-sector
particles and are thus ultimately model-dependent.
However, for those rather general cases in which the ensemble constituents can decay to final states involving 
visible-sector particles, one must ensure that these decay products not disrupt 
big-bang nucleosynthesis~\cite{BBNLimits}, not produce observable distortions in the 
CMB~\cite{HuAndSilkLong,HuAndSilkShort}, not reionize the universe~\cite{Slatyer}, 
and not violate current limits on the fluxes of photons or other cosmic-ray 
particles~\cite{AMSPositron,AMSAntiproton}.  
Indeed, even if the ensemble constituents decay exclusively
into other, lighter dark-sector particles, such decays can nevertheless leave
observable imprints on small-scale structure~\cite{peter,MeiYu}, alter the scale- and 
redshift-dependence of the cosmological gravitational-lensing power 
spectrum~\cite{LensingConstraint}, and affect the luminosity-redshift relation 
for Type~Ia supernovae~\cite{Gong,Savvas}.  
Since these effects all arise from the decays of ensemble constituents,
non-observation of these effects also leads to constraints on the time-variation of
$\Omega_{\rm tot}$. 

Some of these latter constraints
admittedly depend on model-dependent aspects of the decay kinematics of the dark-ensemble constituents.
However the strongest and most general of these constraints effectively amount to limits 
on the variation of $\Omegatot(t)$ within the recent past --- \ie, for redshifts
$0 \lesssim z \lesssim 3$.  Therefore, in addition to our look-back-time constraint
in Eq.~(\ref{constraint2}),
we shall also impose an additional constraint on our effective equation-of-state parameter:
\beq
      \weff(\tnow) ~\leq ~0.05~.
\label{constraint3}
\eeq
Through Eq.~(\ref{eq:wdef}), this thus becomes a constraint on the present-day {\it time-derivative}\/
of $\Omegatot(t)$.
It is important to stress that this constraint is independent of that in Eq.~(\ref{constraint2}):  while
Eq.~(\ref{constraint2}) constrains accumulated changes in $\Omegatot(t)$ over a relatively long interval,
Eq.~(\ref{constraint3}) constrains the time-variation of $\Omegatot(t)$ near the present time. 

Other considerations will also guide our 
interest in certain regions of parameter space.
For example,
from a DDM-inspired standpoint, we are particularly
interested in scenarios for which 
\beq
     \eta(\tnow) ~\sim~ {\cal O}(1)~,
\label{constraint4}
\eeq
\ie, scenarios in which the present-day value of $\eta$ is significantly different
from zero.
This ensures that a sizable number of ensemble constituents continue to survive 
and contribute meaningfully to $\Omegatot$ at the
present time, with dark-matter decays occurring {\it throughout}\/ the present epoch and not 
just in the distant past or future.
Although Eq.~(\ref{constraint4}) is not a strict requirement 
for phenomenological consistency, this condition guides the degree to which we
may regard our ensemble as being fully DDM-like,
with a significant portion of the ensemble
playing a non-trivial role in the phenomenology of the dark sector.
For example, this condition rules out regions of parameter space
in which 
$\tau_n\ll \tau_0 $ for all $n\geq 1$, with $\tau_1 \ll t_{\rm LB}$.
In such regions of parameter space, all excited dark-hadronic states have decayed prior to our look-back
time, leaving us with a single dark-hadronic 
ground state in the present epoch.
Such a scenario trivially satisfies all of our phenomenological constraints on the time-variations
of the total dark-sector abundance,
but is effectively no different from that of a traditional, single-component dark sector.
It is thus less interesting from a DDM perspective.

There are two further phenomenological constraints which will be useful for us to consider 
in the following.
First, we shall demand that $\tau_0 \gg \tnow$.   Although we do not necessarily require $\tau_0\approx 10^9 \tnow$
as in traditional single-component dark sectors,
we generally expect that $\tau_0$ must exceed $\tnow$ by at least several orders of magnitude
in order to satisfy look-back and $w_{\rm eff}$ constraints.
This assumption will be discussed further in Sect.~\ref{sec:Results}.
Likewise, although we have thus far assumed $M_0\geq T_{\rm MRE}$ throughout our analysis, we
actually must 
impose the somewhat stronger bound
$M_0\gsim {\cal O}(10^3) T_{\rm MRE} \approx {\cal O}({\rm keV})$ 
in order to satisfy BBN and structure-formation constraints.
This last requirement implicitly assumes that our lightest ensemble component carries
the largest cosmological abundance (or at least a sizable fraction of the total cosmological
abundance), but we shall see in 
Sect.~\ref{sec:Results}
that this turns out to be true for the vast
majority of phenomenologically interesting cases.

Finally, we shall also make certain simplifying assumptions.
First, for concreteness, we shall 
restrict our attention to situations with $\xi=3$.
In other words, we shall assume that the dominant contributions to the
decay lifetimes $\tau_n$ of our DDM constituents $\phi_n$ scale
as $\tau_n\sim 1/M_n^3$ across the DDM ensemble.
Decay widths of the form $\Gamma_n\sim M_n^3/\Lambda^2$ emerge
naturally from operators 
such as $\phi_n F_{\mu\nu}F^{\mu\nu}/\Lambda$
where $\Lambda$ parametrizes the energy scale associated with such couplings
and where $F^{\mu\nu}$ denotes a field-strength
tensor associated with either the visible-sector (SM)
photon or a dark-radiation photon associated with an additional
Abelian gauge group under which the ensemble
constituents are not charged.
The contributions from such operators 
will dominate the decays of our
DDM constituents in scenarios
in which our DDM ensemble is uncharged with respect
to all SM symmetries,
and in which intra-ensemble decays can be neglected.
Likewise, we shall also make the simplifying assumption that $\kappa=1$ 
in Eq.~(\ref{eq:gnInTermsOffn}).   This restricts us to the bare ``minimal'' case
in which we do not ascribe non-trivial degrees of freedom to our dark-sector quarks,
and thereby focus exclusively on the ensemble of states generated by our infinite 
tower of hadronic resonances.
Finally, throughout our analysis, we shall continue to impose the self-consistency constraints
listed in Eqs.~(\ref{strongconstraint}) [or equivalently (\ref{weakconstraint})], (\ref{eq:BStringConsistCond}),
(\ref{eq:CStringConsistCond}), and (\ref{Tclimit}).

Thus, going forward, the free parameters governing our dark-hadron DDM ensemble may 
be tallied as follows.
First, there are the two parameters $\lbrace B,C\rbrace$ which govern the
individual state degeneracies $\hat g_n$ according to Eq.~(\ref{eq:fnBetterApprox}).
Second, there are the four parameters $\lbrace r,s,M_0, \tau_0\rbrace $ which govern the 
individual abundances $\Omega_n(t)$ in Eqs.~(\ref{expfactor}) through  (\ref{omega0R}).
However, imposing Eq.~(\ref{constraint1}) as an overall normalization condition
allows us to remove $M_0$ as a free parameter.
Thus, for the rest of this paper, we shall consider our DDM ensembles as functions of
their locations within the five-dimensional parameter space
corresponding to the variables $\lbrace B,C, r,s, \tau_0\rbrace$ where
$B\geq 1$, $C^2 \geq 2\pi^2 (4B-3)/3$, and $s\leq 1/C$.


\section{Results\label{sec:Results}}


In general, we seek to determine which values of our defining parameters
$\lbrace B, C, r, s, \tau_0\rbrace$
lead to self-consistent 
and potentially viable 
dark sectors --- \ie, sectors which satisfy our abundance, look-back, and $w_{\rm eff}$ constraints
in Eqs.~(\ref{constraint1}), 
(\ref{constraint2}), and (\ref{constraint3}) respectively,
along with our $M_0> {\cal O}({\rm keV})$ constraint.
For each such set, we also seek to determine the corresponding values of 
relevant mass scales such as the string scale $M_s$.
We also seek to  determine
the extent to which the corresponding ensemble is truly DDM-like,
with a relatively large number of component states playing a significant 
role in the phenomenology of the dark sector and contributing to $\Omega_{\rm tot}$ at the present time.
In general, the larger the value of $\eta(t_{\rm now})$, the more DDM-like the corresponding ensemble.

At first glance, it might seem rather daunting to orient ourselves within 
the five-dimensional
$\lbrace B, C, r, s, \tau_0\rbrace$
parameter space.
However, there are really two separate parts to our analysis --- 
one part which depends only on {\it relative}\/ mass scales,
and one part which makes explicit reference to {\it absolute}\/ mass scales. 
It is clear from Eqs.~(\ref{Omegatot}) and (\ref{omega0timenow})
that once we know $\lbrace B,C, r,s,\tau_0\rbrace$,
we can determine the function $\Omegatot^{\rm (R,NR)}(t)$ up to 
an overall multiplicative constant $\Omega^{\rm (R,NR)}_0(\tnow)$.
Setting   $\Omegatot^{\rm (R,NR)}(\tnow)=\OmegaCDM \approx  0.26$ therefore immediately determines a required
numerical value of $\Omega^{\rm (R,NR)}_0(\tnow)$.
This also determines the corresponding values of $\eta(\tnow)$ and
$\weff(\tnow)$.
Up to this point, we have not yet anchored our results in terms of absolute mass scales.
However, this can also easily be done:  we simply set our required numerical value 
of $\Omega^{\rm (R,NR)}_0(\tnow)$
to the expression in either Eq.~(\ref{omega0NR}) 
or Eq.~(\ref{omega0R}). 
This then determines an absolute value for the mass scale $M_0$, whereupon we find
that $M_s= r M_0$ and $T_c= (s/r) M_0$. 
Thus, in this way, we can extract the values for $M_s$ and $\eta(\tnow)$ corresponding to
every point in the  $\lbrace B,C, r,s,\tau_0\rbrace$ parameter space.

Certain observations can be made rather rapidly.
For example, given Eq.~(\ref{constraint1}), it immediately follows that
       $\Omega_0(t_{\rm now}) \lsim 0.26$ --- a bound which can be saturated only when $\eta(t_{\rm now})=0$.
More generally and more schematically, we might write this constraint in the rough 
order-of-magnitude form 
\beq
       \Omega_0(t_{\rm now}) ~\lsim~ {\cal O}(0.1)~.
\label{omega0constraint}
\eeq
However, let us now consider the expression 
in Eq.~(\ref{omega0R})
for $\Omega_0(t)$ in the relativistic case.
Since $\tau_0$ must significantly exceed $\tnow$ by at least several orders of magnitude,
as discussed in Sect.~\ref{sec:OmegaEtaWeff},
we see that the exponential factor $e^{-t/\tau_0}$ is essentially~1.
Likewise we recall that
$M_0/T_{\rm MRE}\geq {\cal O}(10^3)$, as also discussed in 
Sect.~\ref{sec:OmegaEtaWeff}.
Let us assume that 
this bound is saturated, so that
$M_0/T_{\rm MRE}={\cal O}(10^3)$.
We therefore find that Eq.~(\ref{omega0constraint}) can be satisfied only if
$g_c \sim 10^{4}$.    This would in turn require a mass scale $T_c$ which at the very minimum exceeds the TeV scale 
(thereby introducing a hierarchy between $T_c$ and $M_0$ which is at least a factor of $10^6$) 
and which actually must be so high that there are at least ten times as many effectively relativistic degrees of freedom
below this scale than are known to exist below the TeV scale --- a rather unlikely proposition
resting entirely on currently unknown physics.
Considering 
greater values of $M_0/T_{\rm MRE}$
only worsens this situation and requires even greater values of $g_c$.
Therefore, although there might exist finely tuned slivers of parameter space in which one might
manage to achieve a balancing between $g_c$ and $M_0/T_{\rm MRE}$ sufficient to satisfy Eq.~(\ref{omega0constraint}),
we shall abandon any further consideration of the relativistic case
in what follows.

This situation changes dramatically when we turn to the non-relativistic case
in Eq.~(\ref{omega0NR}).   In this case, we continue to find that $e^{-\tnow/\tau_0}\approx 1$.
However, the presence of the factor 
$(r/s)^{3/2} e^{-r/s}$ 
allows us greater freedom in satisfying
the constraint in Eq.~(\ref{omega0constraint}).
Indeed, the first thing we learn is that our system is going to 
be very sensitive to the ratio $r/s$ --- not surprising, given that this was
already the radio that determined the extent to which our lightest mode
was relativistic or non-relativistic.
However, we now see that $r/s$ is also going to play a large role
in governing the allowed values of the overall
mass scales in our system, with greater (lesser) values of $r/s$ generally corresponding
to higher (lower) absolute mass  scales for our ensemble.

We shall therefore proceed through our parameter space as outlined above,
paying special attention to the values of $r$ and $s$ and in particular to the ratio $r/s$.
Specifically, for each value of $\lbrace B, C, r,s, \tau_0\rbrace$,
we shall determine whether 
our internal consistency constraints 
$B\geq 1$, $C^2 \geq 2\pi^2 (4B-3)/3$, and $s\leq 1/C$ are satisfied
and whether the phenomenological consistency constraints in 
Eqs.~(\ref{constraint2}) and (\ref{constraint3}) are satisfied.
If so, we shall then determine the corresponding values of $M_s$ and $\eta(\tnow)$,
with the overall goal of understanding which regions of parameter space potentially
lead to viable ensembles and which subregions correspond to ensembles 
which are particularly DDM-like.

\begin{figure*}
\centering  
    \includegraphics[width=0.45\textwidth]{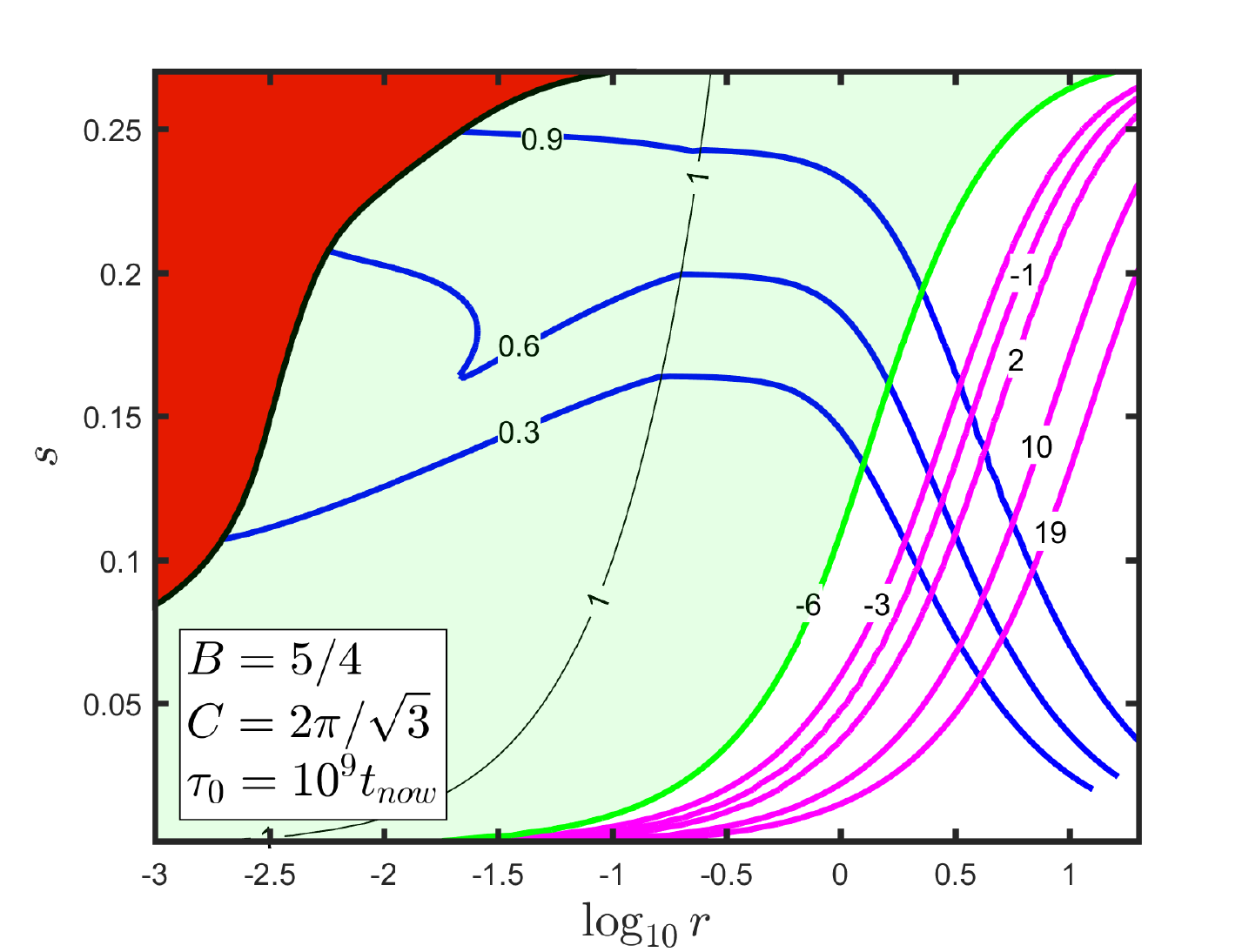}
\hskip 0.05 truein
    \includegraphics[width=0.45\textwidth]{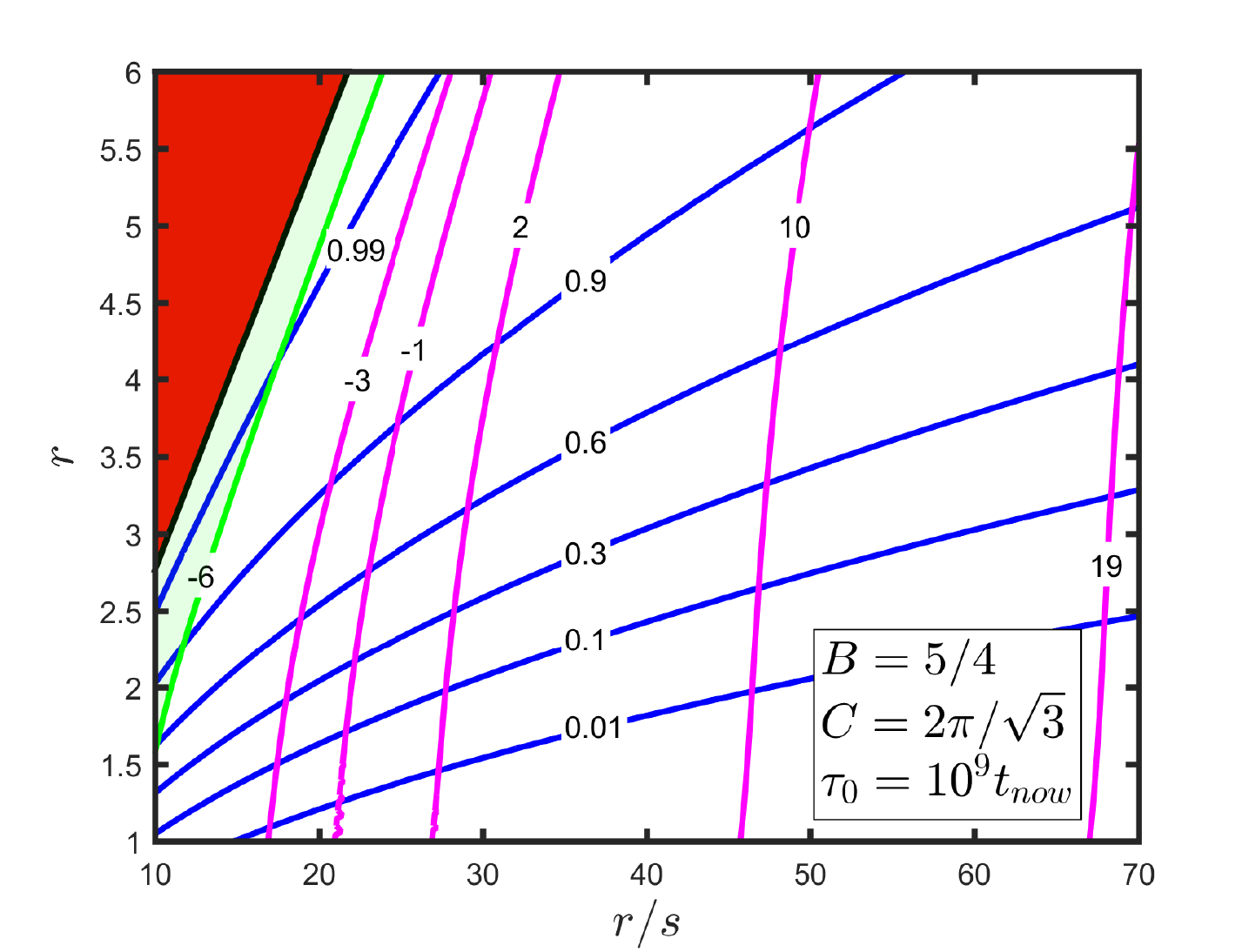}
\caption{ A survey of physics in the $(r,s)$ plane, with $B$, $C$, and $\tau_0$ set to the ``benchmark'' values
shown.  
 {\it Left panel}\/:
The thin black line labeled `1' indicates the contour with 
$r/s=1$;  this is thus the dividing line between the region
in which the lighest state is relativistic (left of this line)
versus non-relativistic (right of this line).
The blue curves indicate contours of $\eta(\tnow)$,
while the magenta lines indicate contours of $M_s$ and are labelled by the value of $\log_{10}(M_s/{\rm GeV})$.
The red region is excluded by look-back and $w_{\rm eff}$ constraints, while the pale green region
is excluded by the constraint $M_0\gsim {\cal O}({\rm keV})$ which is saturated
along the single green contour.
Increasing (decreasing) the value of 
$\tau_0$ does not affect the $M_s$  or $\eta(\tnow)$ contours, and simply shifts the red exclusion region  
to the left (right). 
 {\it Right panel}\/:
Same as left panel, but with features plotted relative to the variables $r$ and $r/s$.
The entire region shown in this panel corresponds to the non-relativistic case.  }
\label{fig3}
\end{figure*}

Because of the somewhat natural and intuitive role played by the $D_\perp=2$ scalar flux tube,
as discussed in Sect.~\ref{sec:DensityOfStates},
we shall adopt the values
\beq
         B~=~ 5/4~,~~~~~ C = 2\pi/\sqrt{3}\approx 3.63~
\label{BCbenchmarks}
\eeq
as ``benchmark'' values and begin our exploration within $(r,s)$ space.
Taking $\tau_0= 10^9 \tnow$, we find the results shown in Fig.~\ref{fig3}.

Let us first concentrate on the left panel
of Fig.~\ref{fig3}.
The red region indicates those values of $(r,s)$
which are excluded by look-back and $w_{\rm eff}$ constraints,
while the pale green region is excluded by the requirement that 
$M_0\gsim {\cal O}({\rm keV})$.
The blue curves indicate contours of $\eta(\tnow)$ and the 
magenta curves indicate contours of $M_s$, labelled by 
values of $\log_{10} (M_s/{\rm GeV})$.
The single green curve indicates the contour with $M_0=1$~keV.~
The thin black curve indicates the contour with $r/s=1$, and thus
serves as the nominal dividing line between the regions in which
the lowest ensemble state is relativistic (above and to the left)  or
non-relativistic (below and to the right). 

Several things are immediately apparent from this figure.
First, we see that the portion of the parameter space
corresponding to the relativistic case is excluded
by our constraint on $M_0$.   This is entirely in keeping with
our conclusions already reached above.
Nevertheless, we also see that beyond this region
there exists an entire area of parameter space in which all of our
constraints are satisfied.  Moreover, within this region we see that
$M_s$ varies from the keV/MeV-range all the way to the Planck
scale.  Likewise, $\eta(\tnow)$ varies through all of its possible
values.
This is therefore not only an allowed region, but one which
is likely to be exceedingly rich in phenomenology.
Indeed, given the contours plotted in this figure,
we see that the ``sweet spot'' within the $(r,s)$ parameter space 
lies roughly within the range 
\beq
    \begin{cases}
    ~1\lsim r\lsim 6\cr
    ~0.05 \lsim s \lsim 0.18~.\cr
    \end{cases}
\label{sweetspot}
\eeq
This is the region of $(r,s)$ parameter space where
the plotted blue and magenta
contours intersect each other and form a  ``cross-hatched'' region,
as illustrated in the left panel of Fig.~\ref{fig3}.
This sweet spot is therefore the region that will be of maximum  interest to us.
Indeed, within this region,
we observe from the left panel of Fig.~\ref{fig3}
that $\eta(\tnow)$ increases if either $r$ or $s$ is increased,
while $M_s$ increases in the former case but decreases in the latter.

The right panel of Fig.~\ref{fig3} focuses on this
sweet-spot region and shows  the same $M_s$ and $\eta$ contours,
only now plotted with respect to the variables $r/s$ and $s$
using a linear rather than logarithmic axis.
The fact that the $M_s$ contours are approximately vertical in this region
indicates that $M_s$ is dominantly determined by the ratio $r/s$, 
exactly as anticipated above, with increasing values of $r/s$ corresponding
to increasing values of $M_s$.   Indeed, we see from the right panel of Fig.~\ref{fig3} that $M_s$ increases
extremely rapidly as a function of $r/s$, in keeping with the 
exponential dependence in Eq.~(\ref{omega0NR}).   
Likewise, increasing the value of $r/s$ while holding $r$ fixed  
tends to {\it decrease}\/ the value of $\eta(\tnow)$.
Thus, for fixed $r$, we find that $M_s$ and $\eta(\tnow)$ tend to
vary inversely with respect to each other as functions of $r/s$, 
with our ensembles becoming less
DDM-like at higher mass scales and more DDM-like at lower mass scales.
Likewise, for fixed $r/s$, we find that increasing $r$ tends to increase $\eta(\tnow)$,
as already evident from the left panel of Fig.~\ref{fig3}.

It is easy to understand these results physically.
For fixed $r$, increasing $r/s$ corresponds to decreasing $s$.
This lowers the critical temperature $T_c$ at which our initial cosmological
abundances are established,
which has the effect of decreasing the abundances of the heavier states
relative to the lighter states.
This therefore decreases the value of $\eta(\tnow)$.
By contrast, holding $r/s$ fixed and increasing $r$ corresponds 
to increasing $s$ as well.    The increase in $r$ 
renders all of the ensemble states more massive but
provides a smaller proportional mass increase 
for the heavier states
than for the lighter states.
Thus the mass {\it ratios}\/ between heavier and lighter states decreases,
which tends to increase the value of $\eta(\tnow)$.
Likewise, as discussed above, increasing $s$ also tends to
increase the value of $\eta(\tnow)$.
These two effects then tend to reinforce each other, as evident
in Fig.~\ref{fig3}.

\begin{figure*}
\centering  
    \includegraphics[width=0.45\textwidth]{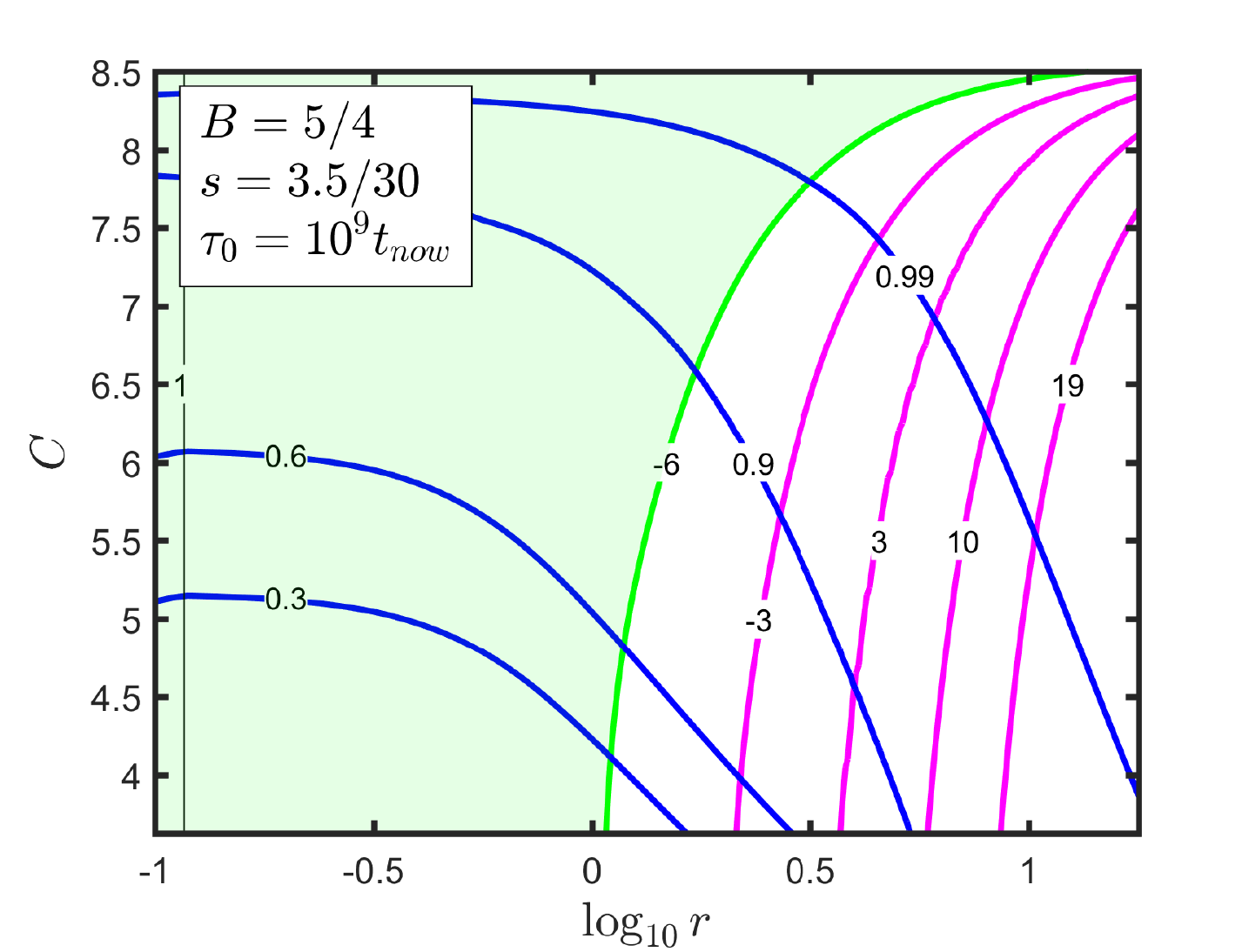}
\hskip 0.05 truein
    \includegraphics[width=0.45\textwidth]{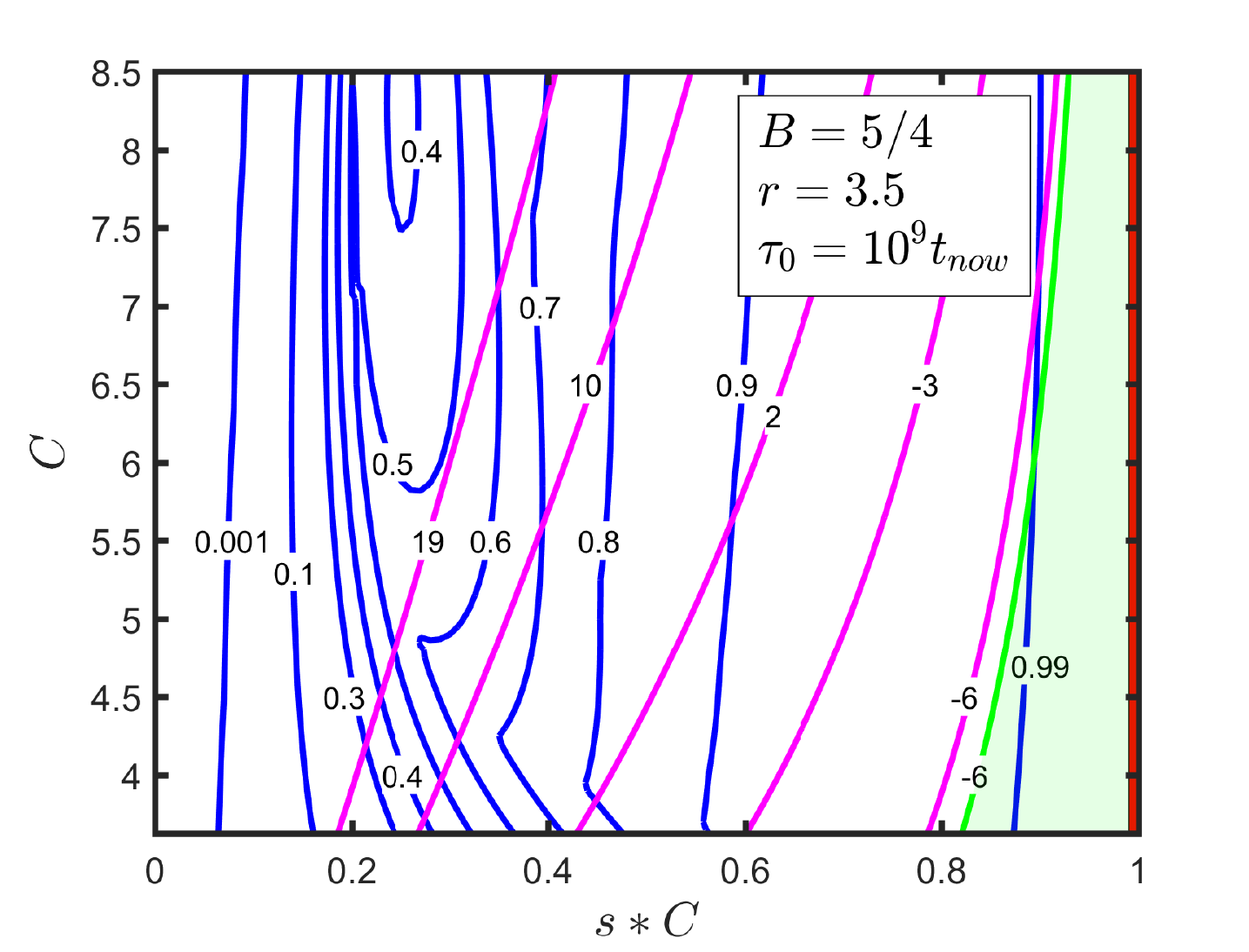}
    \includegraphics[width=0.45\textwidth]{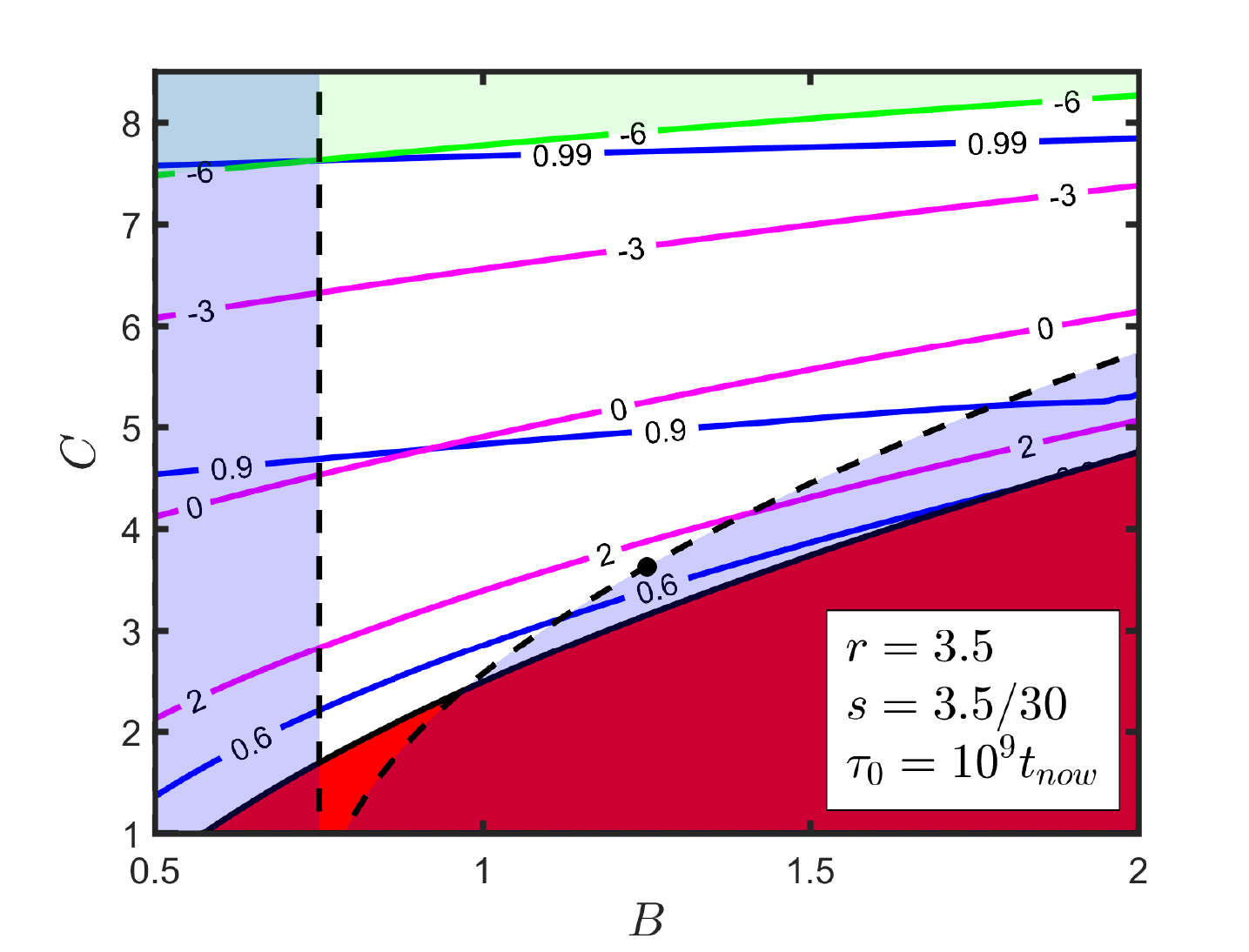}
\caption{
Contours of $\eta(\tnow)$ (blue curves) and $M_s$ (magenta curves), labelled as in Fig.~\ref{fig3} and
plotted in three different planar ``slices''
through the $(B,C,r,s)$ parameter space.
The top two panels show these contours plotted in the  $(r,C)$ and $(s,C)$ planes, respectively,
while the bottom panel shows these contours plotted in the $(B,C)$ plane.
In all panels, colored shaded regions are excluded by either string consistency constraints (blue shaded regions),
internal consistency constraints (red region in lower panel), or phenomenological look-back, $w_{\rm eff}$, or $M_0\gsim {\cal O}({\rm keV})$ 
constraints (pale green regions as well as the red region along the right edge of the upper right panel). 
As in Fig.~\ref{fig3},
the thin black vertical
 $r/s=1$ contour (visible at the extreme left of the upper left panel) continues to represent 
the boundary between the regions in which the lightest
state is either relativistic (left of the line) or non-relativistic (right of the line).} 
\label{fig4}
\end{figure*}

Having identified our sweet-spot region in $(r,s)$ parameter space,
we now investigate how these values of $M_s$ and $\eta(\tnow)$ vary
as our other parameters $B$, $C$, and $\tau_0$ are varied.
To do this, we study variations in these parameters
relative to an $(r,s)$ ``benchmark''
\beq
       r= 3.5~,~~~~~
       r/s = 30~,
\label{rsbenchmarks}
\eeq
which we henceforth take as representative of our sweet-spot region
in the $(r,s)$ plane. 
In Fig.~\ref{fig4} we illustrate the effects of 
variations in $B$ and $C$ relative to this benchmark, plotting 
contours of $M_s$ and $\eta(\tnow)$ in the $(r,C)$ plane (upper left panel),
the $(s, C)$ plane (upper right panel), and the $(B,C)$ plane (lower panel).
Note that since we must always have $s\leq 1/C$,
it is actually the normalized product $s\cdot C$ which captures the dependence
on $s$ in situations where $C$ might also be varied.
In the upper right panel we therefore plot our contours relative to
$s\cdot C$ rather than $s$ alone.  
Likewise, in the lower panel of Fig.~\ref{fig4} we have continued
to indicate our allowed regions of $B$ and $C$ as in Fig.~\ref{fig:g1BCExclusion},
where the dot continues to represent the $D_\perp=2$ scalar-string 
benchmark values in Eq.~(\ref{BCbenchmarks}).

Together, the three panels of Fig.~\ref{fig4}
tell a consistent story.
First, with $r$ and $s$ held fixed, 
we see from the upper left and lower panels of Fig.~\ref{fig4} 
that increasing $C$ generally tends to increase $\eta(\tnow)$.
This result makes sense:
increasing $C$ corresponds
to increasing the {\it degeneracies}\/ of the heavier states  
relative to the lighter states.   However, with $s$ held constant, each of
these heavier states continues to accrue the same abundance as before.
Thus increasing $C$ 
increases the total abundance carried by the heavier states
relative to that carried by the lighter states,
thereby increasing $\eta(\tnow)$.
Second, we see from the lower panel of Fig.~\ref{fig4} that 
while our values of $M_s$ and $\eta(\tnow)$ are quite sensitive to $C$,
they are far less sensitive to $B$.~   This too makes sense, since
$C$ governs the exponential rate of growth in the state degeneracies
while $B$ governs only the subleading polynomial behavior.
Third, in each of the above two cases, we also note that
increasing $C$ while holding $r$ or $B$ fixed
also corresponds to decreasing $M_s$.   Thus, once again, 
we see that $M_s$ and $\eta(\tnow)$ tend to vary inversely with each other,
giving rise to more DDM-like ensembles at lower energy scales and
more traditional ensembles at higher energy scales.

Finally, we see from the upper right panel of Fig.~\ref{fig4} that
our values of $\eta(\tnow)$ are largely {\it insensitive}\/ to variations in $C$  
as long as $s\cdot C$ is held fixed.
However, this too is easy to understand.   Increasing $C$ 
while holding $s\cdot C$ fixed corresponds to decreasing $s$ as we increase $C$.~
Increasing $C$ induces an exponential increase in the degeneracy of each massive
state, while decreasing $s$ decreases the critical temperature $T_c$, thereby inducing  a corresponding 
exponential decrease in the abundance associated with each such state.
Thus, to first approximation,
these two effects tend to mitigate each other:
they produce more states, but also cause each state to carry a correspondingly smaller abundance.
 
\begin{figure*}
\centering  
  \includegraphics[width=0.45\textwidth]{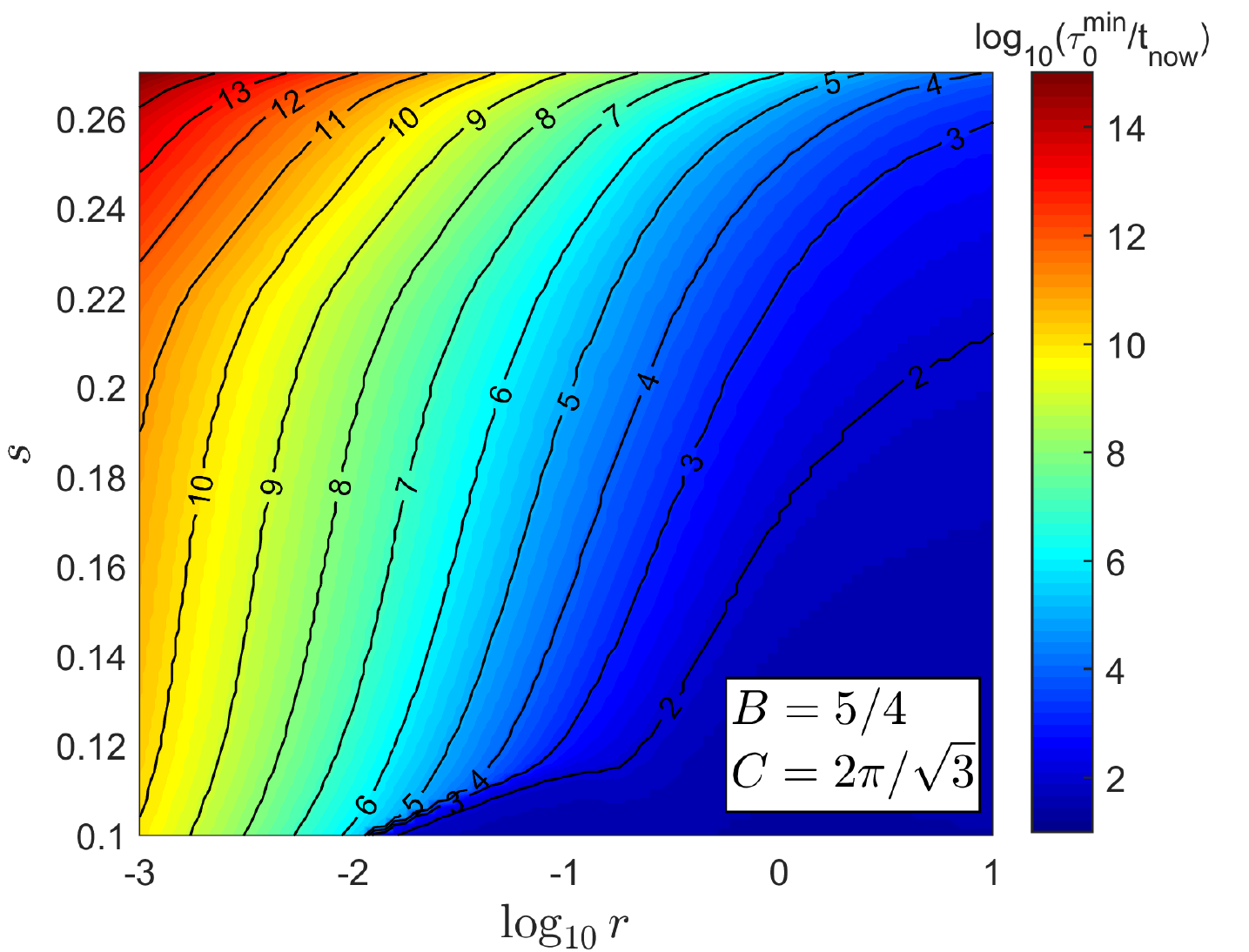}
\hskip 0.05 truein
  \includegraphics[width=0.45\textwidth,keepaspectratio]{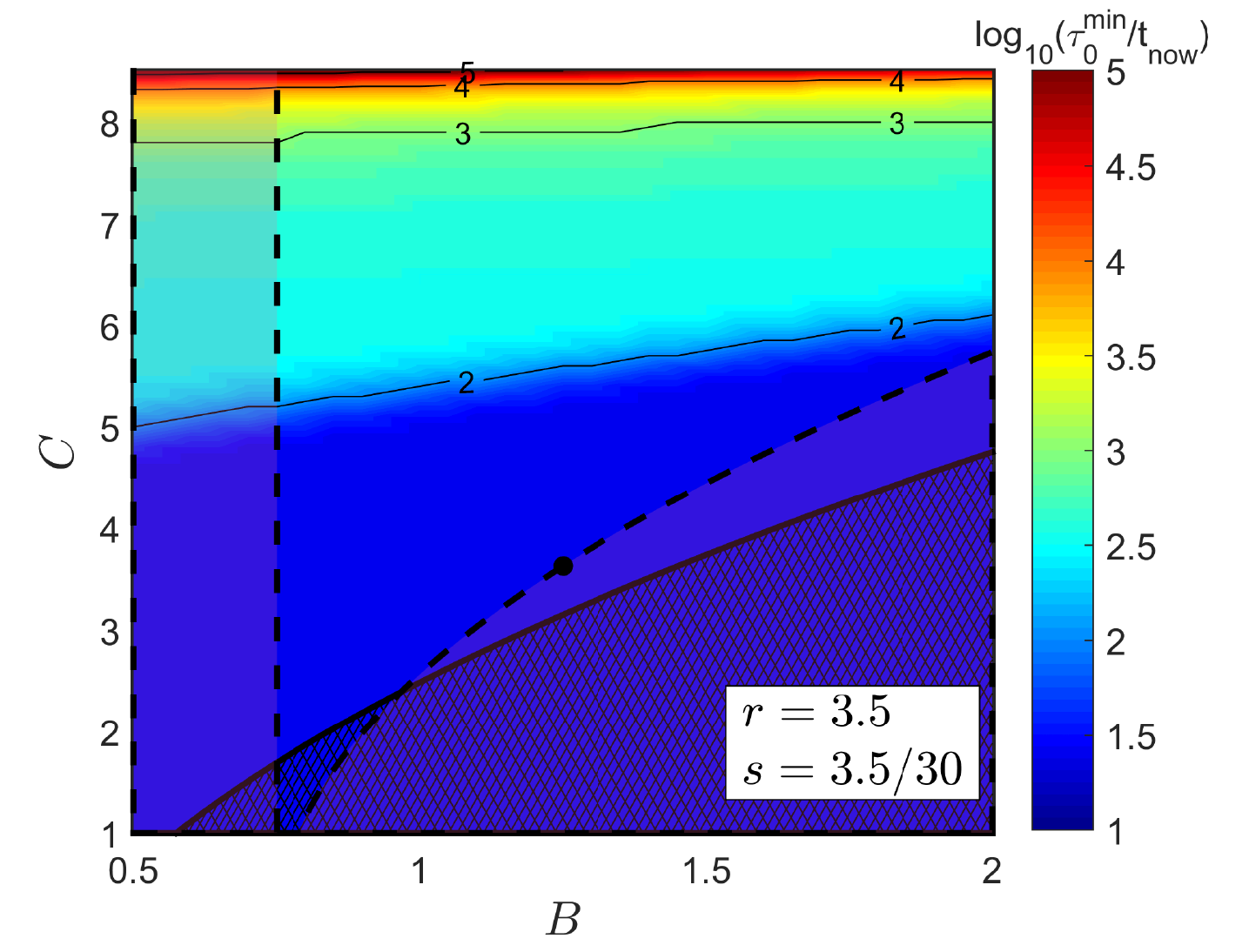}
\caption{ 
 Contours of the minimum value of $\tau_0^{\mathrm{min}}$ consistent with the 
   look-back and $\weff$ constraints discussed in the text,
 plotted in the $(r,s)$ plane (left panel) and in the $(B,C)$ plane (right panel).} 
\label{fig5}
\end{figure*}

Thus far we have not discussed the effects of varying our remaining free parameter $\tau_0$.
Varying $\tau_0$ does not affect the degeneracies of states or their cosmological abundances.
Indeed, variations in $\tau_0$ affect only the {\it lifetimes}\/ of these states.
In principle, this has the potential to affect the values of quantities such as $\eta(\tnow)$
since the determination of $\eta(\tnow)$ requires totalling the abundances of only those
states which have not yet decayed at the present time.
However, under the assumption that $\tau_0\gg \tnow$ (or under the equivalent assumption
that our scenario already satisfies the look-back and $w_{\rm eff}$ constraints), we know that
$\Omegatot(\tnow)$ is not changing rapidly at the present time. 
In other words, the total abundances   
of those states which are decaying at the present time is relatively small.
In such cases, the $M_s$  and $\eta(\tnow)$ contours are therefore largely insensitive to $\tau_0$.
Indeed, in Fig.~\ref{fig3}, the sole effect of varying $\tau_0$ 
is therefore merely to ``slide'' the red exclusion regions in Fig.~\ref{fig3} horizontally relative to the rest
of the plot:   these exclusion regions move to the right (and therefore become more 
threatening to our sweet-spot region) 
if $\tau_0/\tnow$ is decreased, and move to the left (and therefore become
even less of a concern) if $\tau_0/\tnow$ is increased.

While this is entirely as expected, the natural question then arises:   for any values of $\lbrace B,C,r,s\rbrace$, 
what is the minimum value of $\tau_0$ that can be tolerated before violating our look-back and $\weff$
constraints?
Contours indicating the resulting minimum values $\tau_0^{\rm min}$ are plotted in Fig.~\ref{fig5}
in both the $(r,s)$ and $(B,C)$ planes, taking our ``benchmark'' values in Eqs.~(\ref{rsbenchmarks})
and~(\ref{BCbenchmarks}) respectively.
In general, we see from Fig.~\ref{fig5} that a wide variety of values of $\tau_0^{\rm min}$ are possible,
depending on our specific location in parameter space, with larger values of $\tau_0^{\rm min}$ corresponding to
very small values of $r$ or relatively large values of $s$ or $C$.~
However, for our sweet-spot benchmark values in Eqs.~(\ref{BCbenchmarks}) and~(\ref{rsbenchmarks}),
we see from Fig.~\ref{fig5} that $\tau_0^{\rm min}$ can be as small as approximately $10^{2}\tnow$.

This, too, is not entirely a surprise.
After all, a bound on the lifetime of the longest-lived DDM constituent on the order 
$\tau_0/t_{\mathrm{now}} \sim \mathcal{O}(100)$ 
is roughly on the same order as the most conservative bounds on the lifetime $\tau_\chi$ of a 
traditional
single-component dark-matter candidate which decays into other purely dark-sector states.  
Indeed, model-independent bounds on decaying dark matter in traditional single-component models in 
which the dark-matter particle carries essentially all of the observed dark-matter abundance and decays into dark radiation have been derived by a number of groups (see, \eg, Refs.~\cite{DeLopeAmigo,Audren,Aubourg,Blackadder}).  Depending on the assumptions inherent in the various analyses and on the breadth of cosmological data incorporated, 
such studies place a bound on the lifetime of such a dark-matter candidate on the order of $\tau_\chi/t_{\mathrm{now}} \gtrsim \mathcal{O}(10 - 100) $.  Thus, a bound on $\tau_0$ in this range is {\it a priori}\/ reasonable  ---
especially since our analysis in Fig.~\ref{fig5} determines the value of $\tau_0^{\rm min}$ based only on 
cosmological look-back and $\weff$ constraints.
Of course, if the ensemble constituents decay into {\it visible}\/-sector particles with a non-negligible branching fraction, the constraints on $\tau_0$ are expected to increase significantly.
Indeed, the most stringent bounds on a single dark-matter 
particle $\chi$ which decays primarily into visible-sector radiation require that this
particle be hyperstable, with $\tau_\chi \sim 10^9 \tnow$.

Despite the possibilities for lowering $\tau_0$ afforded by the results in Fig.~\ref{fig5}, 
we shall continue to retain our benchmark value 
$\tau_0 = 10^9 t_{\mathrm{now}}$.
We do this in order to be consistent with the most conservative decay scenarios possible.
Although this value for $\tau_0$ is quite large, we emphasize that 
this is only the lifetime of the lightest ensemble constituent, and
that a significant fraction of the ensemble constituents will generally have lifetimes much less than $\tau_0$.  
Moreover, even in cases for which the majority of the ensemble is long-lived, 
DDM ensembles can nevertheless yield striking astrophysical signatures~\cite{DDMindirect,DDMboxes1,DDMboxes2} 
which differ from 
those of traditional dark-matter candidates.  
Thus, even with such values of $\tau_0$, the phenomenology of the resulting
ensemble can differ significantly from that of traditional dark-matter candidates.

Having explored the relevant 
$\lbrace B,C,r,s,\tau_0\rbrace$ 
parameter space of our ensemble and identified
our sweet-spot region,
we now examine the characteristics of the corresponding ensembles in more detail.
In particular, we seek to understand what these ensembles look like, and how their
overall structure evolves with time.
As discussed in Sect.~\ref{quantities},
the most relevant aggregate properties of any dark-sector ensemble are 
its total cosmological abundance $\Omegatot(t)$,
its effective equation-of-state parameter $\weff(t)$,
and  its tower fraction $\eta(t)$, each of which is generally time-dependent.
We therefore begin by examining how each of these quantities evolves with time for
ensembles in and near our sweet spot.

\begin{figure*}
\centering
  \includegraphics[width=0.30\textwidth, keepaspectratio]{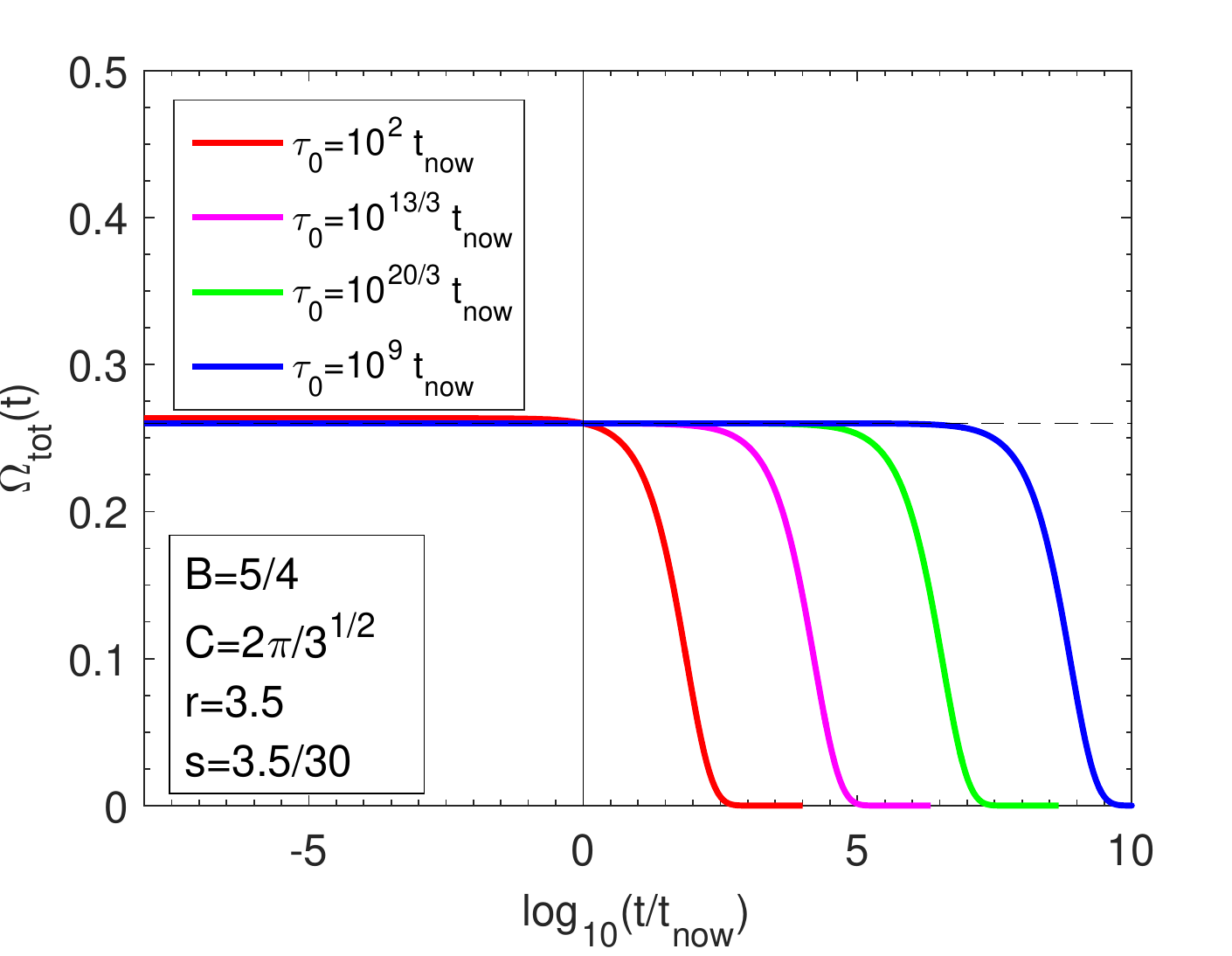}
  ~~\includegraphics[width=0.30\textwidth, keepaspectratio]{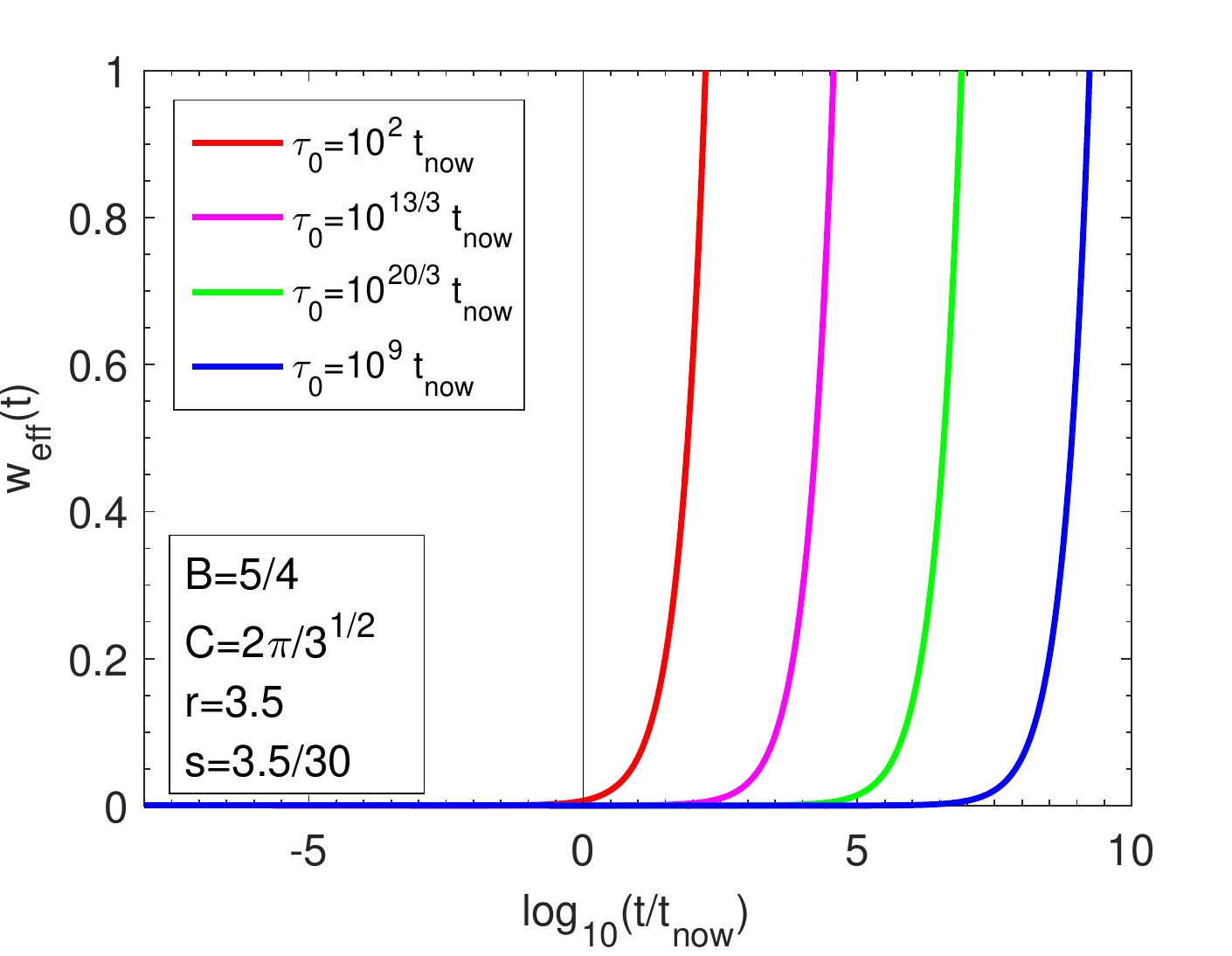}
  ~~\includegraphics[width=0.30\textwidth, keepaspectratio]{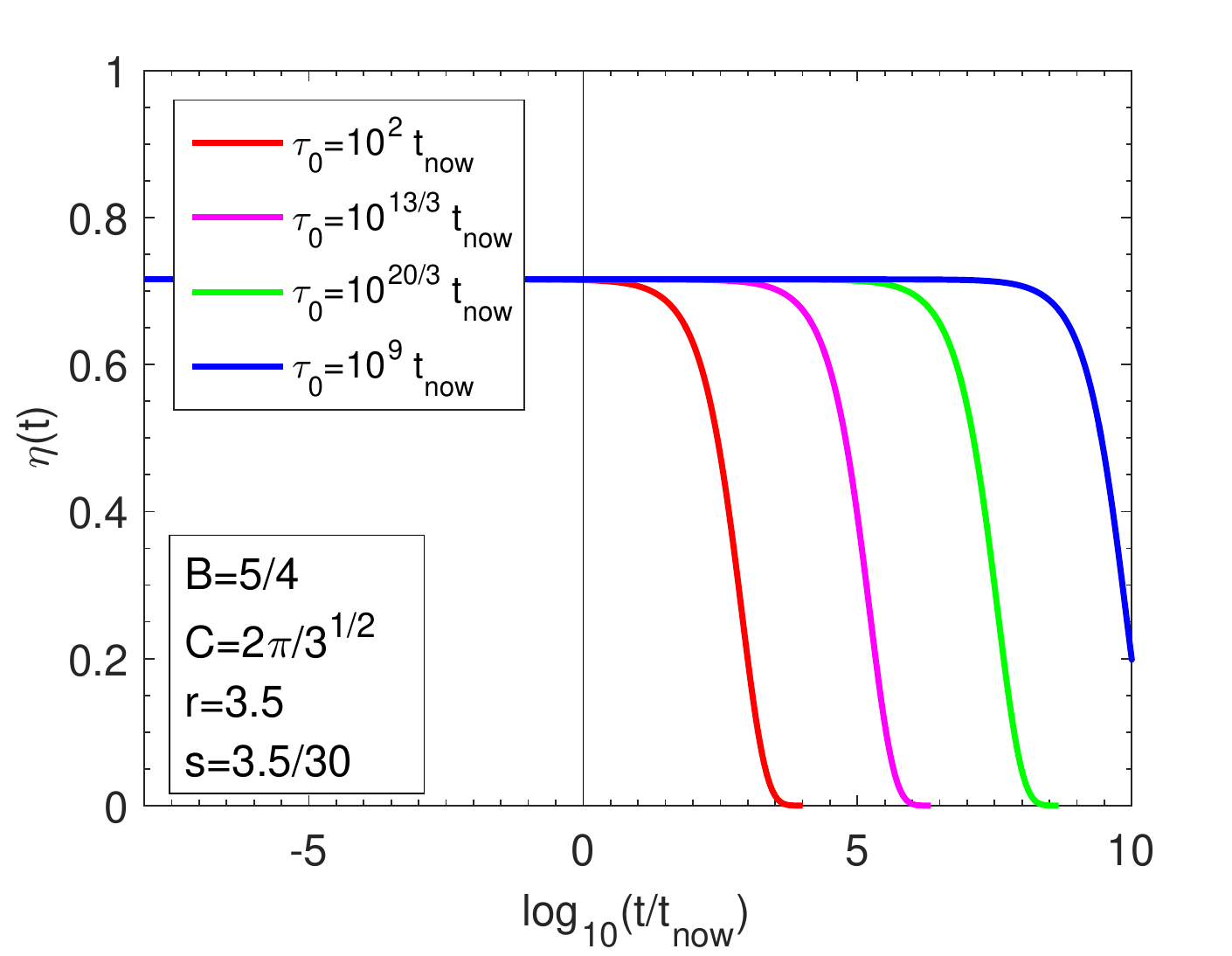}
  \includegraphics[width=0.30\textwidth, keepaspectratio]{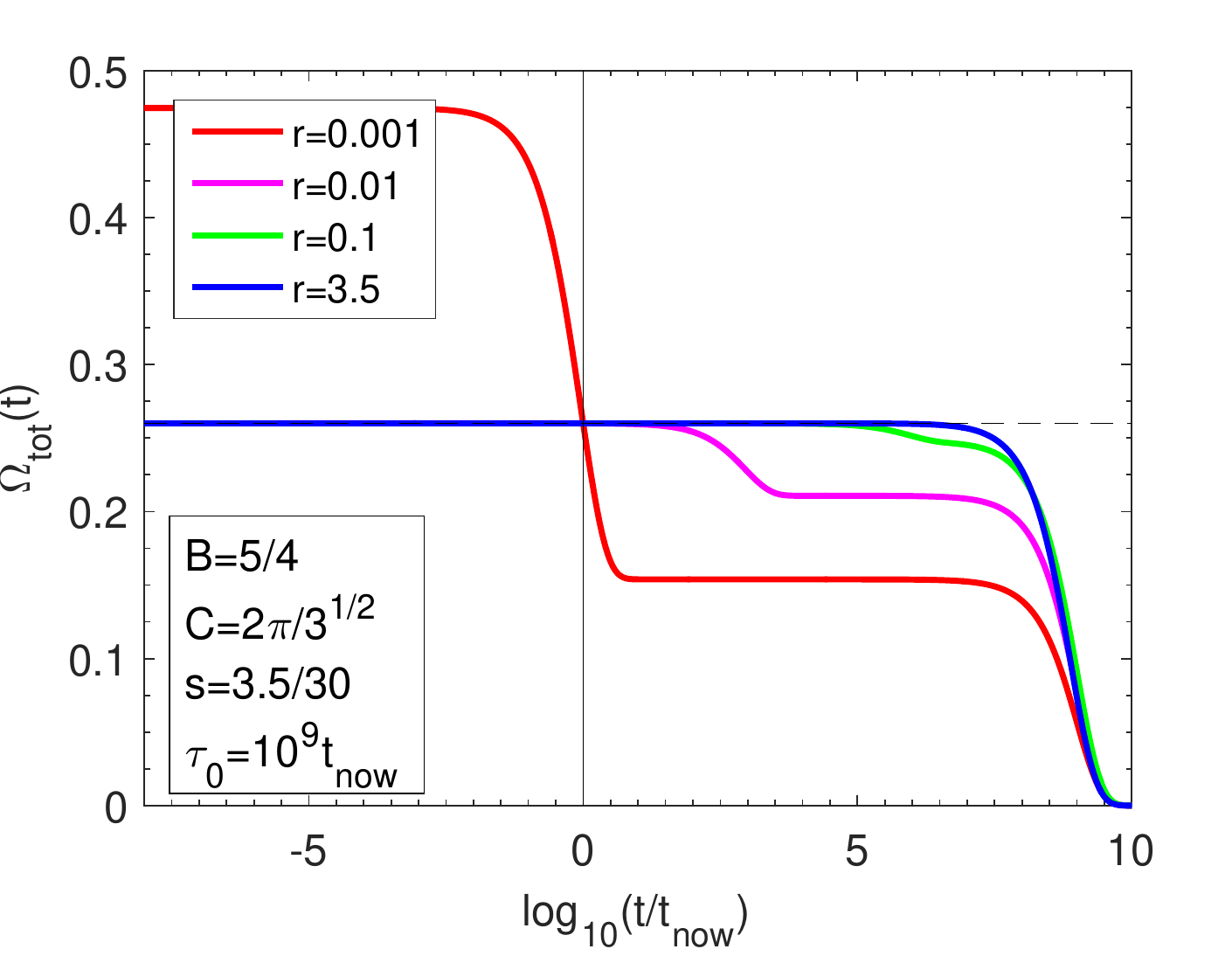}
  ~~\includegraphics[width=0.30\textwidth, keepaspectratio]{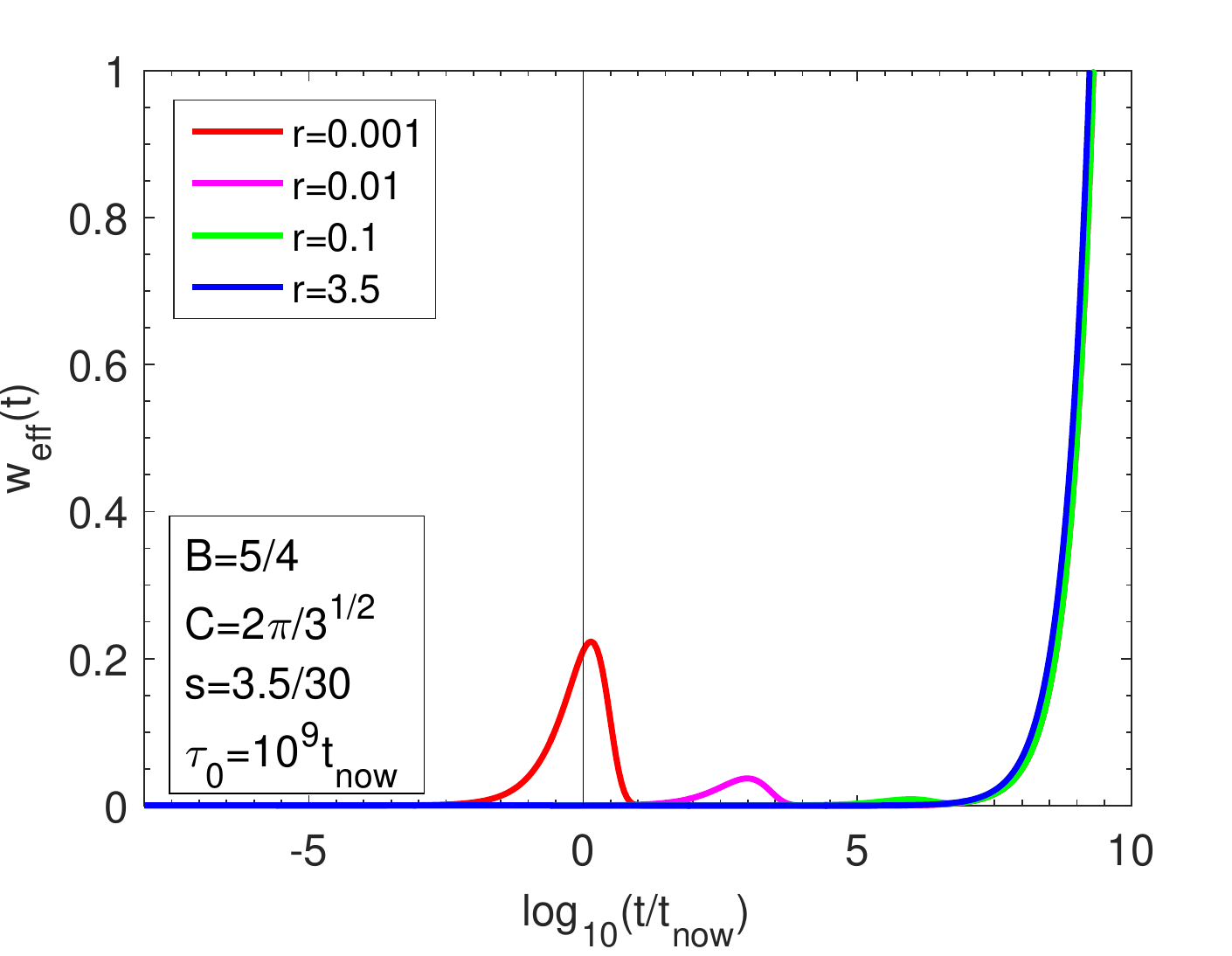}
  ~~\includegraphics[width=0.30\textwidth, keepaspectratio]{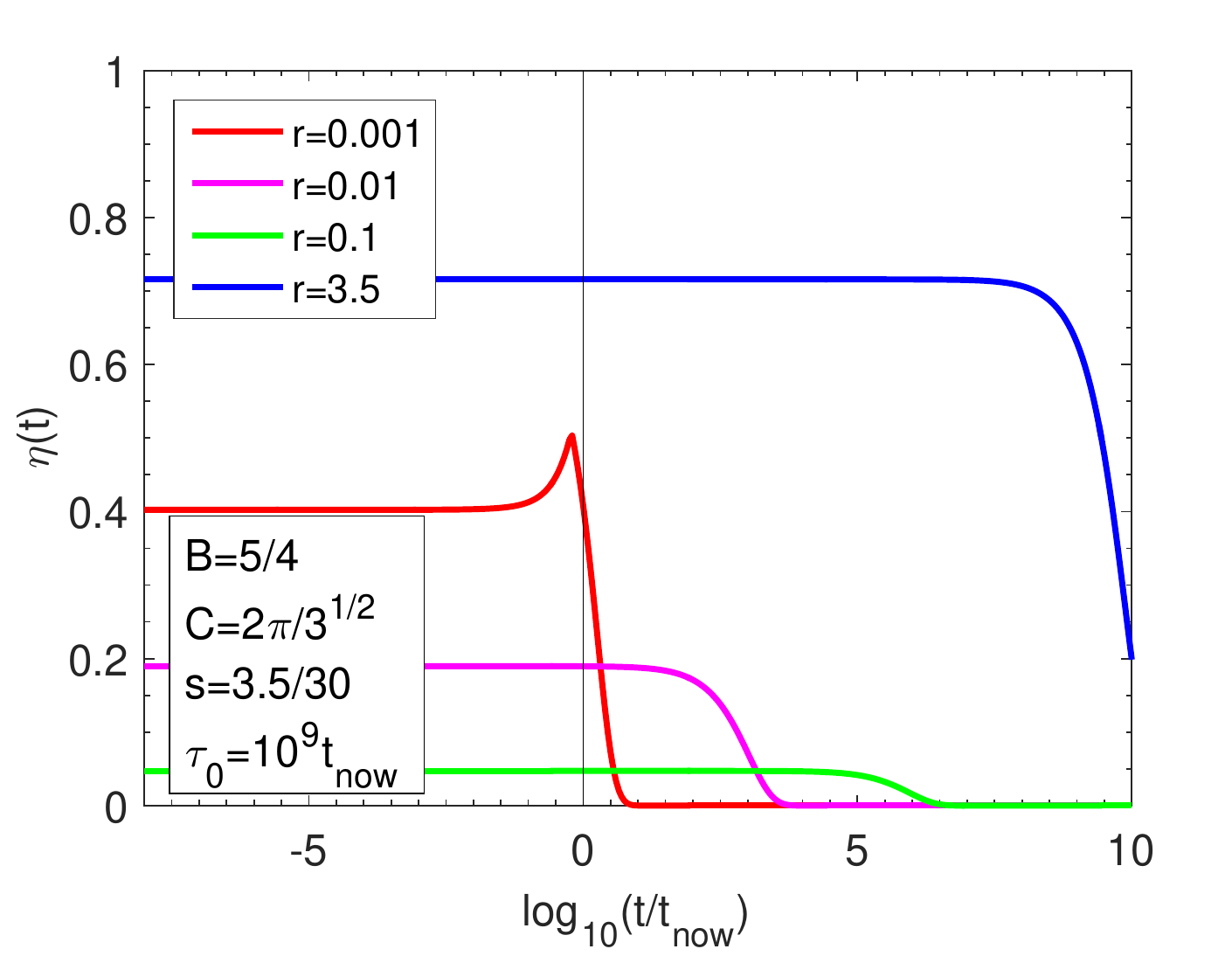}
  \includegraphics[width=0.30\textwidth, keepaspectratio]{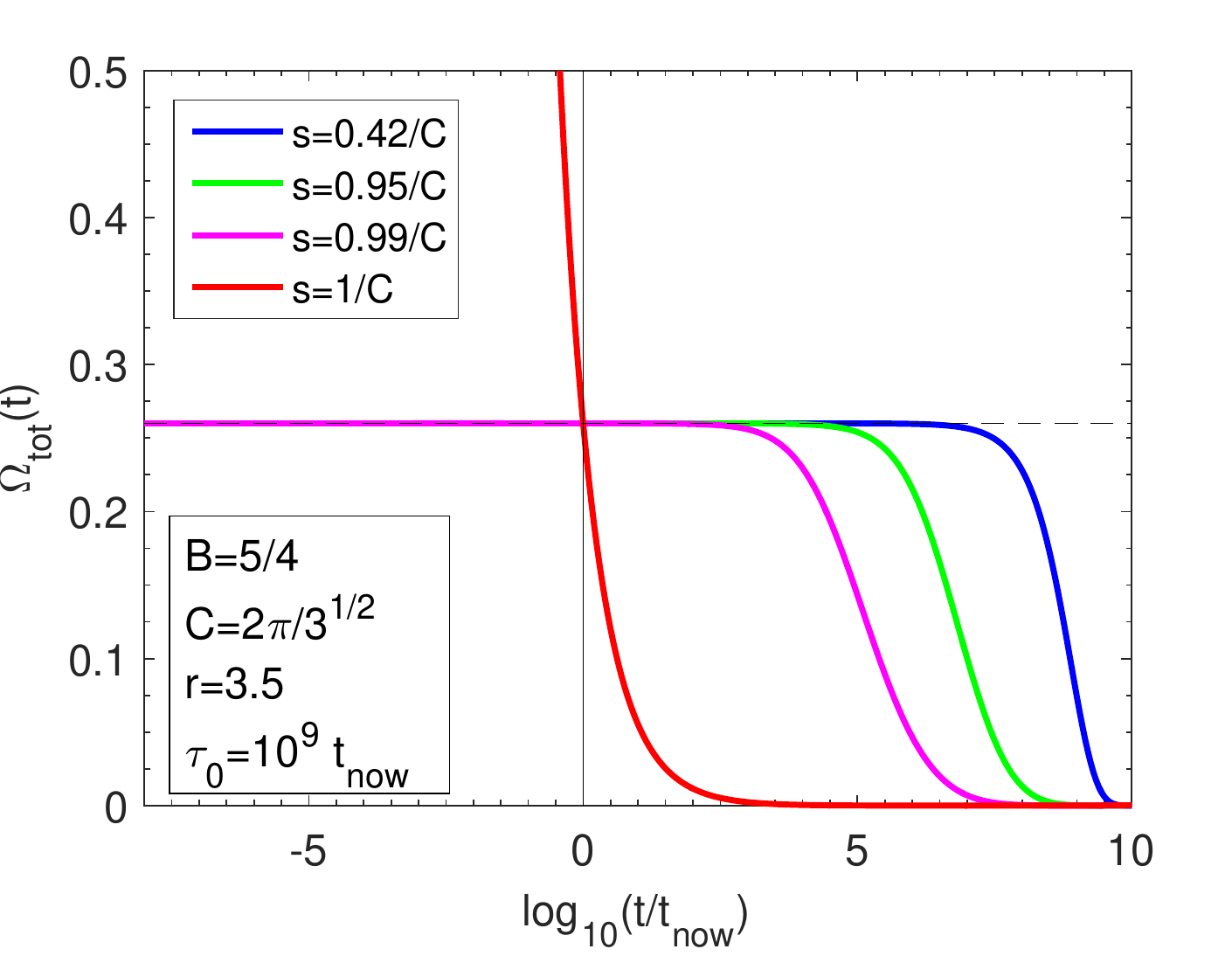}
  ~~\includegraphics[width=0.30\textwidth, keepaspectratio]{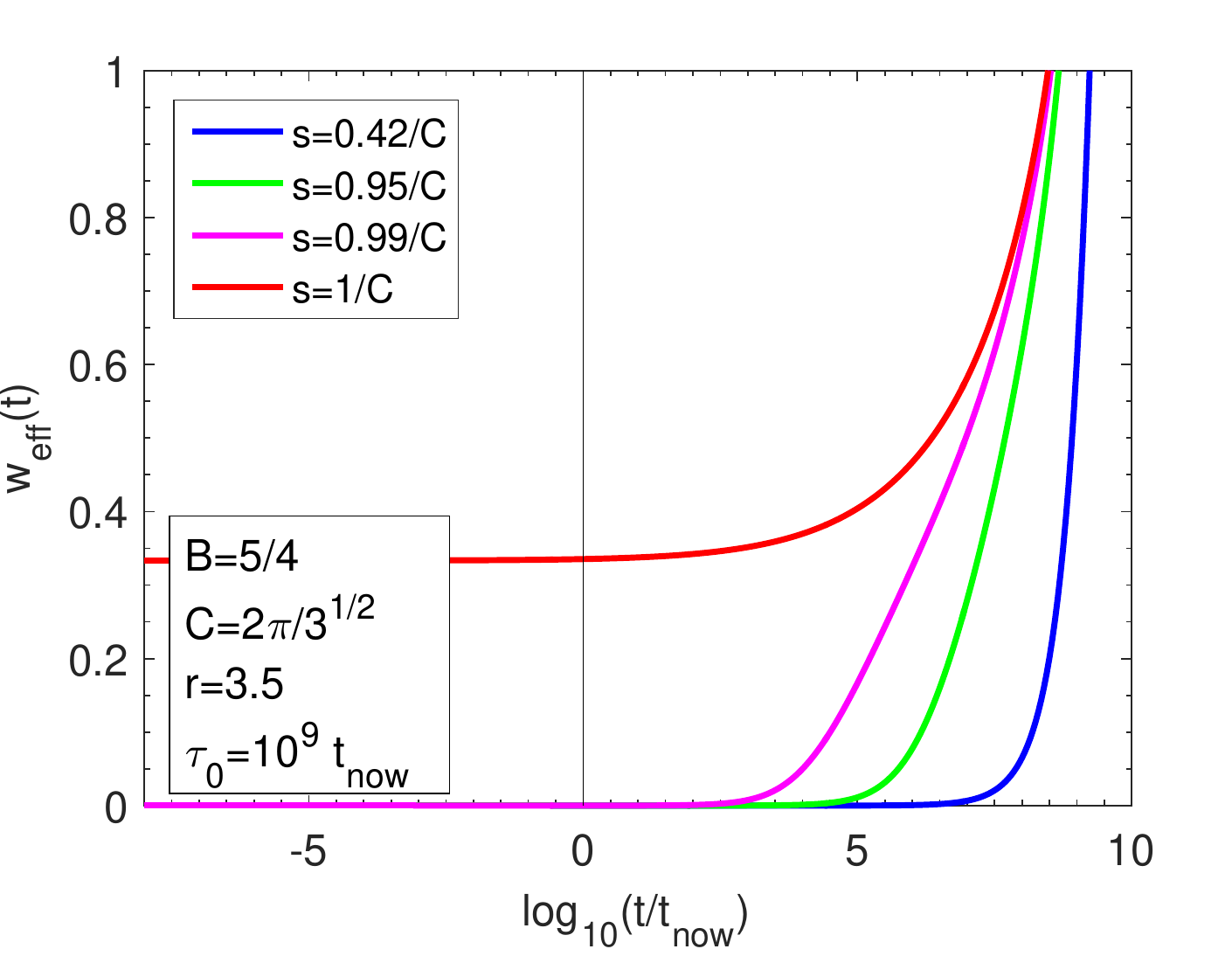}
  ~~\includegraphics[width=0.30\textwidth, keepaspectratio]{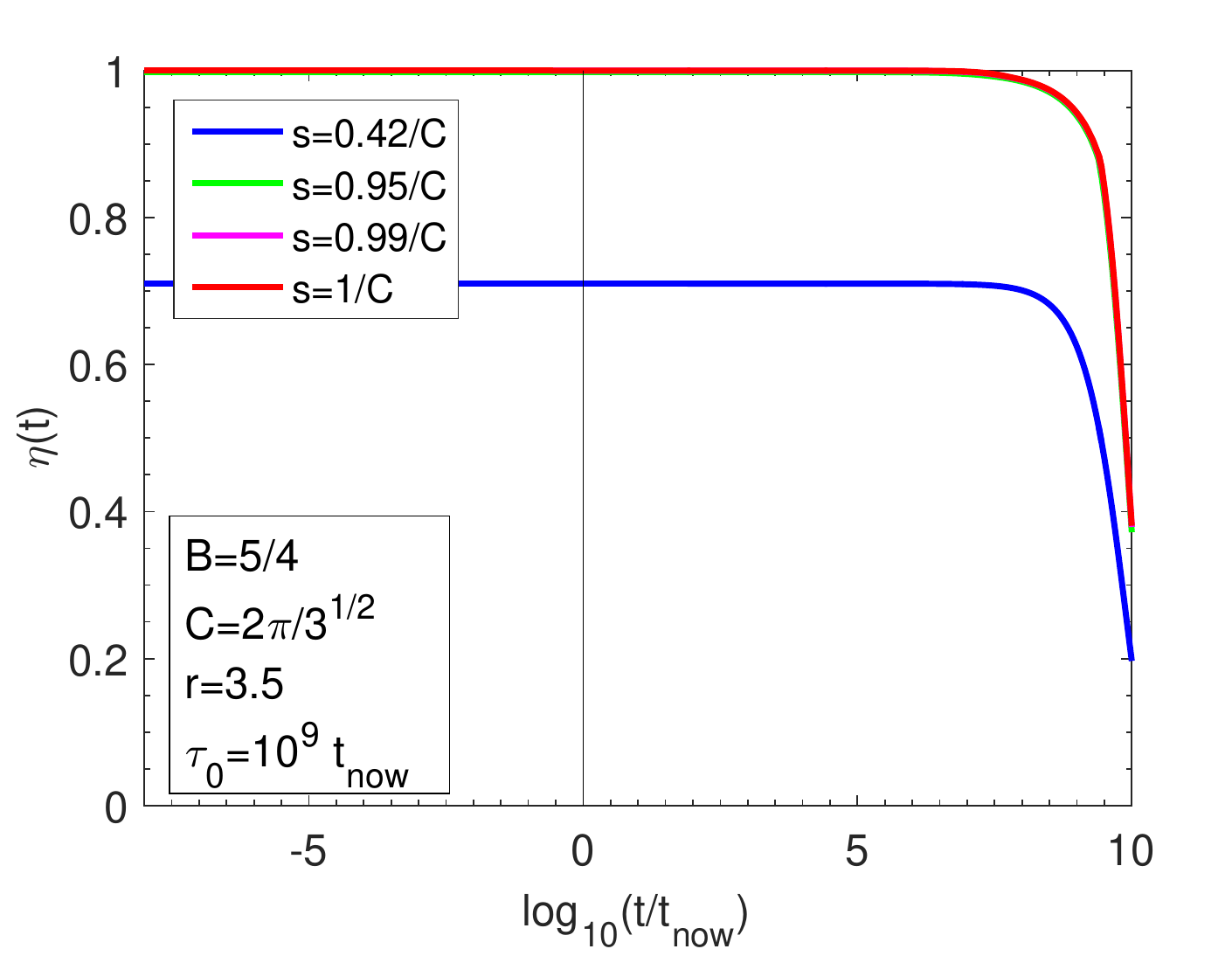}
  \includegraphics[width=0.30\textwidth, keepaspectratio]{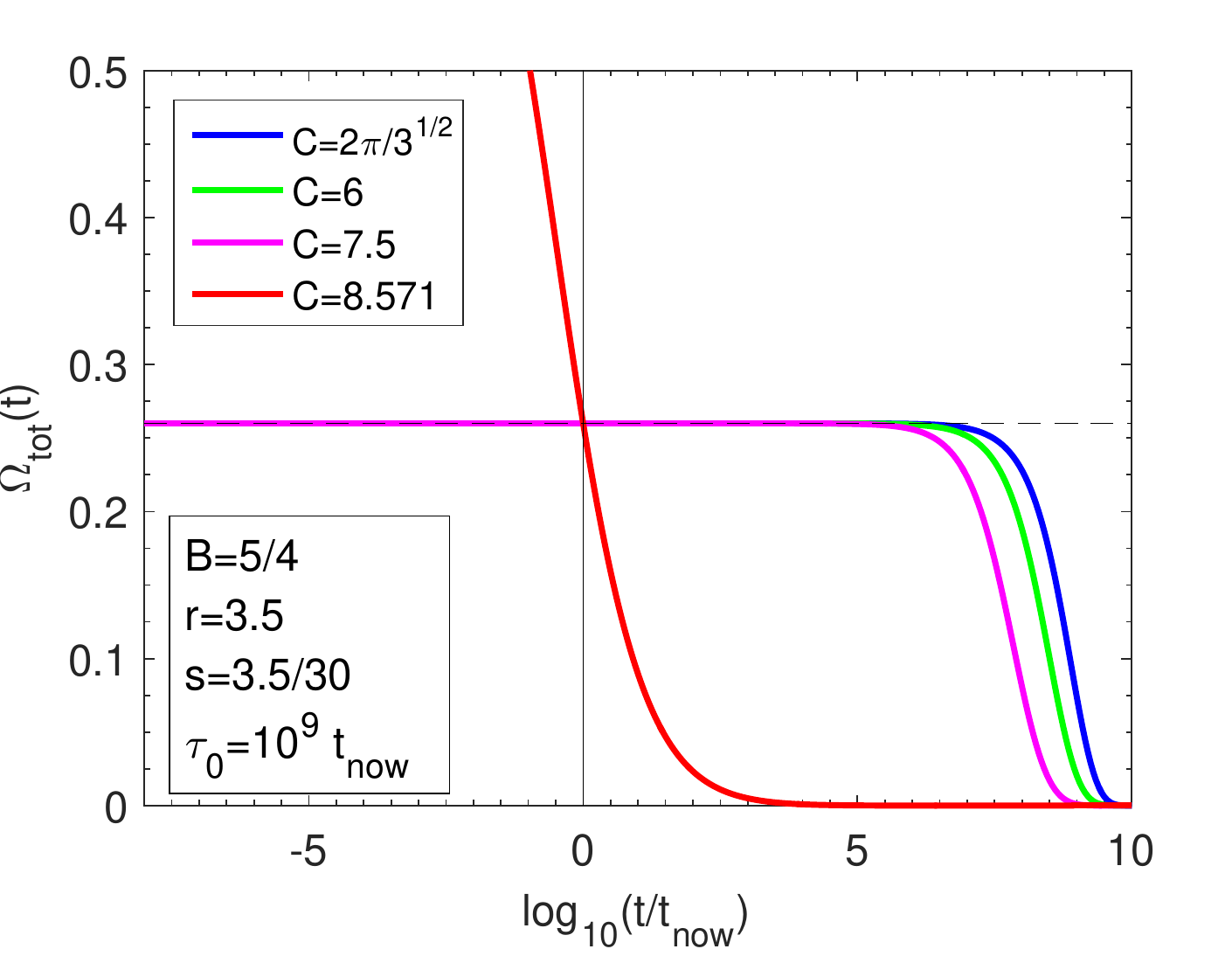}
  ~~\includegraphics[width=0.30\textwidth, keepaspectratio]{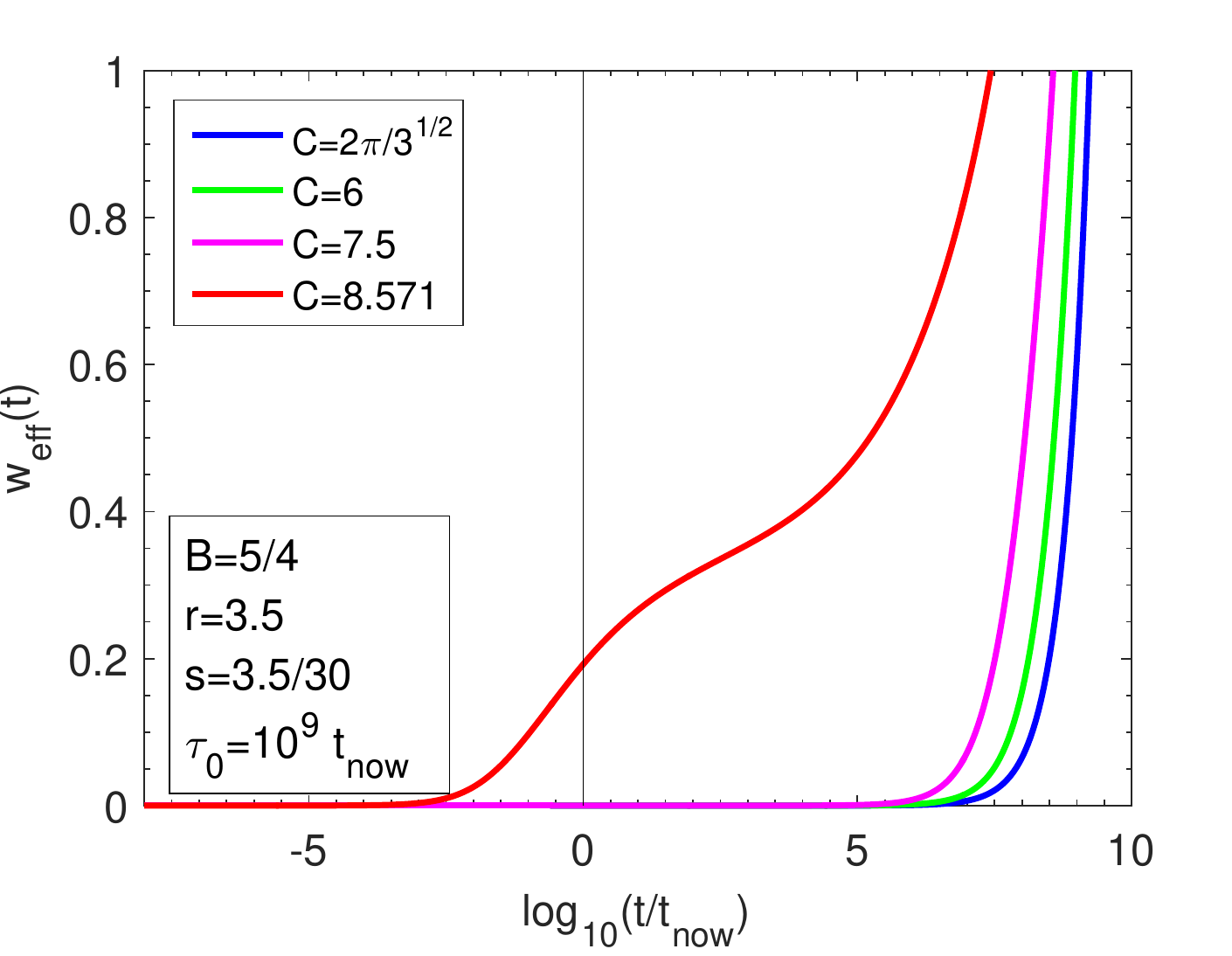}
  ~~\includegraphics[width=0.30\textwidth, keepaspectratio]{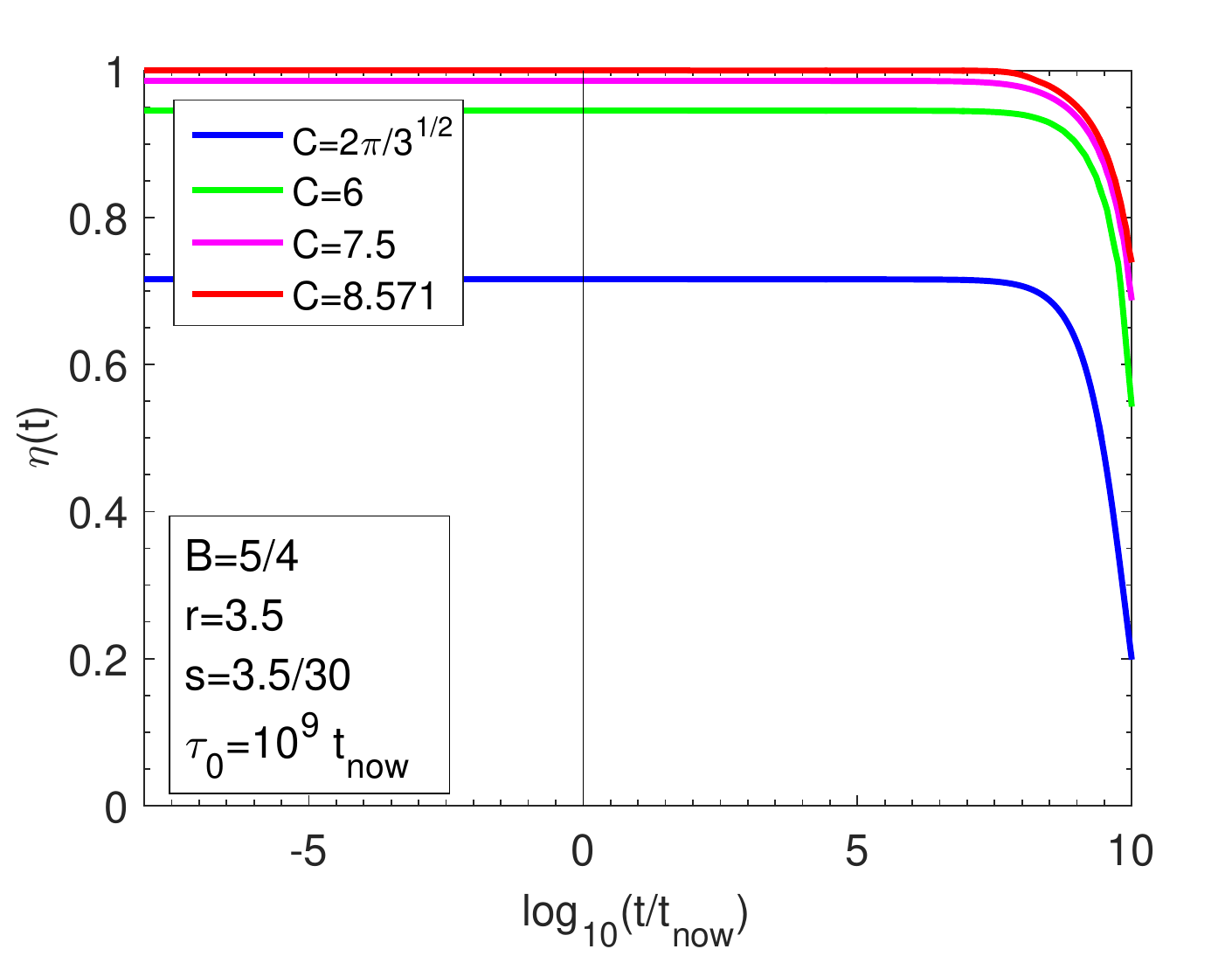}
  \includegraphics[width=0.30\textwidth, keepaspectratio]{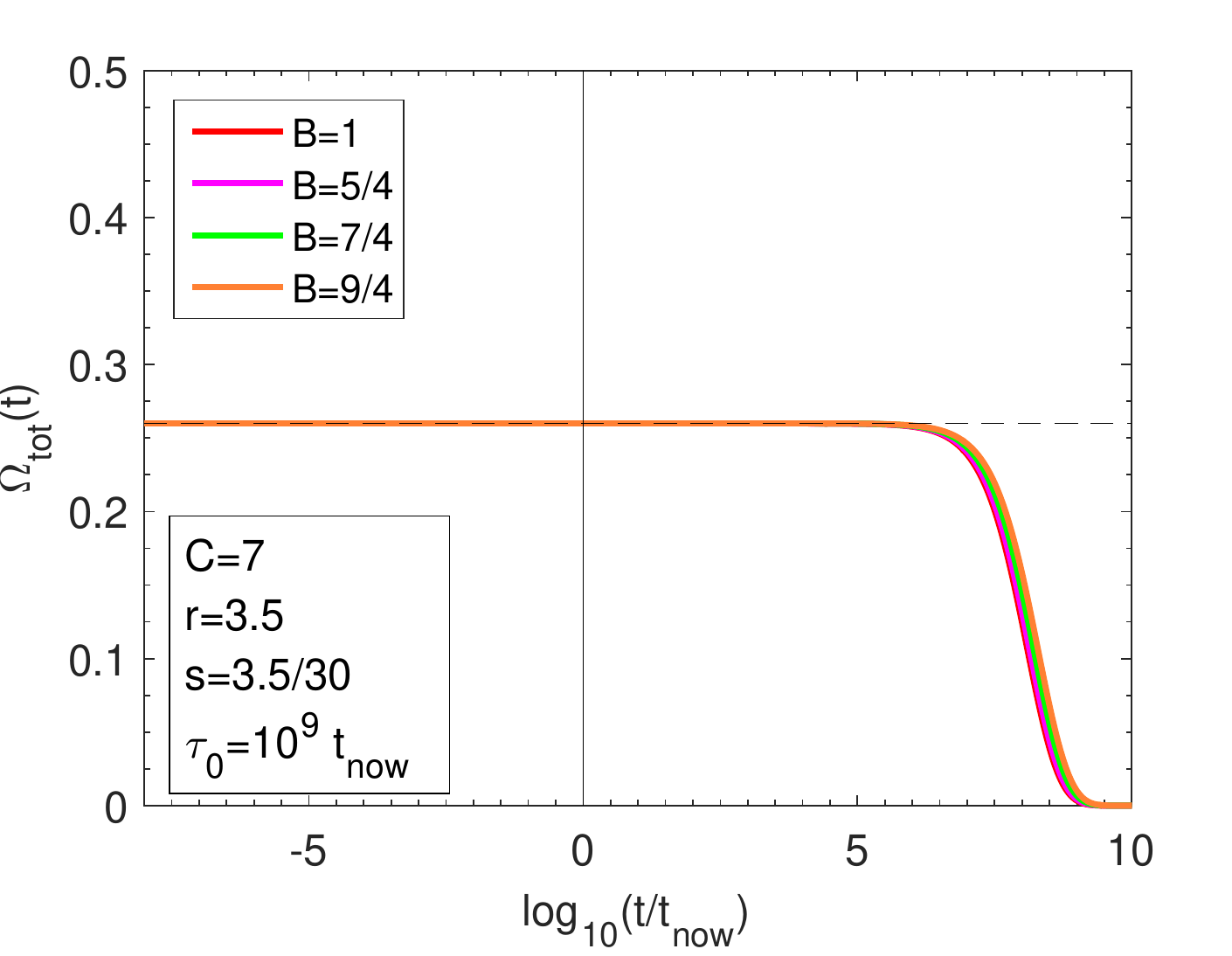}
  ~~\includegraphics[width=0.30\textwidth, keepaspectratio]{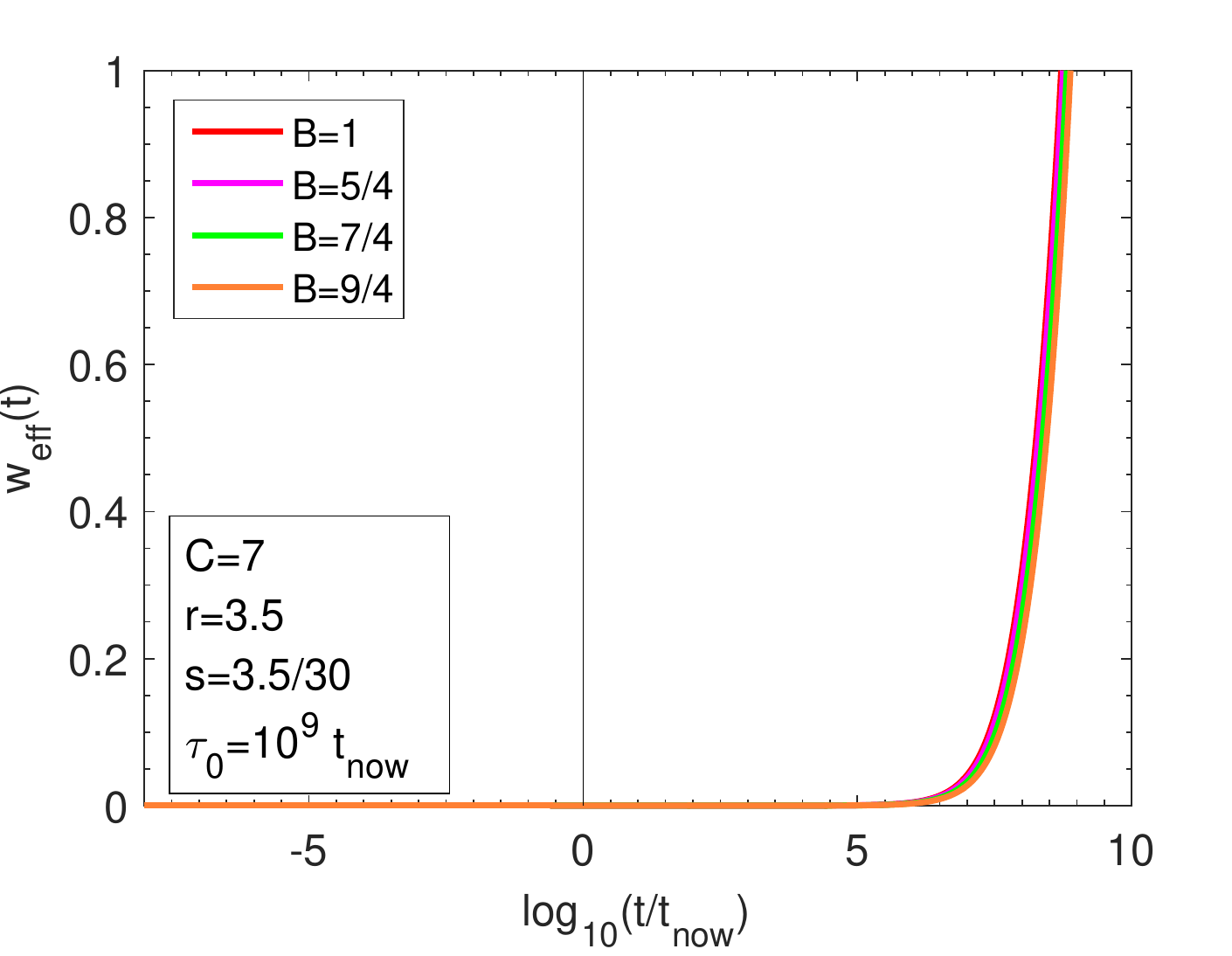}
  ~~\includegraphics[width=0.30\textwidth, keepaspectratio]{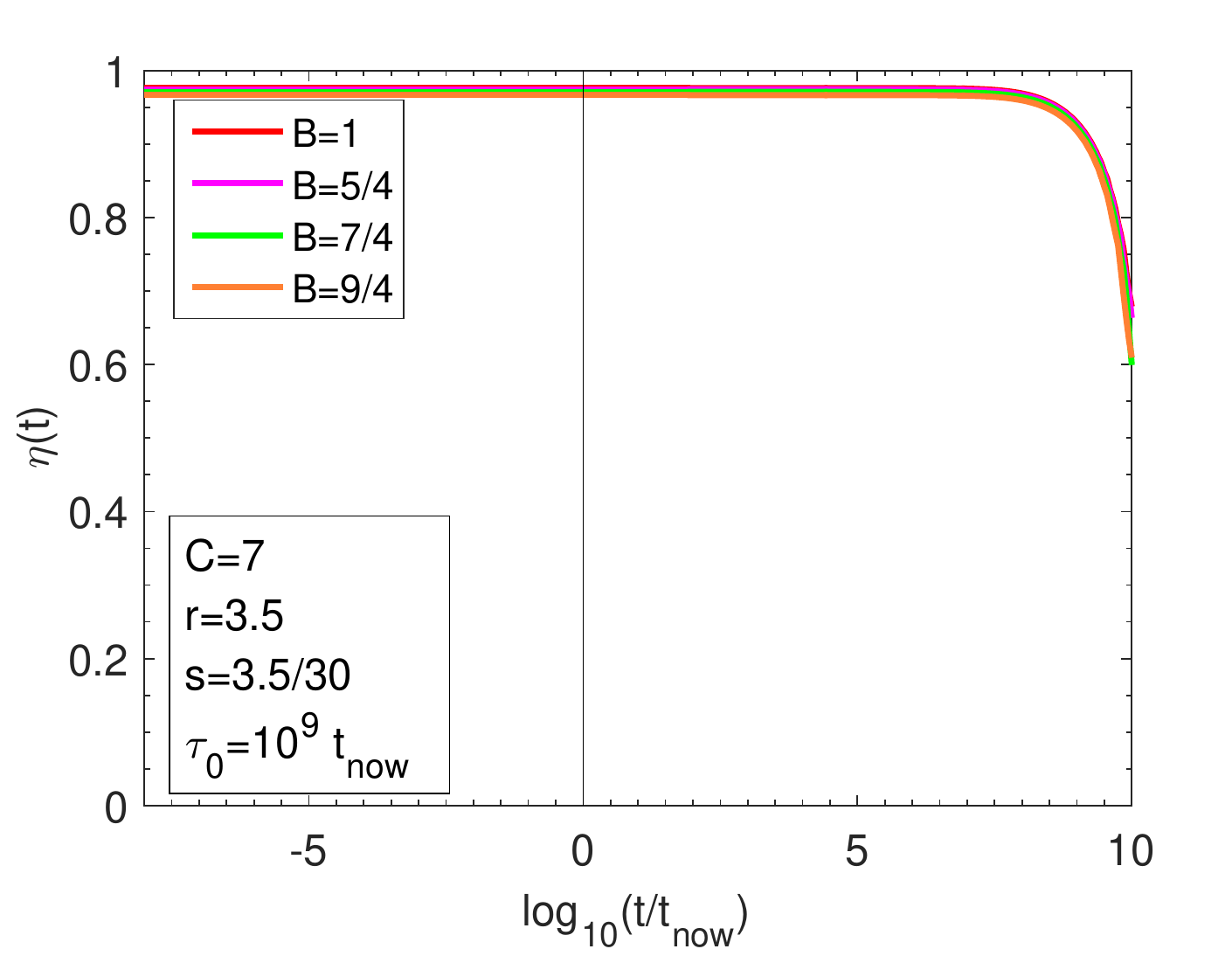}
\caption{Total cosmological abundances $\Omega_{\rm tot}$ (left column), 
    equation-of-state parameters 
     $w_{\rm eff}$ (middle column), 
     and tower fractions $\eta$ (right column) for
        our DDM ensembles, 
     plotted as functions of time when all input variables 
     are held fixed at their ``benchmark'' values except for 
     $\tau_0$ (top row), $r$ (second row), $s$ (third row), 
      $C$ (fourth row), and $B$ (bottom row). 
     In all panels the blue curve corresponds to our ``benchmark'' point
     with $B=5/4$, $C=2\pi/\sqrt{3}$, $r=3.5$, $s=3.5/30$, and $\tau_0=10^9 \tnow$,
     while the curves of other colors indicate departures away from this point.
     For reasons discussed in the text, the bottom row 
     illustrates variations in $B$ along a line that does not include
     the benchmark point.
     Note that, as expected, some variations away from the benchmark point violate 
     our look-back, $\weff$, or $M_0$ constraints.
     However, our internal self-consistency constraints are always satisfied, with
     $\Omegatot(\tnow)=\OmegaCDM\approx 0.26$ in all cases.}
\label{fig6}
\end{figure*}

This information is shown in Fig.~\ref{fig6}.
In this figure, we consider a ``benchmark'' 
ensemble with $B=5/4$, $C=2\pi/\sqrt{3}$, 
$r=3.5$, $s=3.5/30$, and $\tau_0=10^9 \tnow$,
as well as nearby ensembles
in which $\tau_0$ is varied (top row),
$r$ is varied (second row), 
$s$ is varied (third row),
$C$ is varied (fourth row),
and $B$ is varied (fifth row).
In each case, we plot
the corresponding 
total cosmological abundance $\Omega_{\rm tot}$ (left column),
    equation-of-state parameter $w_{\rm eff}$ (middle column),
     and tower fraction $\eta$ (right column) as functions of time.
Note that in each case the overall abundance is normalized through an appropriate choice of
$M_s$ such that $\Omega(t_{\rm now}) = \OmegaCDM \approx 0.26$, as required.

In each panel of Fig.~\ref{fig6} (except for those along the bottom row),
the blue curve corresponds to our ``benchmark'' point.
We therefore begin by focussing on these benchmark curves.
The curve for $\Omega_{\rm tot}(t)$ appears nearly constant at $\OmegaCDM\approx 0.26$ for all of the 
cosmological history plotted (which we assume to have been matter-dominated), 
including the present time $t_{\rm now}$.   Indeed, this behavior continues
all the way into the future until $t\approx 10^9 t_{\rm now}$, at which point $\Omega_{\rm tot}(t)$ begins
to decline gently to $\Omega_{\rm tot}=0$.   
This behavior is matched by $w_{\rm eff}(t)$, which remains near zero for most
its cosmological evolution before gently rising  to $w_{\rm eff}>0$ at 
$t\approx 10^9 t_{\rm now}$.    This makes sense, since Eq.~(\ref{eq:wdef}) tells us that
$w_{\rm eff}(t)$ is proportional to the time-derivative of $\Omega_{\rm tot}(t)$.
Finally, we see that $\eta(t)$ remains more or less fixed at approximately $\eta\approx 0.72$ 
during most of its cosmological history before smoothly dropping to $\eta=0$.

This behavior is easy to understand.   If this has been a traditional ensemble with a
single dark-matter component whose decay we could model as essentially instantaneous (just
as we are assuming for the individual components of our dark-matter ensembles), 
our curve for $\Omega_{\rm tot}(t)$ would have been fixed precisely at 
its present value $\OmegaCDM\approx 0.26$ over the entire range shown until suddenly dropping 
(essentially discontinuously) to $\Omega_{\rm tot}=0$ 
when the single dark-matter particle decays at $t\approx 10^9 t_{\rm now}$.
Likewise, $w_{\rm eff}(t)$ would have been strictly fixed at $w_{\rm eff}=0$ during the cosmological
evolution, while $\eta(t)$ would have been fixed at zero all along.
However, this is not a traditional dark-matter setup:   this is a DDM ensemble in which the present-day
cosmological abundance $\Omega_{\rm tot}(t_{\rm now})\approx 0.26$ is spread across a
relatively large number of individual components with different masses and different lifetimes.    
It is thus the continued, ordered, sequential decays
of these different components
which produce the softer, gentler drop in $\Omega_{\rm tot}(t)$ as $t$ approaches $t\approx 10^9 t_{\rm now}$.   
In fact, $\Omega_{\rm tot}(t)$
is actually falling slightly {\it throughout}\/ the cosmological evolution shown;  this behavior is not visible
in Fig.~\ref{fig6} only because at early times prior to $t\approx 10^9 t_{\rm now}$ the states which are decaying
are extremely heavy and thus carry extremely small abundances.   
By contrast, at late times approaching $t\approx 10^9\tnow$, the states which are decaying are relatively
low-lying and carry more significant abundances.
This is also evident in our curve for $\eta(t)$:   
for most of the cosmological history, the value $\eta\approx 0.72$ tells us that only approximately 
$28\%$
of the total dark-sector cosmological abundance 
is carried by the dominant (lightest) state in the ensemble, even at early times, 
while the remaining $72\%$
of the abundance is carried by the more massive states ---  particularly those which, though more massive, 
are nevertheless relatively low-lying.
As a result of the sequential decays of such states,
$\eta(t)$ --- like $\Omegatot(t)$ --- is also actually falling slightly  
 {\it throughout}\/ the cosmological evolution shown.  
It is only due to the decays of the relatively low-lying 
states near $t\approx 10^9 t_{\rm now}$ that $\eta(t)$ ultimately falls gently but noticeably to zero.

At first glance, it may seem surprising that all three of our primary quantities 
$\Omega_{\rm tot}$, $w_{\rm eff}$, and 
$\eta$ are nearly constant at $t\approx t_{\rm now}$.   
However, this is ultimately the direct consequence of our benchmark choice $\tau_0=10^9 t_{\rm now}$:
with this choice, those states within the ensemble which are decaying today are all extremely massive
and thus carry very little abundance.    The DDM nature of such an ensemble is nevertheless clear from 
its $\eta$-value, which is as high as $0.72$ even at the present time.
In this connection, we again emphasize that taking $\tau_0=10^9 t_{\rm now}$ was merely a conservative choice
which is not by itself intrinsic to the DDM framework;
indeed we learned from Fig.~\ref{fig5}
that we could easily have chosen $\tau_0$ as small as $\tau_0 \approx 10^2\tnow$  
without running afoul of our look-back and $\weff$ constraints.
Indeed, without further details concerning the precise nature of these ensembles 
(including, most critically, the ultimate decay products of their constituents),
such small values for $\tau_0$ would have been equally viable.

This observation is illustrated along the top row of Fig.~\ref{fig6}, where we show
the evolution of our blue ``benchmark'' curves as we vary $\tau_0$ between our conservative
value $\tau_0\approx 10^9\tnow$ and the more extreme value
$\tau_0\approx 10^2\tnow$.
In general, changing $\tau_0$ 
does not affect the internal structure of the ensemble --- 
it merely affects the lifetimes of the individual ensemble constituents,
rescaling them all up or down together.   
Since it is these lifetimes 
which produce the non-trivial time-dependence for $\Omegatot$, $\weff$, and $\eta$, we expect 
that changing $\tau_0$ should preserve the general shapes of these curves 
and merely translate these curves along the time axis.
This behavior is verified in the panels along the top row of Fig.~\ref{fig6}. 
Indeed, we can even see from these panels why $\tau_0 \approx 10^2 \tnow$ is
the minimum value of $\tau_0$ that may be chosen for our benchmark point:   choosing
$\tau_0$ any smaller would shift our curves even further towards earlier times, whereupon
$\Omegatot(t)$ would begin to experience significant variations 
within the interval $10^{-6}\tnow \lsim t \lsim \tnow$
and $\weff(\tnow)$ would begin to deviate significantly from zero.
Such behavior would then violate our look-back and $\weff$ constraints, respectively.

Let us now turn to the behavior of our $\Omegatot$, $\weff$, and $\eta$ curves
as we vary $r$, as shown in the panels along the second row of Fig.~\ref{fig6}.
Two observations underlie the behavior shown.
First, we note 
that changing $r$ changes the lifetimes of the states at each mass level
according to Eq.~(\ref{eq:Lifetimes}), 
with $\tau_n/\tau_0\to 0$ as $r\to 0$.
This result is simple to understand:   as $r\to 0$, the $n=0$ states become hierarchically
lighter than the $n\geq 0$ states and thus the $n>0$ states have hierarchically shorter lifetimes.  
Second, we note that changing $r$ also changes the relative abundances which are generated at $t_c$
according to
\beq
      \frac{ \Omega_n(t_c)}{\Omega_0(t_c)} ~=~
       \frac{ (n+r^2)^{5/4}}{r^{5/2}} \exp\left( - \frac{ \sqrt{n+r^2} -r }{s}\right)~.
\label{nonmon}
\eeq
This quantity is non-monotonic as a function of $r$, 
first dropping as $r$ is reduced from large values and ultimately hitting a minimum
before increasing again and diverging as $r\to 0$.
Indeed, for $n=1$ and $s$ set to its benchmark value $s=3.5/30\approx 0.117$,
this minimum occurs at $r\approx 0.4$. 

These two effects are responsible for the behaviors shown in the second row of 
Fig.~\ref{fig6}.   As $r$ decreases from its benchmark value with $\tau_0$ held fixed, 
the excited states with $n>0$ start decaying earlier and earlier.   Rescaling our overall
abundances in order to keep $\Omegatot(\tnow)=\OmegaCDM$ produces the effects shown in
the left panel.
Indeed, we see from this panel that the case with 
$r= 0.001$ actually violates our look-back and $\weff$ constraints, as already
evident from Fig.~\ref{fig3}.
Even the $\Omegatot(t)$ curve with $r=0.01$ is
tightly constrained:   shifting $\tau_0$ towards any smaller values 
below $10^9\tnow$
(\ie, shifting this curve further towards the left)
also leads to violations of our look-back and $\weff$ constraints,
as already anticipated in the left panel of Fig.~\ref{fig5}. 
Likewise, as a result of the 
observations below Eq.~(\ref{nonmon}),
the relative sizes of the abundances $\Omega_n$ associated with the excited $n>0$ states
relative to the abundance $\Omega_0$ associated with the $n=0$ ground state 
vary non-monotonically with $r$, shrinking as $r$ drops
from 3.5 to approximately 0.4, and then growing again as $r$ drops still further. 
This then explains the non-monotonic behavior for $\eta(t)$ as a function of $r$, 
as shown in the right panel.

By contrast, the effects of varying $s$ and $C$ are shown along 
the third and fourth rows of Fig.~\ref{fig6}, respectively.
While the quantity $s$ governs the exponential rate at which the Boltzmann suppression of 
the abundances of the ensemble constituents {\it decreases}\/ with $n$,
the quantity $C$ governs the exponential rate at which the 
the degeneracy of states for the ensemble {\it grows}\/ with $n$.  
As a result, the effects of decreasing $s$ or increasing $C$ are largely
similar to each other as far as $\Omegatot(t)$ is concerned, as evident in Fig.~\ref{fig6}:
both tend to increase the primordial aggregate abundances
$\widehat{\Omega}_n$ of the heavier states in the ensemble.
This effect 
causes $\Omegatot(t)$ to begin to decline earlier and earlier as these heavier states
are the first to decay.
By contrast, it is important to note that increasing $C$ and decreasing $s$ nevertheless have
{\it opposite}\/ effects on the value of $\eta(\tnow)$:   the former increases $\eta(\tnow)$,
as anticipated in Fig.~\ref{fig4}, while the latter decreases $\eta(\tnow)$,
as anticipated in Fig.~\ref{fig3}.   
This difference occurs because increasing $C$ merely increases the state degeneracies
$\hat g_n$ of the heavy states, thereby injecting more abundance into the heavy states relative
to the light states, while decreasing $s$ has the effect of increasing the abundances of
{\it all}\/ of our states, including the abundance of the dominant abundance-carrier at $n=0$. 
This causes the total abundance of the ensemble to grow more rapidly than the
abundances of the excited $n>0$ states alone, thereby decreasing $\eta(\tnow)$.

One important feature to note from these plots is the appearance of a Hagedorn instability 
as $s\to 1/C$ (or equivalently as $C\to 1/s$).   
In these limiting cases, the total energy density $\Omegatot$ injected into the system through our confining phase
transition at $t=t_c$ diverges, violating the constraint in Eq.~(\ref{finitetotal}).
Such cases therefore violate our look-back and $\weff$ constraints,
as evident in Fig.~\ref{fig6}.
Indeed, the Hagedorn instability is a 
critical feature of theories with
exponentially growing degeneracies of states~\cite{Hagedorn}.

Finally, we turn to the fifth and final row of Fig.~\ref{fig6}. 
Note that in order to remain within 
the self-consistency bound in Eq.~(\ref{eq:CStringConsistCond}),
it is not possible to increase $B$ above our benchmark value $5/4$ when $C=2\pi/\sqrt{3}$.
For this reason, we have chosen to hold $C$ fixed at a greater value, specifically $C=7$,
when exploring the effects of varying $B$.
Unfortunately, we see that variations in $B$ are barely distinguishable in these plots,
even when $B$ is varied all the way from $B=1$ (corresponding to $D_\perp=1$)
to $B=9/4$ (corresponding to $D_\perp= 6$).
This tells us that the sorts of abundance-based or equation-of-state-based analyses we are doing 
here are relatively insensitive to the number of uncompactified transverse spacetime directions
into which our dark-sector flux tube can vibrate,
as long as $C$ (related to the total central charge of the degrees of freedom on the flux-tube worldsheet)
is held fixed.
Of course, in a realistic setting, there are likely to be many other more specific probes of $D_\perp$,
including probes that are based on specific properties of the dark-sector dynamics.
Our result here merely indicates 
that studies based on cosmological abundances alone are not likely to be the most useful in this regard.

\begin{figure*}
\centering
  \includegraphics[width=0.6\textwidth, keepaspectratio]{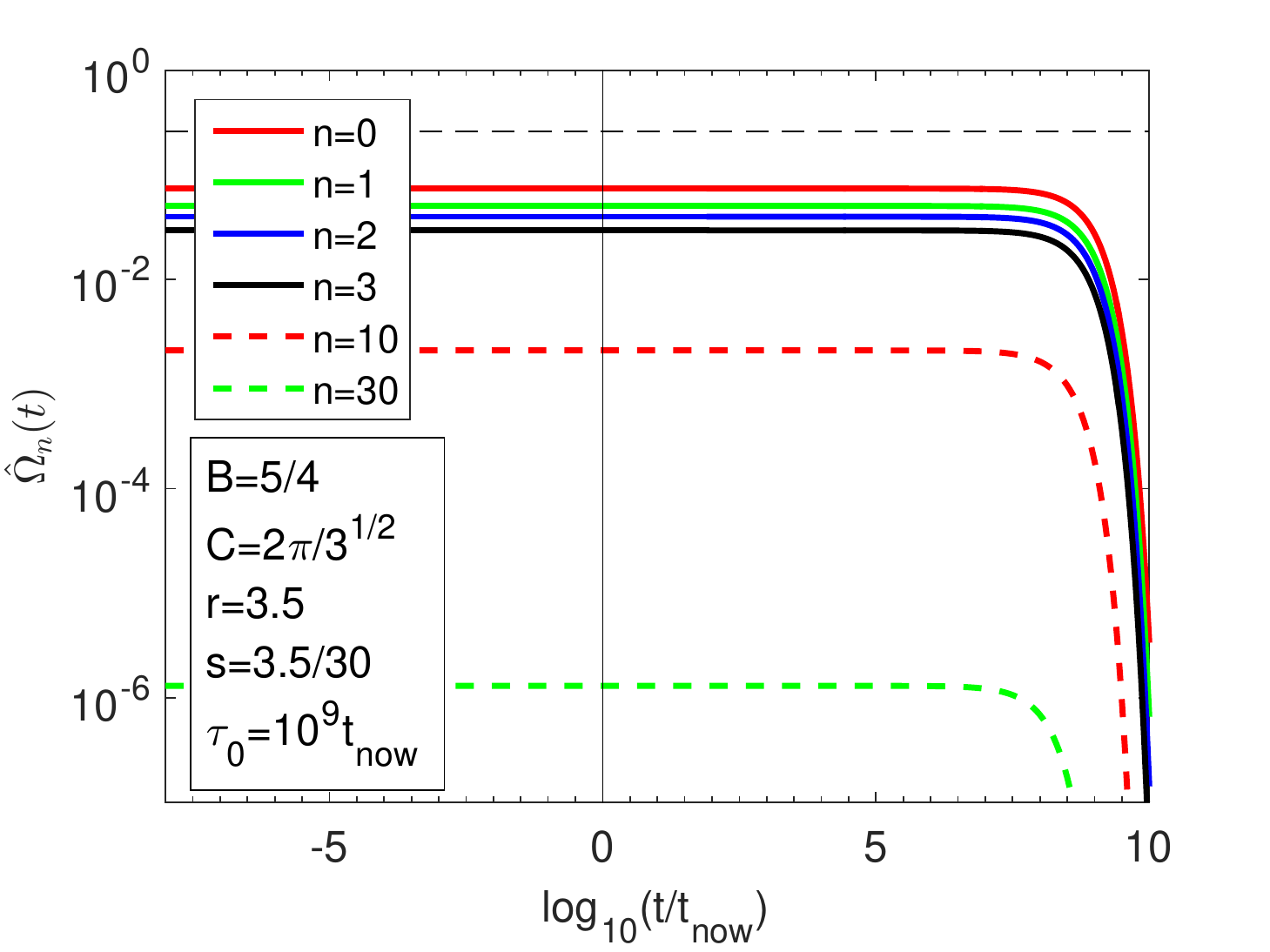}
\caption{The level-by-level aggregate cosmological 
abundances $\widehat\Omega_n\equiv g_n \Omega_n$ of our benchmark DDM model, 
plotted as functions of time for a series of low-lying mass levels $n$.
We see that the lightest states decay later and carry the largest cosmological
abundances, while the heavier states decay earlier and carry smaller cosmological
abundances --- a key feature of the  DDM framework.
As required, the sum of all abundance contributions at $t=\tnow$ is 
$\Omega_{\rm tot}(t_{\rm now}) = \OmegaCDM \approx 0.26$.  }
\label{fig7}
\end{figure*}

\begin{figure*}
\centering  
    \includegraphics[width=0.45\textwidth]{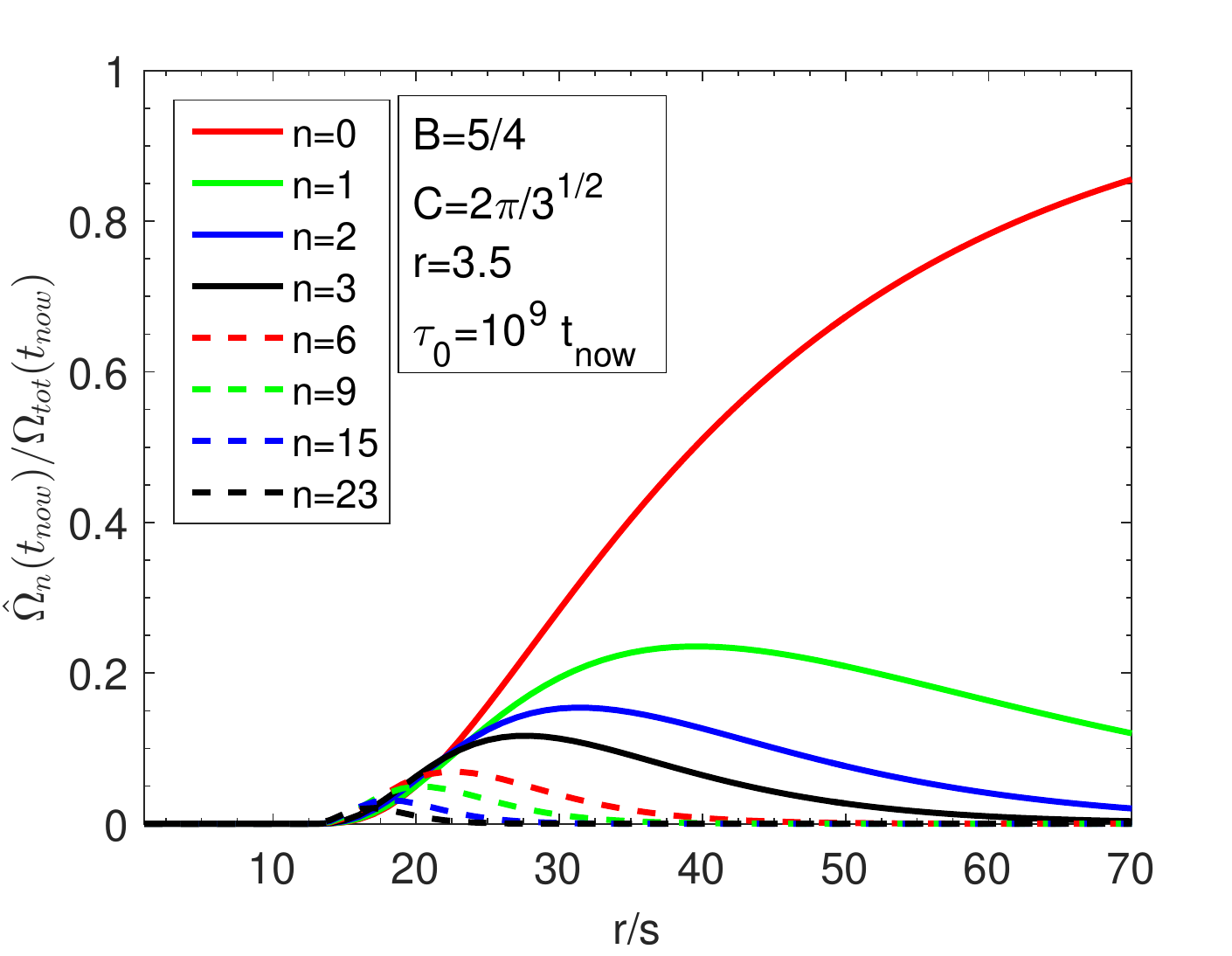}
\hskip 0.05 truein
    \includegraphics[width=0.475\textwidth]{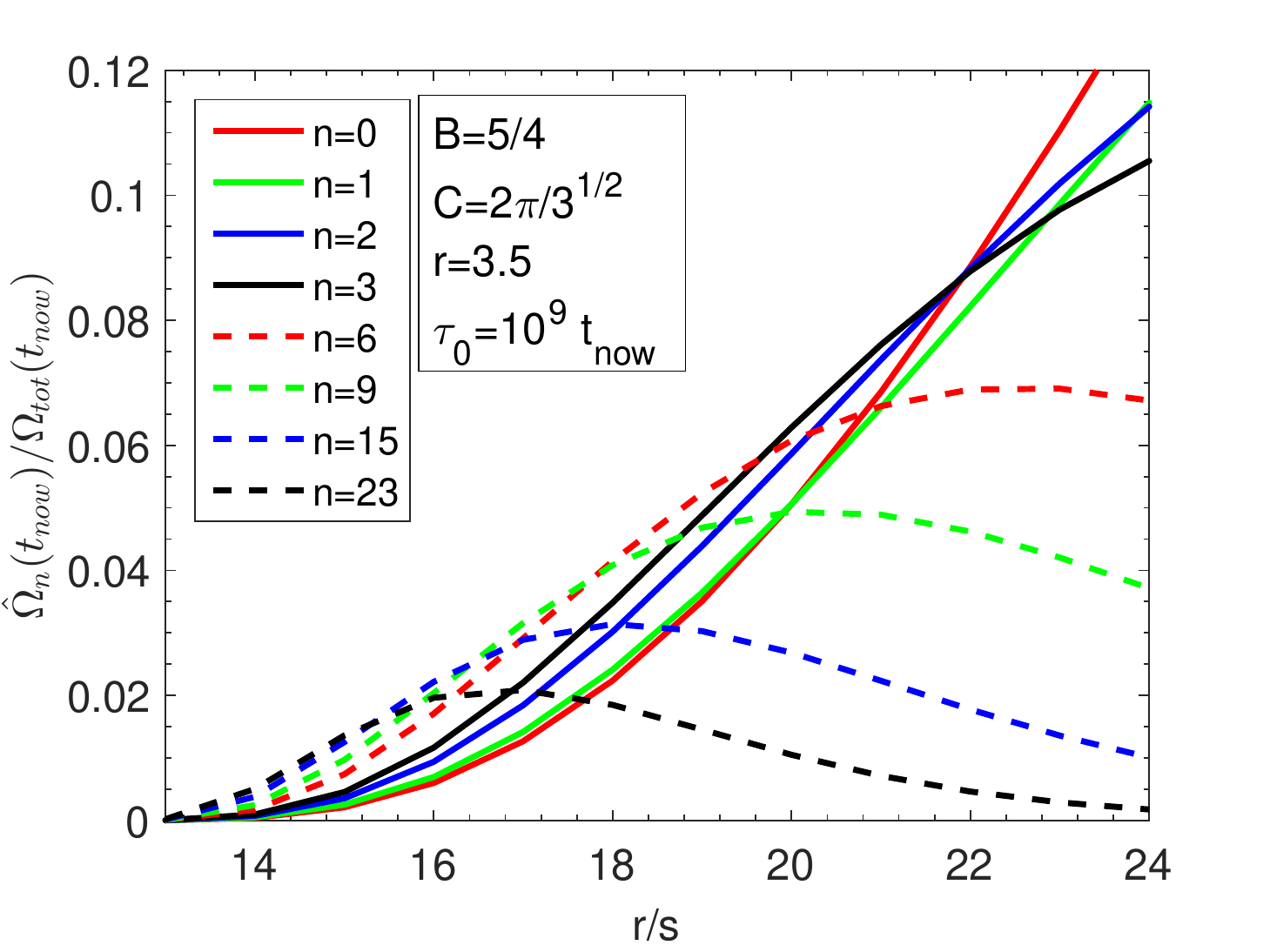}  
\caption{{\it Left panel}\/:  Present-time aggregate abundance 
fractions $\widehat \Omega_n(t_{\rm now})/\Omega_{\rm tot}(t_{\rm now})$, 
plotted as functions of $r/s$.    As $r/s$ increases, the $n=0$ state
carries an increasingly large
fraction of the total abundance,
resulting in scenarios which have smaller values of $\eta$
and which are therefore less DDM-like.
By contrast, for smaller $r/s$, we see that the lightest state
carries a smaller proportional fraction of the total abundance
and in fact may not even be the dominant state for sufficiently small $r/s$.
{\it Right panel}\/:   A zoom-in of the left panel, illustrating 
how the level $n$ of the 
states carrying the largest collective abundance $\Omega_n(\tnow)$ shifts as a function of $r/s$.
For example, for $r/s=15$, all states carry relatively small abundances and 
it is actually the $n=23$ states which collectively carry
the largest collective abundance at the present time.  Such scenarios are therefore extremely DDM-like.}
\label{fig8}
\end{figure*}

We have seen in Fig.~\ref{fig6} how the total abundances $\Omegatot$ of our DDM ensembles
vary as a function of time.
However, it is also interesting to understand how the {\it individual}\/ aggregate abundances $\widehat\Omega_n(t)$
at each mass level $n$ contribute to this behavior.
The result is shown in Fig.~\ref{fig7} for our benchmark DDM model.
As we see from Fig.~\ref{fig7}, there are  many mass levels $n$ 
whose states 
contribute to $\Omegatot(\tnow)$:  states with smaller values of $n$ carry larger abundances
and have longer lifetimes, persisting into later times before decaying,
while those with larger values of $n$ carry smaller abundances
and have shorter lifetimes, decaying earlier.
Indeed, this balancing between lifetimes and abundances is a fundamental hallmark 
of the DDM framework.
Although the sum of these abundances at $t=\tnow$ is fixed at $\Omegatot(\tnow)=\OmegaCDM\approx 0.26$,
we see that even states with relatively large values of $n$  
have lifetimes $\tau_n$ exceeding $t_{\rm now}$   
and thus contribute non-trivially to $\Omegatot(\tnow)$.
Indeed, for our benchmark model, we find that 
there are no fewer than seven distinct mass levels 
contributing more than 0.01 to $\Omegatot(\tnow)$
and no fewer than ten distinct mass levels 
contributing more than $1\%$ of $\Omegatot(\tnow)$.

\begin{turnpage}
\begin{figure*}
\centering  
    \includegraphics[width=0.32\textwidth]{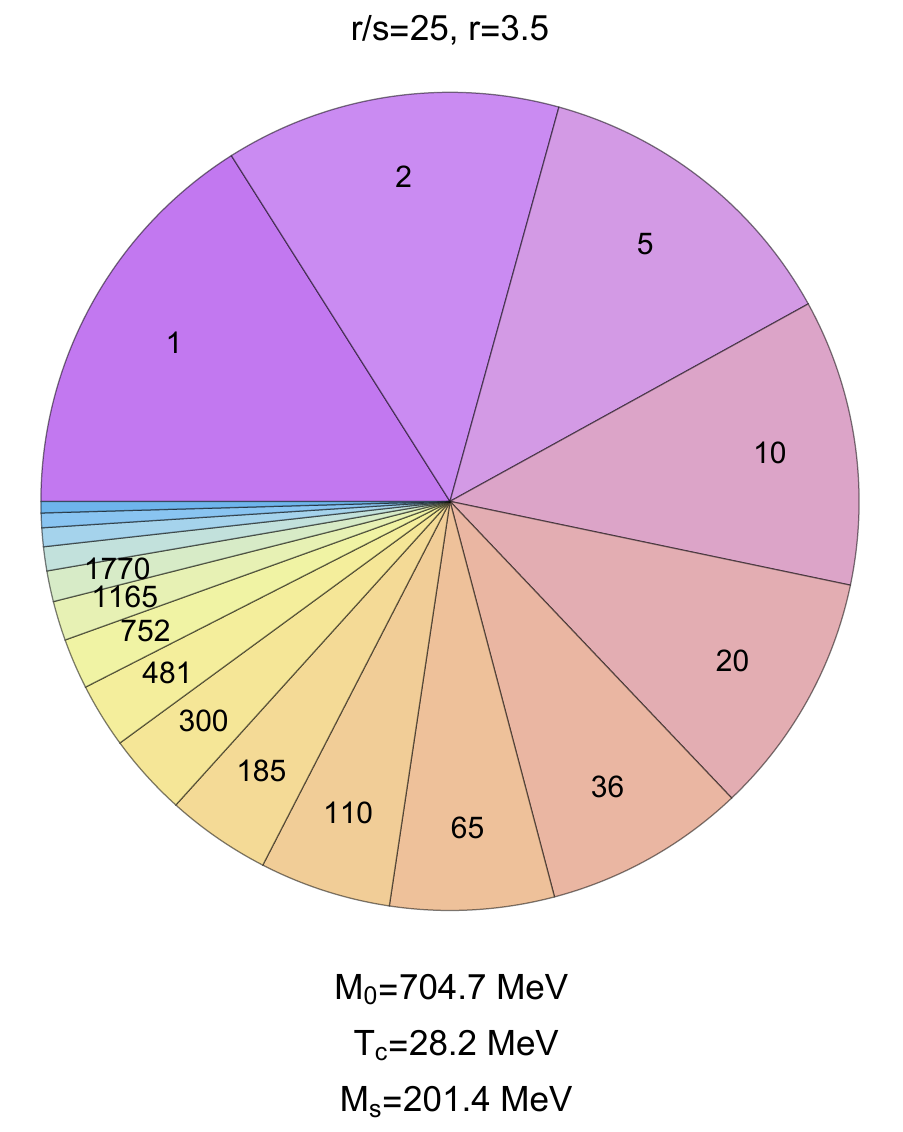}
    ~\includegraphics[width=0.32\textwidth]{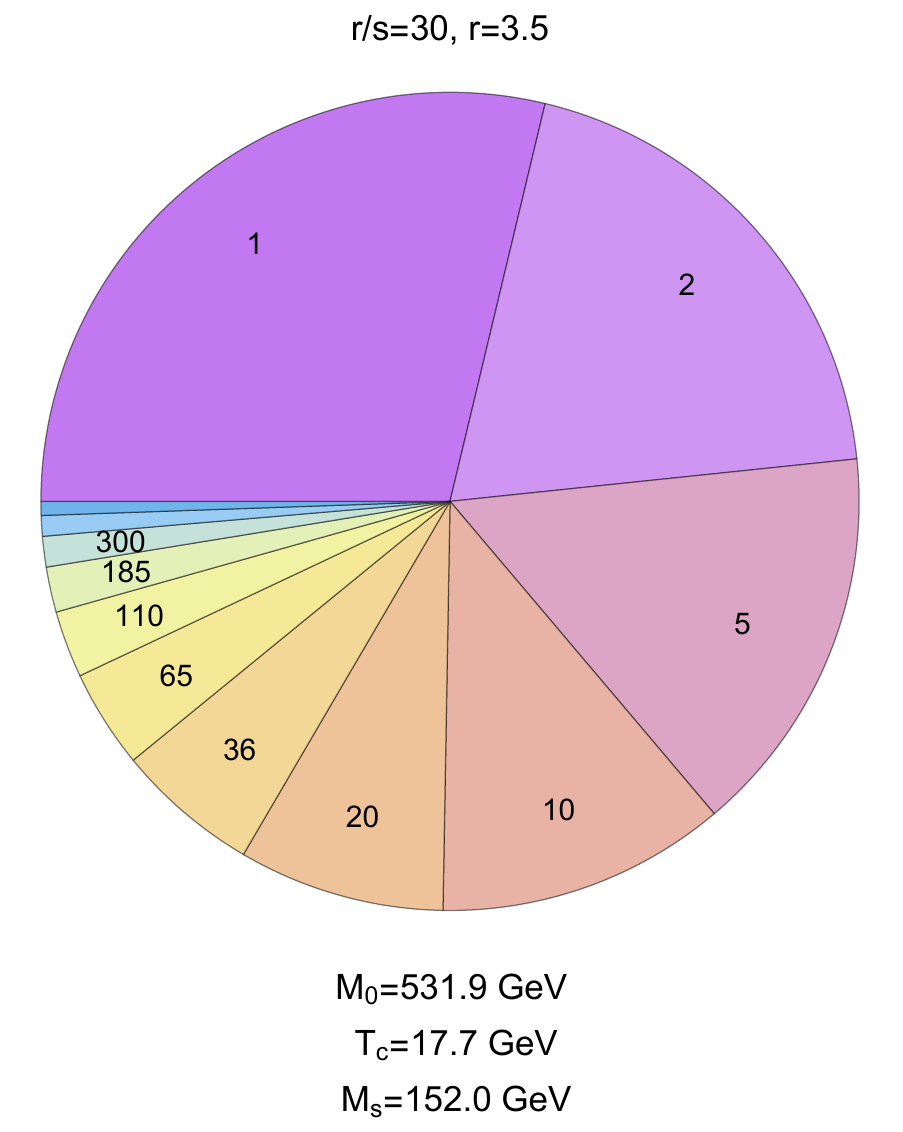}
    ~\includegraphics[width=0.32\textwidth]{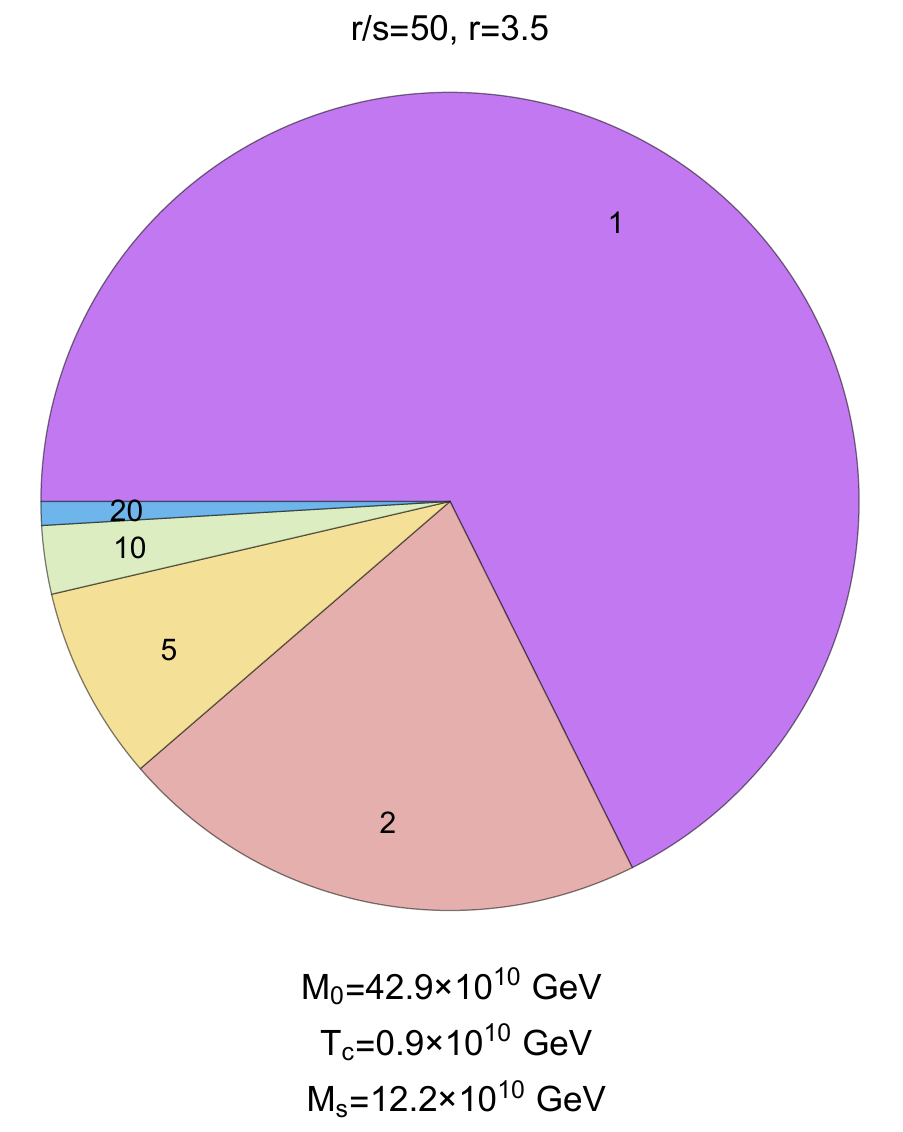}
    ~\includegraphics[width=0.32\textwidth]{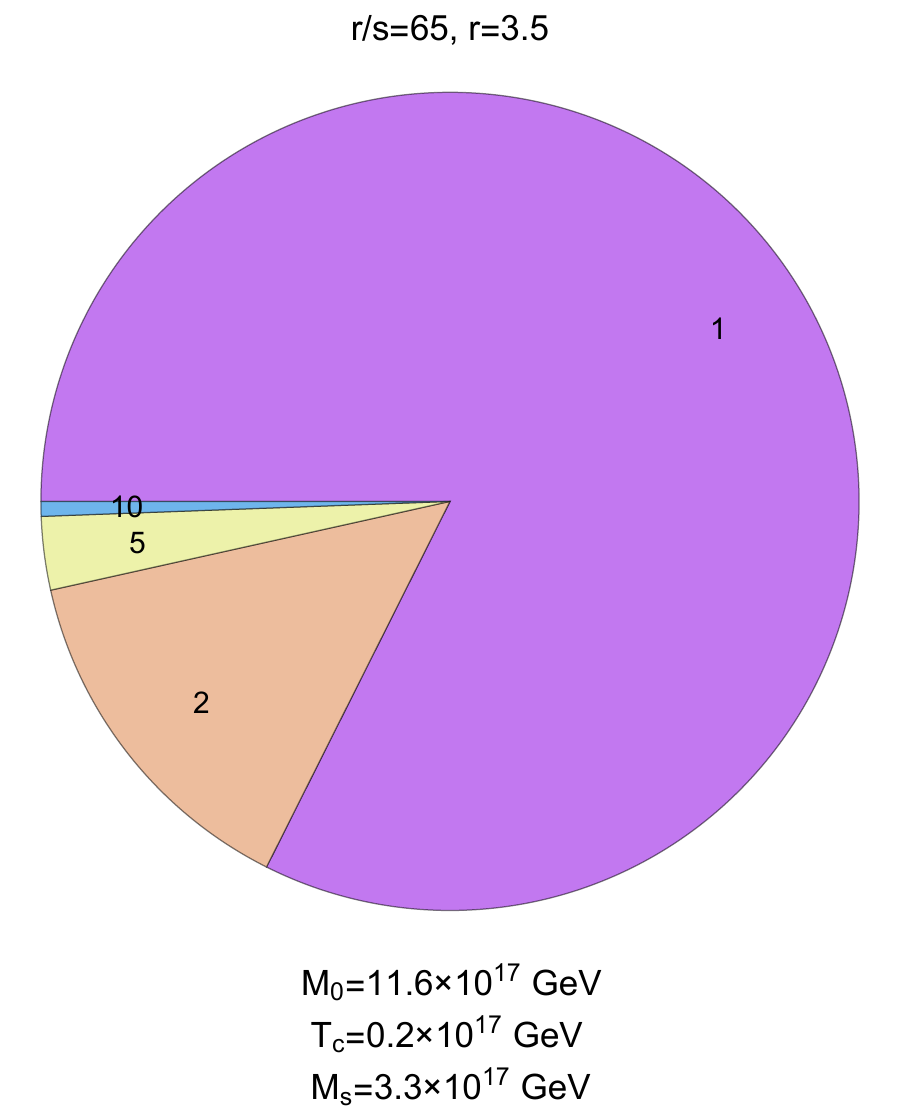}
\vskip 0.5 truein
    \includegraphics[width=0.32\textwidth]{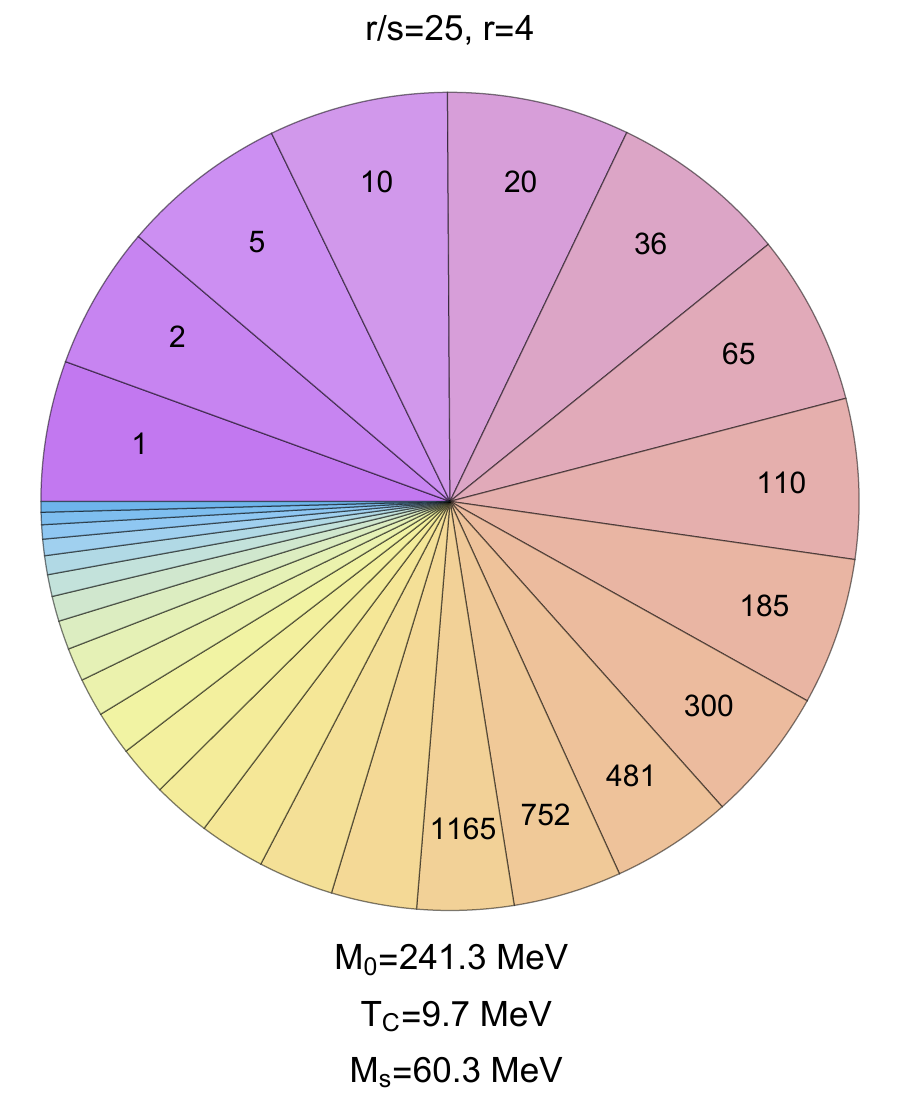}
    ~\includegraphics[width=0.32\textwidth]{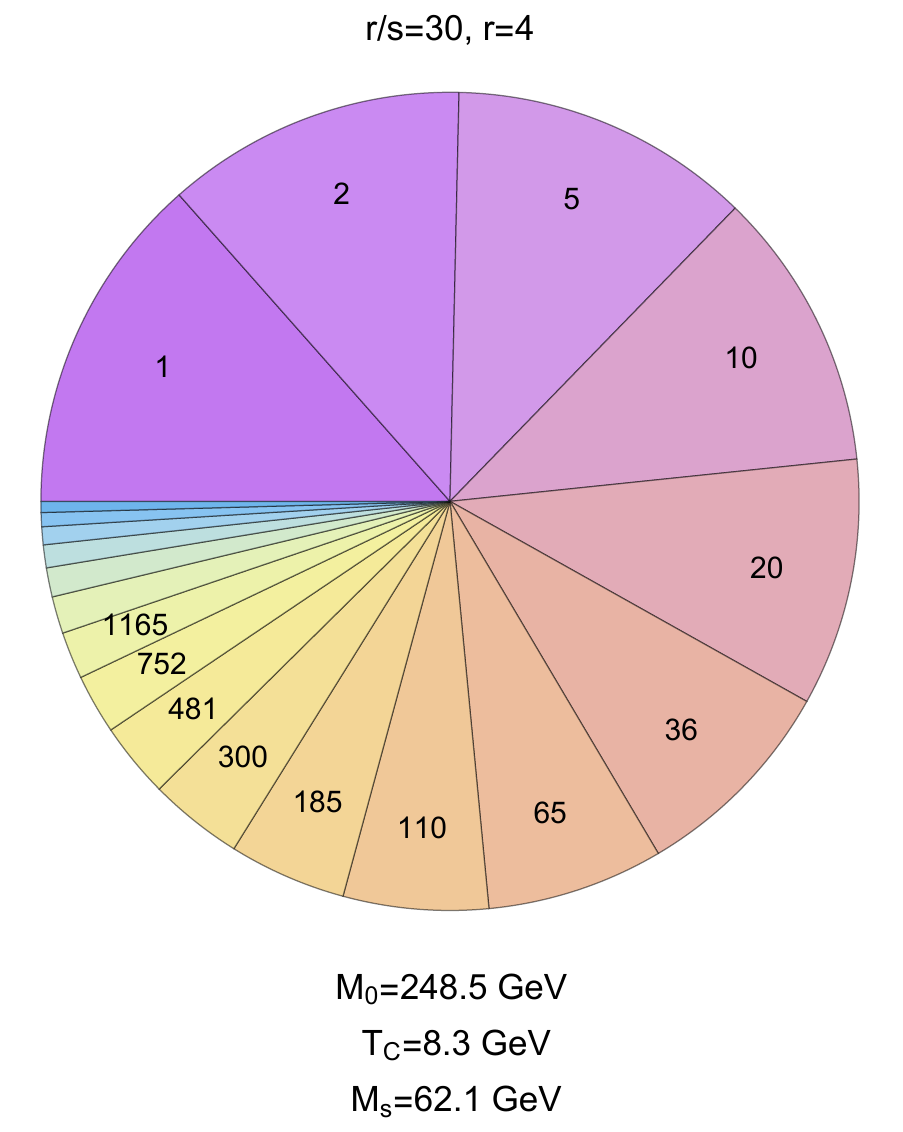}
    ~\includegraphics[width=0.32\textwidth]{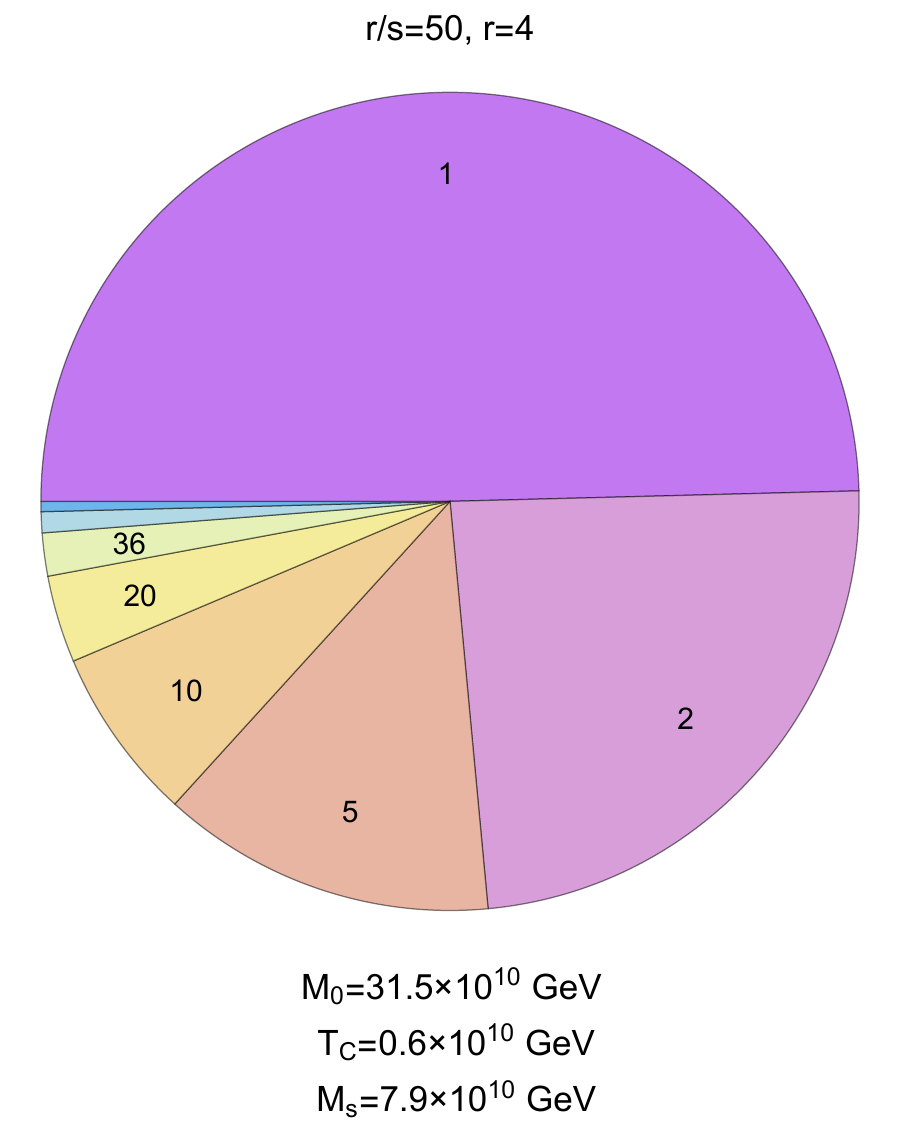}
    ~\includegraphics[width=0.32\textwidth]{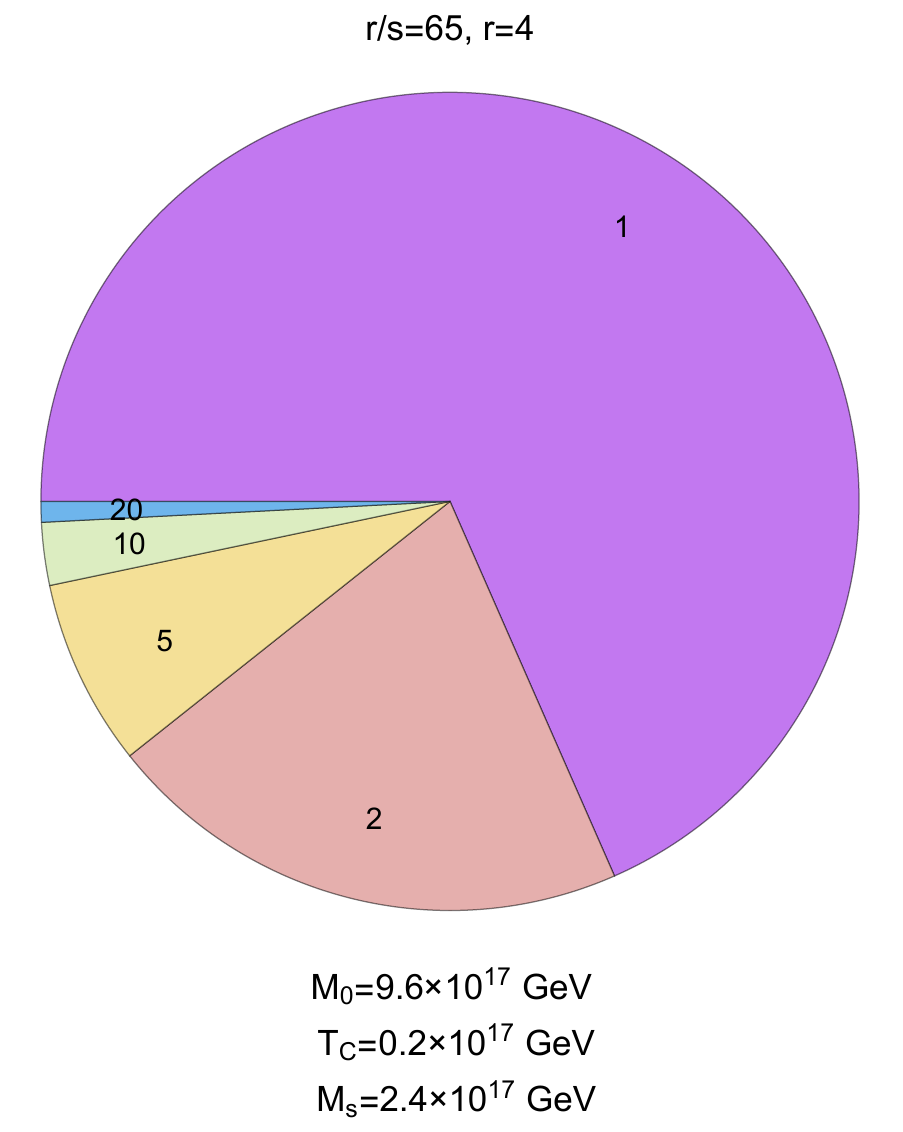}
\caption{Pie for everyone!
Dark-matter pie charts showing the relative contributions to 
$\Omega_{\rm tot}(t_{\rm now})=\OmegaCDM \approx 0.26$ from 
the lowest-lying states
for $r=3.5$ (top row) and $r=4$ (bottom row),
with $r/s= \lbrace 25, 30, 50, 65\rbrace$ across each row
and with $B=5/4$ and $C=2\pi/\sqrt{3}$ held fixed.
In the majority of cases (but not all cases), the largest pie slice corresponds to the abundance contribution from 
the $n=0$ mass level, and the successively smaller pie
slices (progressing in a clockwise fashion within the pie chart) 
correspond to the aggregate abundance contributions from 
successively higher mass levels.    
Within each pie slice we have also indicated the degeneracy $\hat g_n$ 
of individual states whose cosmological  abundances comprise the contribution from that slice.
Note that in each case, the pie shown only corresponds  to the dark-matter 
slice $\OmegaCDM \approx 0.26$ 
of the bigger ``cosmic pie'' which also includes contributions from dark energy and visible matter. 
 }
\label{fig9}
\end{figure*}
\end{turnpage}

\begin{turnpage}
\begin{figure*}
\centering  
    \includegraphics[width=0.32\textwidth]{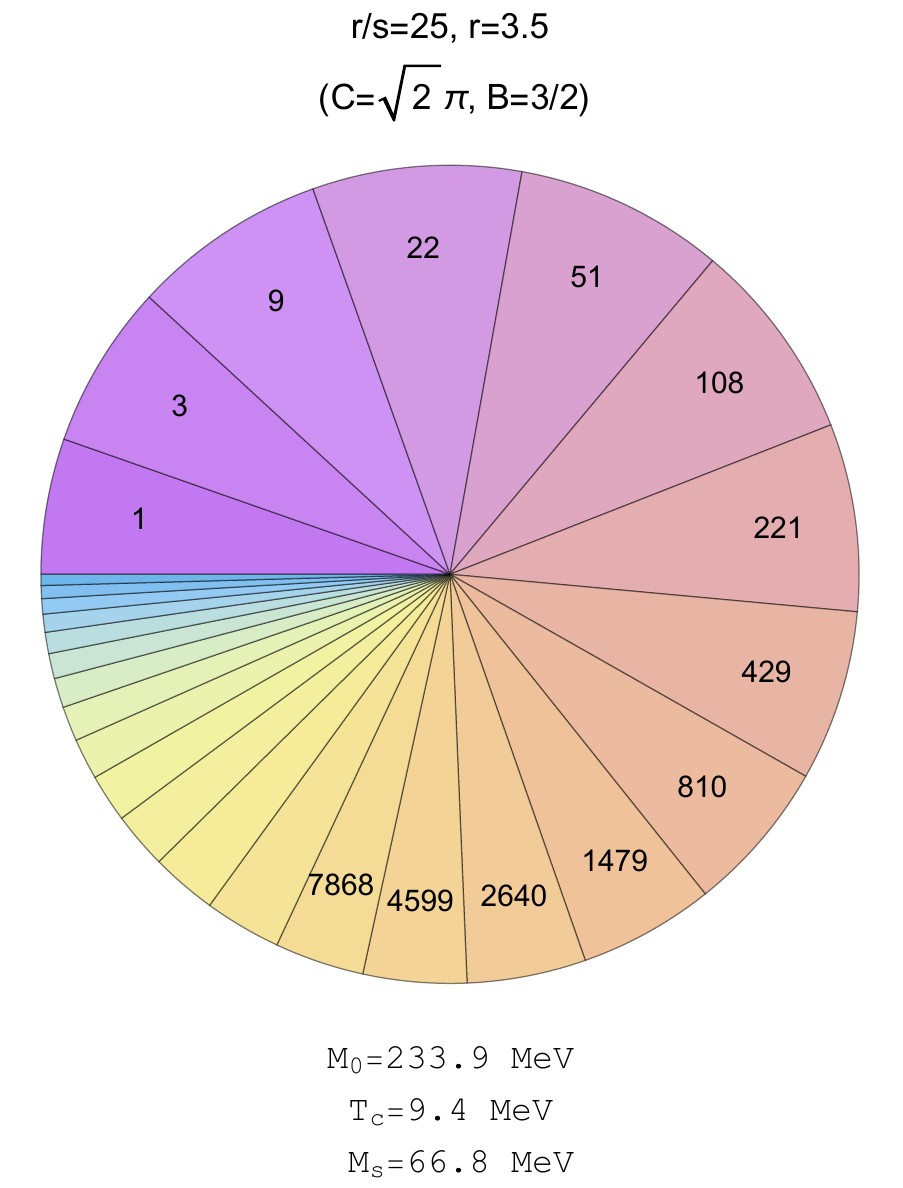}
    ~\includegraphics[width=0.32\textwidth]{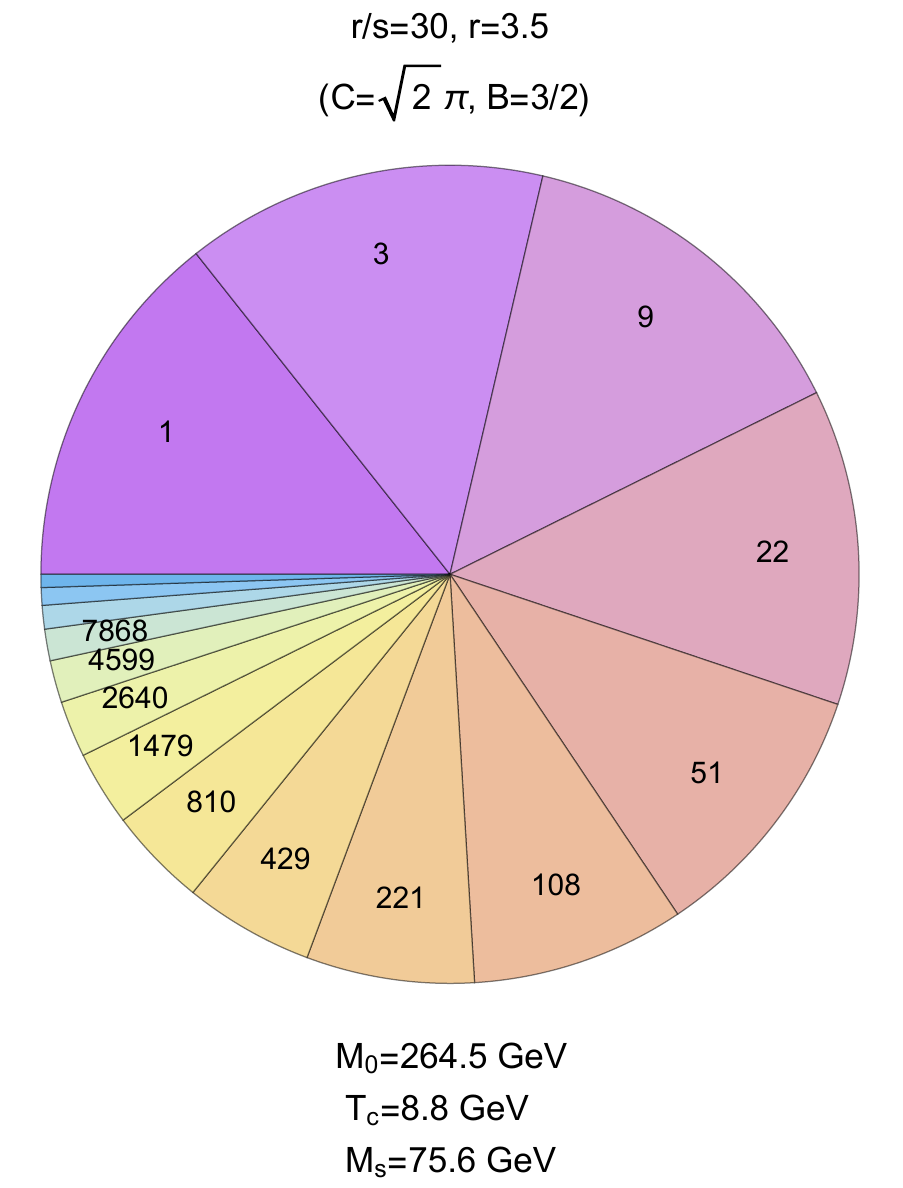}
    ~\includegraphics[width=0.32\textwidth]{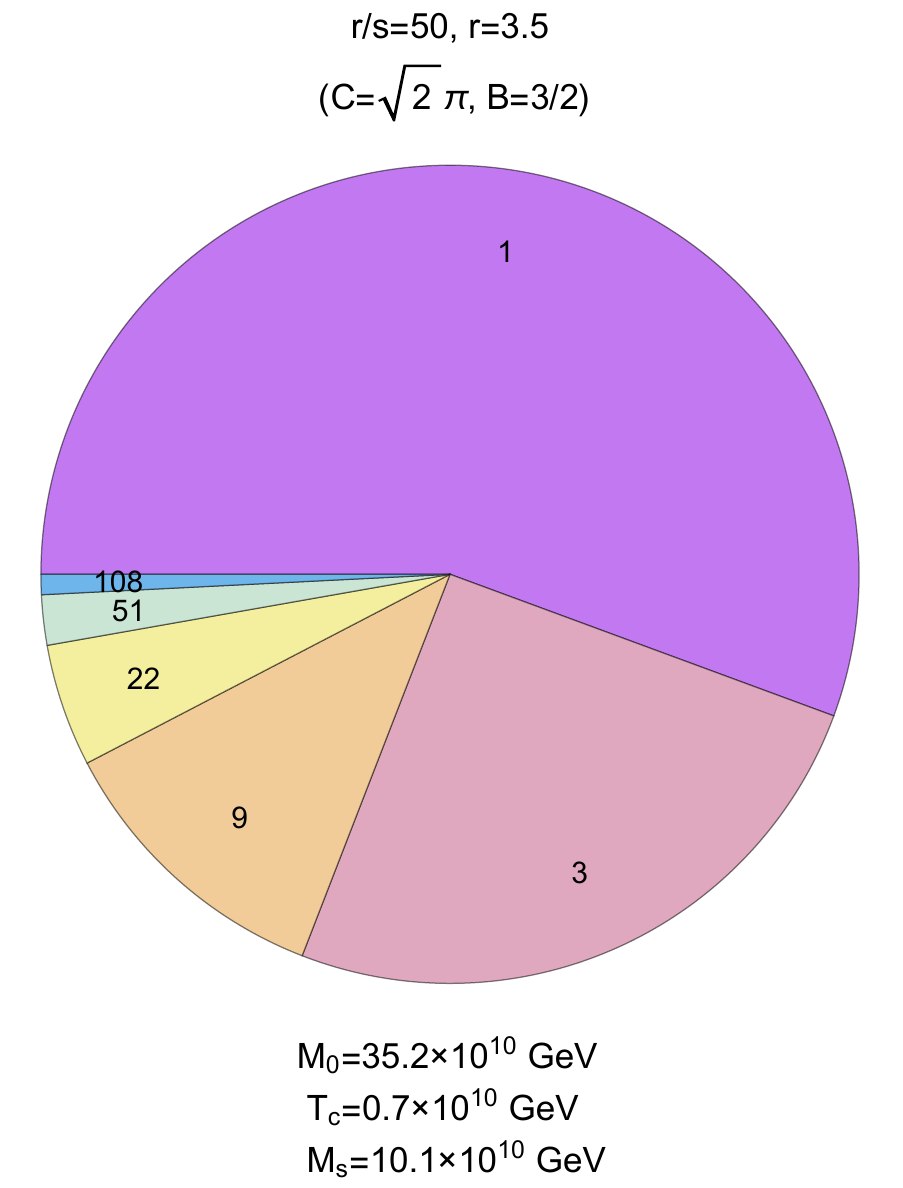}
    ~\includegraphics[width=0.32\textwidth]{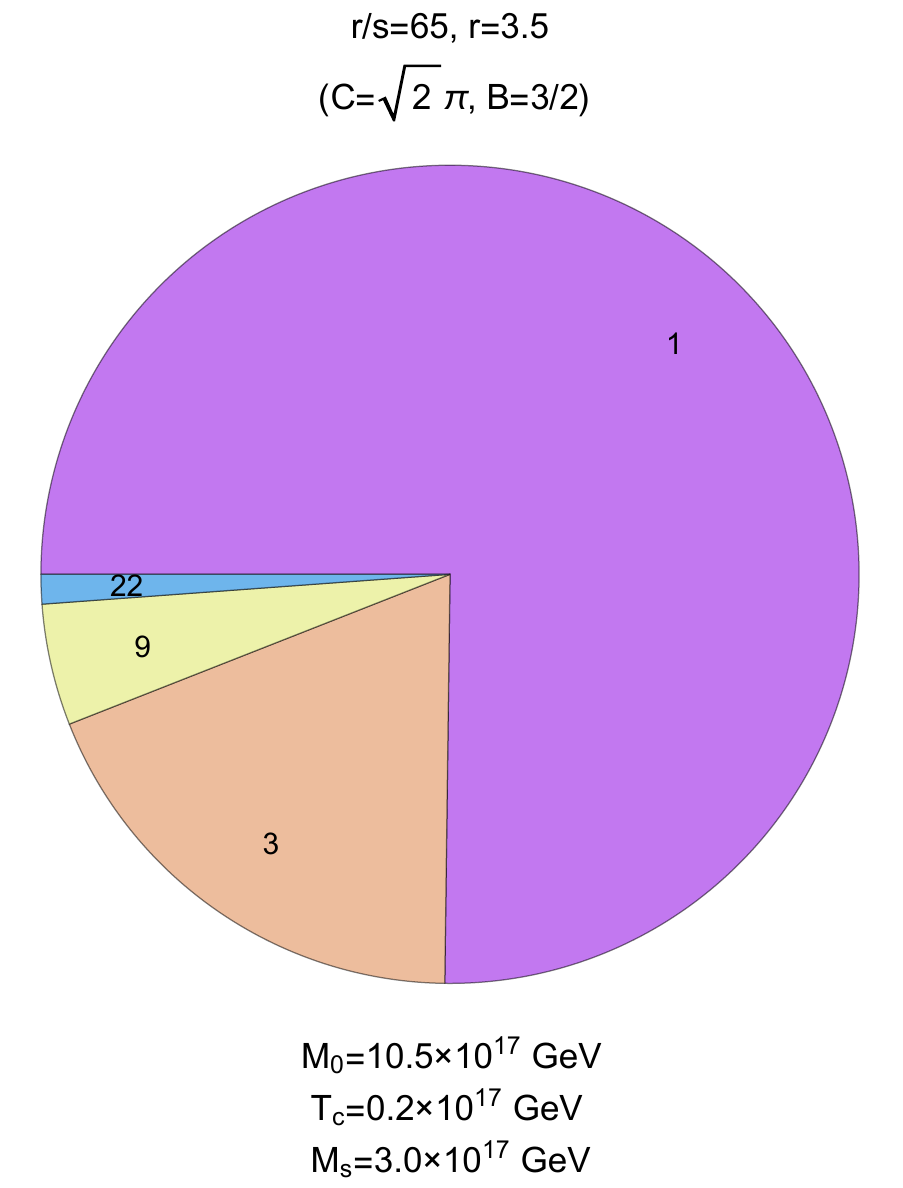}
\vskip 0.3 truein
\caption{Seconds on pie!
Same as the top row of Fig.~\protect\ref{fig9} 
except that we have now taken $C=\sqrt{2}\pi$ and $B= 3/2$,
corresponding to the $D_\perp =3$ scalar string.
These changes in $C$ and $B$ change the degeneracies $\hat g_n$ of states at each mass level,
as indicated within the corresponding pie slices,
and result in ensembles which are even more DDM-like and which 
have correspondingly smaller mass scales than those along the top row of Fig.~\protect\ref{fig9}.} 
\label{fig10}
\end{figure*}
\end{turnpage}

It is also interesting to examine how these results vary
as a function of the ratio $r/s$ which, as we  have seen, governs the overall mass  scales
associated with these DDM ensembles.
The results are shown in Fig.~\ref{fig8}, where we plot
the aggregate fractions $\widehat \Omega_n(\tnow)/\Omegatot(\tnow)$ 
for a variety of different mass levels $n$ as a function of $r/s$.
As evident in Fig.~\ref{fig8}, 
the lighest state carries a larger and larger fraction of the total abundance
as $r/s$ increases, 
resulting in scenarios which have smaller values of $\eta$
and which are therefore less DDM-like.
By contrast, the lightest state
carries a smaller proportional fraction of the total abundance as $r/s$ decreases,
and in fact may not even be the dominant state for sufficiently small $r/s$.
Indeed, for $r/s=15$, we find that all states carry relatively small abundances, and 
it is actually the states at the $n=23$ mass level which collectively carry
the largest individual abundance at the present time.  
Such scenarios are therefore extremely DDM-like.

Putting all the pieces together,
we can summarize our results as in Figs.~\ref{fig9} and \ref{fig10}.~
Fig.~\ref{fig9} consists of 
a sequence of dark-matter pie charts showing the relative contributions to 
$\Omega_{\rm tot}(t_{\rm now})=\OmegaCDM \approx 0.26$ from 
the lowest-lying states
for $r=3.5$ (top row) and $r=4$ (bottom row),
with $r/s= \lbrace 25, 30, 50, 65\rbrace$ across each row.
Within each pie, 
we illustrate the corresponding collective abundances 
$\widehat\Omega_n(\tnow)$ 
as separate slices, one for each value of $n$,
while the numbers listed within each slice indicate the number 
of individual states $\hat g_n$ contributing at that mass level. 
For each pie chart we have also shown the corresponding values of $M_0$, 
$T_c$, and $M_s$. 
For these calculations we have used the input values
$T_{\rm MRE}=0.7756$~eV,
$g_{\rm MRE}=3.36$,
and $ g_c=\lbrace 10.75, 61.75, 106.75, 106.75\rbrace$, respectively, for $r/s=\lbrace 25, 30, 50, 65\rbrace$.
We have also assumed our standard benchmark values 
$B=5/4$, $C=2\pi/\sqrt{3}$, and $\tau_0=10^9\tnow$.

Let us begin by focusing on the ``benchmark'' pie chart within Fig.~\ref{fig9} corresponding 
to $r=3.5$ and $r/s=30$.  For this pie chart, we see that the
largest pie slice corresponds to the abundance contribution from 
the $n=0$ mass level, while the successively smaller pie
slices progressing in a clockwise fashion within the pie chart 
correspond to the abundance contributions from 
successively higher mass levels.    
For this pie chart, we find that $M_0\approx 532$~GeV,
$T_c\approx 18$~GeV, and $M_s\approx 152$~GeV.
Note that this value for $M_s$ is in agreement with the $M_s$ contours shown in Fig.~\ref{fig3}. 
We also see geometrically from this pie chart that $\eta\approx 0.72$, in agreement with the results
shown in Figs.~\ref{fig3}, \ref{fig4}, and \ref{fig6}.

Given this, we can now investigate how this benchmark pie chart deforms as a function of $r/s$ and $r$.
Results are illustrated in the other pie charts shown in Fig.~\ref{fig9}.
We see in general that increasing $r$ from 3.5 to 4.0 (\ie, passing from the 
top row of pie charts in Fig.~\ref{fig9} to the bottom row)
has the net effect of shifting cosmological abundance away from
the ground state, thereby increasing $\eta$ 
and generally making each pie slice smaller while simultaneously lowering the corresponding mass scales.
This is in complete accord with the results shown in Fig.~\ref{fig3}. 
Likewise, decreasing or increasing $r/s$ (\ie, moving left or right along either row) 
has the effect of increasing or decreasing $\eta$ while decreasing or increasing our corresponding mass scales.
Indeed, we see that the variable $r/s$ allows us to interpolate between two extremes:   traditional ensembles with
high mass scales at large $r/s$ versus DDM-like ensembles with smaller mass scales at small $r/s$.
We further observe that for sufficiently small $r/s$, the largest pie slice is no longer the $n=0$ slice 
(labelled `1' in each pie chart) --- as $r/s$ decreases,
this honor gradually shifts towards the pie slices corresponding
to higher mass levels.
This is in accordance with the results in Fig.~\ref{fig8}.

Fig.~\ref{fig10} is similar to the top row of Fig.~\ref{fig9}, except that we have now 
increased our values of $C$ and $B$ to $\sqrt{2}\pi$ and $B=3/2$, respectively.
These new values maintain $c_{\rm int}=0$ and correspond to the $D_\perp=3$ scalar string.
These changes in $C$ and $B$ increase the degeneracies $\hat g_n$ of states at each mass level, with
the new values indicated within the corresponding pie slices.
Although the cosmological abundances per state are not affected by the changes in $C$ and $B$,
these increased degeneracies result in ensembles which are even more DDM-like and which 
have correspondingly smaller mass scales than those along the top row of Fig.~\protect\ref{fig9}.
These results are consistent with those shown in Fig.~\ref{fig4}.

We see, then, that a tremendous variety of DDM ensembles 
exist which have the two fundamental features outlined in the Introduction ---
Regge trajectories and exponentially rising degeneracies of states.
These ensembles are consistent with our 
look-back and $\weff$ constraints, and thus satisfy the 
zeroth-order
constraints that may be imposed on such ensembles
on the basis of their total energy densities and equations of state alone.
We also observe an important feature, a inverse correlation between
the tower fraction $\eta$ (which governs the extent to which
our ensemble is truly DDM-like) and the magnitude of its underlying mass scales.
Indeed, we have seen that 
while traditional ensembles typically have high corresponding mass scales,
our ensembles become increasingly DDM-like for lower mass scales --- all while 
remaining consistent with our look-back and $\weff$ constraints.
These observations will likely be an important guide and ingredient 
in any future attempts to build realistic dark-matter models of this type.


\section{Conclusions\label{sec:Conclusion}}

 
In this paper, we have investigated the properties of a hitherto-unexplored class of 
DDM ensembles whose constituents are the composite states which emerge in the confining phase of 
a strongly-coupled dark sector.  In ensembles of this sort, the masses of the
constituent particles lie along well-defined Regge trajectories
and the density of states 
within the ensemble grows exponentially as a function of the constituent-particle mass.
This exponential growth is ultimately compensated by a Boltzmann suppression factor
in the primordial abundances of the individual constituents, resulting in a finite total
energy density $\Omega_{\rm tot}(t)$.
We also showed that such ensembles can naturally exhibit a balancing between lifetimes
and cosmological abundances of the sort required by the DDM framework.

For each such ensemble, we calculated the corresponding effective equation-of-state
parameter $\weff(t)$ as well as the tower fraction $\eta(t)$.
We also imposed a number of zeroth-order model-independent phenomenological constraints 
which follow directly from knowledge of $\Omegatot(t)$, $\weff(t)$, and $\eta(t)$.
In general, we found that the imposition of such constraints
tends to introduce {\it correlations}\/ between the different underlying variables which 
parametrize our DDM ensembles, so that an increase in one variable (such as, \eg,  the exponential
rate of growth in the state degeneracies) requires a corresponding shift in another variable (in this case,
an increase in the lifetime of the lightest state in the ensemble, as indicated in the right
panel of Fig.~\ref{fig5}).  
Perhaps one of our most important results is the existence of
an inverse correlation 
between the tower fraction $\eta(t)$ 
associated with a given a DDM ensemble and its corresponding fundamental mass scales,
so that the present-day cosmological abundance of the dark sector must be distributed across an increasing number of
different states in the ensemble as these fundamental mass scales 
are dialed from the Planck scale down to the GeV scale.

We are certainly not the first to consider dark-matter scenarios 
in which the dark matter is composite.  
Indeed, within the context of traditional dark-matter models, it has been appreciated for
some time that the dark-matter particle could be a composite state.  For example, the 
lightest technibaryon in technicolor theories was long ago identified as a promising 
dark-matter candidate~\cite{NussinovTechnibaryon,BarrTechnibaryon}, and 
mechanisms~\cite{GudnasonTechnibaryon} were advanced by which this particle could 
be rendered sufficiently light so as to be phenomenologically viable.  Indeed,
several explicit models~\cite{RyttovTIMP} have been developed along these lines.  
Other more exotic baryon-like composites 
have also been advanced as potential dark-matter candidates~\cite{GUTzilla}.
Lattice studies of baryon-like states in the confining phases of both $SU(3)$ and 
$SU(4)$ gauge theories have also been 
performed~\cite{AppelquistLattice1,AppelquistLattice2,AppelquistLattice3}.
  
A variety of scenarios in which a long-lived meson-like state which appears in 
the confining phase of a strongly-coupled hidden sector have been
developed as well (for a review see, \eg, Ref.~\cite{KribsReview}).  These include
scenarios in which the dark-matter particle is a pseudo-Nambu-Goldstone boson (PNGB)
stabilized by a dark-sector analogue of flavor 
symmetry~\cite{KilicDarkPion,HurKoPion,HolthausenPion,HatanakaPion,AmetaniPNGB} or 
$G$-parity~\cite{BaiPion}, or alternatively by some other symmetry of the 
theory with no SM analogue~\cite{Ectocolor,BhattacharyaPion,CarmonaPNGB,FrigerioHeavyEta}. 
Complementary lattice studies of strongly-coupled dark-sector scenarios in which 
the dark-matter candidate is a PNGB have been performed as 
well~\cite{LewisLattice,HietanenLattice}.  Scenarios 
in which the dark-matter candidate is not a PNGB, but rather a bound state of one
heavy quark and one light quark, have 
also received recent attention~\cite{CiDM,CiDMParity,CiDMCosmology}, primarily due
to the non-standard direct-detection phenomenology to which they give rise, as
have scenarios in which the dark-matter candidate is a bound state of heavy
quarks alone~\cite{KribsQuirks}.  More general studies of composite hidden-sector 
theories which give rise to meson-like or baryon-like dark-matter candidates within 
different regions of parameter space have also been 
performed~\cite{AntipinCompositeHS1,AntipinCompositeHS2}. 

Composite hidden-sector states consisting of non-Abelian gauge fields alone 
(so-called ``glueball'' states) have also long been recognized as promising dark-matter 
candidates~\cite{OkunThetons1,OkunThetons2} --- a possibility which has received
renewed attention~\cite{SoniGlueballs,ForestellGlueballs} as well.
Indeed, hidden sectors involving cosmologically stable dark glueball states arise
naturally in a variety string constructions~\cite{FaraggiGlueballs,HalversonGlueballs},
as well as in certain anomaly-mediated supersymmetry-breaking 
scenarios~\cite{FengShadmiWIMPless}. 
 
In addition, the possibility that composite states in the dark sector 
could themselves form bound states (so-called ``dark nuclei'') has also been
studied~\cite{DetmoldDarkNuclei1,KrnjaicDarkNuclei}, as has the possibility 
that these nuclei themselves could combine to form dark ``atoms'' or even 
dark ``molecules''~\cite{ClineDarkMolecules,BoddyDarkHydrogen}.  Indeed, lattice 
studies~\cite{DetmoldDarkNuclei1,DetmoldDarkNuclei2} corroborate the
existence of stable dark nuclei states even within simple, two-flavor models 
with $SU(2)$ as the confining gauge group.  In such models, a dark-sector
equivalent of BBN serves as the mechanism for abundance
generation.  Such models can have interesting phenomenological consequences,
especially in the regime in which a significant fraction of the dark-matter 
abundance is contributed by nuclei with large nucleon 
numbers~\cite{LargeDarkNuclei1,LargeDarkNuclei2}.

Composite dark-matter models are interesting from a phenomenological perspective
as well.  For example, the states of a strongly-coupled hidden sector provide
a natural context~\cite{SIMPParadigmWacker} for strongly-interacting massive particle 
(SIMP) dark matter~\cite{SIMPParadigmCarlson,SIMPParadigmdeLaix} models,
in which $3\rightarrow 2$ processes rather than $2\rightarrow 2$ processes
play a dominant role in determining the dark-matter abundance.  Indeed, a 
number of explicit models along these lines have 
been constructed~\cite{SIMPModelHochberg,SIMPModelHansen,SIMPModelBernal,SIMPModelKamada}. 
One of the most interesting
ramifications of SIMP models is that they naturally give rise to dark-matter 
self-interactions with cross-sections sufficiently large that dark-matter scattering 
can have an observable impact on structure formation~\cite{BoddySIDM1}.
Such composite dark-matter models can have other phenomenological consequences as
well, both at indirect-detection experiments~\cite{BoddySIDM2,DarkShower} and at 
colliders~\cite{SIMPPhenoLee,SIMPPhenoHochberg,EnglertSIMPCollder,BruggisserSIMPCollider}.
Finally, the presence of additional non-Abelian gauge sectors, each with their own
analogue of the QCD $\Theta$-angle, could have potential implications for the 
physics of axions and axion-like particles~\cite{MultipleThetaAngles}.

While all of these represent theoretically viable possibilities for the dark sector,
the dark ensemble we have considered in this paper is unique for several important reasons.
In traditional composite dark-matter models, it is usually a single bound 
state (usually the lightest bound state) 
which serves as the primary dark-matter candidate and which therefore carries the full
dark-matter abundance $\OmegaCDM$.
While there may be several other dark states to which this bound state couples
--- and which may
play a role in determining the abundance 
of the dark-matter candidate ---
it is nevertheless true that only one (or a few) composite states carry the dark-matter abundance
$\OmegaCDM$ and thereby play a significant role in dark-sector phenomenology.
By contrast, within the DDM framework, the dark-matter abundance is potentially spread
across a relatively large set of composite states with various masses and lifetimes.
Thus the usual required stability of the traditional dark-matter candidate is
not a required feature of the DDM ensemble, 
thereby allowing the associated dark-matter abundance $\OmegaCDM(t)$ and 
dark-matter equation-of-state parameter $\weff(t)$
to vary with time --- even during the current, matter-dominated era.

Moreover, because the DDM framework requires an enlarged viewpoint in which the entire
spectrum of composite states are potentially relevant for determining the
properties of the dark sector,
features that describe the entire composite spectrum suddenly become relevant for
determining dark-sector phenomenology  --- features which would not have
been relevant for previous studies within more traditional frameworks.
These features include the fact that the masses of such bound states 
actually lie along Regge trajectories, and that the densities of such bound states
experience a Hagedorn-like exponential growth as a function of mass.
Indeed, these features do not play a role within traditional studies
of composite dark states, but they have been the cornerstones
of the analysis we have presented here.
In this context, we note that a similar approach was also adopted in 
Ref.~\cite{LargeDarkNuclei1} with regard to ensembles of dark nuclei whose abundances 
are generated via a dark-sector analogue of BBN.~  This is indeed
another context in which the full ensemble of dark-sector states plays an important
role in dark-matter phenomenology.

Given the initial steps presented here, there are many avenues for future research.
For example, in this paper we have primarily focused on the phenomenology associated
with the ``sweet-spot'' region in Eq.~(\ref{sweetspot}),
as this region gives rise to
a rich spectrum of associated mass scales and DDM-like behaviors.
However, other regions may also be relevant for 
different situations,
including the case of dark ensembles emerging from 
the bulk sectors of actual critical Type~I string theories.
Indeed, such theories typically have significantly larger central charges and values of $D_\perp$
than those corresponding to the $D_\perp=2$ flux tube,
and thus correspond to values of $(B,C)$ which are very far from 
the ``benchmark'' values in Eq.~(\ref{BCbenchmarks}).
Such strings also likely correspond to 
values of $(r,s)$ which are far from those in Eq.~(\ref{sweetspot}).
Likewise, in our analysis we have taken
$\kappa=1$ and $\xi=3$.  Although these simple choices were well-motivated and conservative, 
it would certainly be interesting to explore the consequences of alternative choices.
It would also be of interest to explore the ramifications of relaxing some of
the approximations we have made in our analysis.
These include 
       the ``instantaneous freeze-out'' approximation
        that underpins the Boltzmann suppression factor
in Eq.~(\ref{eq:OmeganPrimordial}),
as well as our implicit assumption 
that the Hubble expansion within 
which our calculations have taken place
is unaffected by potential gravitational backreaction from our continually evolving dark sector.
While these approximations may certainly be justified to first order,
a more refined calculation is still capable of altering our results numerically if not qualitatively.

It would also be interesting to subject the DDM ensembles 
we have studied here to more detailed phenomenological constraints.
The constraints we have studied here, such as our look-back and $\weff$ constraints,
are those that 
follow directly (and in a completely model-independent manner) from 
knowledge of $\Omegatot(t)$ and $\weff(t)$ alone, and as such
we have seen that they are sufficient
to rule out vast regions of parameter space.
It is nevertheless true that a plethora of additional
constraints could be formulated once a particular scenario with a particular particle content
is specified,
and that imposing such additional constraints could 
potentially narrow our viable parameter space still further.

Finally, and perhaps most importantly,
in this paper we have assumed that the effects of {\it intra-ensemble}\/ decays on the decay widths of the
ensemble constituents are negligible.  Such an assumption is certainly consistent
with our other assumptions about the structure of the theory.
In general, following our string-based approach to understanding the dyamics of these
bound-state flux tubes, we may regard the strength of the interactions among the different dark hadrons 
in our DDM model
as being governed by an additional parameter, a so-called ``string coupling'' $g_s$,
which we have not yet specified but which does not impact any of the results
we have presented thus far.  
In general, $g_s$ can be 
different from the coupling which governs the decays of our ensemble states to SM states
and which is thus embedded within $\tau_0$.
In an actual string construction, the value of 
$g_s$ is determined by the vacuum expectation value (VEV) of the dilaton field,
but the dynamics that determines this VEV is not well understood.
In general, however,
intra-ensemble decays will provide an additional
contribution to the total decay widths $\Gamma_n$, especially for the heavier ensemble 
constituents, and the decays of these heavier constituents can serve as an 
additional source for the abundances of the lighter constituents. 
The effects of such intra-ensemble decays will be discussed in more
detail in Ref.~\cite{toappear}.


\begin{acknowledgments}


We would like to thank Emilian Dudas and Eduardo Rozo for discussions.  
The research activities of KRD, FH, and SS were supported in part by the Department of Energy 
under Grant DE-FG02-13ER41976 / DE-SC0009913;
the research activities of KRD were also supported in part by
the National Science Foundation through its employee IR/D program.
The opinions and conclusions 
expressed herein are those of the authors, and do not represent any funding agencies.

\end{acknowledgments}


\bigskip



\begin{references}


\bibitem{DDM1} 
  K.~R.~Dienes and B.~Thomas,
  Phys.\ Rev.\ D {\bf 85}, 083523 (2012)
  [arXiv:1106.4546 [hep-ph]].
  
\bibitem{DDM2} 
  K.~R.~Dienes and B.~Thomas,
  Phys.\ Rev.\ D {\bf 85}, 083524 (2012)
  [arXiv:1107.0721 [hep-ph]].

\bibitem{DDMcolliders}
    K.~R.~Dienes, S.~Su and B.~Thomas,
  Phys.\ Rev.\ D {\bf 86}, 054008 (2012)
  [arXiv:1204.4183 [hep-ph]].


\bibitem{DDMcolliders2}
   K.~R.~Dienes, S.~Su and B.~Thomas,
  Phys.\ Rev.\ D {\bf 91}, no. 5, 054002 (2015)
  [arXiv:1407.2606 [hep-ph]].


\bibitem{DDMdirect}
  K.~R.~Dienes, J.~Kumar and B.~Thomas,
  Phys.\ Rev.\ D {\bf 86}, 055016 (2012)
  [arXiv:1208.0336 [hep-ph]].

\bibitem{DDMindirect}
   K.~R.~Dienes, J.~Kumar and B.~Thomas,
  Phys.\ Rev.\ D {\bf 88}, no. 10, 103509 (2013)
  [arXiv:1306.2959 [hep-ph]].



\bibitem{DDMboxes1}
K.~K.~Boddy, K.~R.~Dienes, D.~Kim, J.~Kumar, J.~C.~Park and B.~Thomas,
  arXiv:1606.07440 [hep-ph].

\bibitem{DDMboxes2}
K.~K.~Boddy, K.~R.~Dienes, D.~Kim, J.~Kumar, J.~C.~Park and B.~Thomas,
  arXiv:1609.09104 [hep-ph].





\bibitem{DDMAxion} 
  K.~R.~Dienes and B.~Thomas,
  Phys.\ Rev.\ D {\bf 86}, 055013 (2012)
  [arXiv:1203.1923 [hep-ph]].


\bibitem{DesigningDDM}
  K.~R.~Dienes, J.~Fennick, J.~Kumar and B.~Thomas, to appear.
  

\bibitem{RandomDDM}
  K.~R.~Dienes, J.~Fennick, J.~Kumar and B.~Thomas,
  Phys.\ Rev.\ D {\bf 93}, 083506 (2016)
  [arXiv:1601.05094 [hep-ph]].


\bibitem{Hagedorn} 
  R.~Hagedorn,
  Nuovo Cim.\ Suppl.\  {\bf 3}, 147 (1965).

\bibitem{StringReviews}
  For reviews, see, \eg:\\
  M.~B.~Green, J.~A.~Schwarz and E.~Witten, {\it Superstring Theory, Vols. I and II} 
 (Cambridge University Press, 1987);
  J.~Polchinski, {\it String Theory, Vols. I and II} 
      (Cambridge University Press, 1998).

\bibitem{Cudell} 
  K.~R.~Dienes and J.~R.~Cudell,
  Phys.\ Rev.\ Lett.\  {\bf 72}, 187 (1994)
  [hep-th/9309126].


\bibitem{HR}
  G.~H.~Hardy and S.~Ramanujan,
  Proc.\ London Math.\ Soc.\  {\bf 17}, 75 (1918).

\bibitem{KV}
  I.~Kani and C.~Vafa,
  Commun.\ Math.\ Phys.\  {\bf 130}, 529 (1990).

\bibitem{missusy}
  K.~R.~Dienes,
  Nucl.\ Phys.\ B {\bf 429}, 533 (1994)
  [hep-th/9402006].



\bibitem{Nambu}
    Y.~Nambu (unpublished, 1970).

\bibitem{Ramond}
  P.~Ramond,
  Phys.\ Rev.\ D {\bf 3}, 2415 (1971).

\bibitem{NS}
  A.~Neveu and J.~H.~Schwarz,
  Nucl.\ Phys.\ B {\bf 31}, 86 (1971).


\bibitem{Polyakov}
  A.~M.~Polyakov,
  Nucl.\ Phys.\ B {\bf 268}, 406 (1986).

\bibitem{Green}
  M.~B.~Green,
  Phys.\ Lett.\ B {\bf 266}, 325 (1991).


\bibitem{PolchinskiStrominger}
  J.~Polchinski and A.~Strominger,
  Phys.\ Rev.\ Lett.\  {\bf 67}, 1681 (1991).



\bibitem{Planck} 
  P.~A.~R.~Ade {\it et al.}  [Planck Collaboration],
  arXiv:1502.01589 [astro-ph.CO].
  
\bibitem{BOSS} 
  L.~Anderson {\it et al.}  [BOSS Collaboration],
  Mon.\ Not.\ Roy.\ Astron.\ Soc.\  {\bf 441}, no. 1, 24 (2014)
  [arXiv:1312.4877 [astro-ph.CO]].

\bibitem{SCP} 
  N.~Suzuki, D.~Rubin, C.~Lidman, G.~Aldering, R.~Amanullah, K.~Barbary, L.~F.~Barrientos and J.~Botyanszki {\it et al.},
  Astrophys.\ J.\  {\bf 746}, 85 (2012)
  [arXiv:1105.3470 [astro-ph.CO]].

\bibitem{BBNLimits} 
  R.~H.~Cyburt, J.~Ellis, B.~D.~Fields, F.~Luo, K.~A.~Olive and V.~C.~Spanos,
  JCAP {\bf 0910}, 021 (2009)
  [arXiv:0907.5003 [astro-ph.CO]].

\bibitem{HuAndSilkLong}
  W.~Hu and J.~Silk,
  Phys.\ Rev.\  D {\bf 48}, 485 (1993).

\bibitem{HuAndSilkShort}
  W.~Hu and J.~Silk,
  Phys.\ Rev.\ Lett.\  {\bf 70}, 2661 (1993).

\bibitem{Slatyer} 
  T.~R.~Slatyer,
  Phys.\ Rev.\ D {\bf 87}, no. 12, 123513 (2013)
  [arXiv:1211.0283 [astro-ph.CO]].

\bibitem{AMSPositron} 
  L.~Accardo {\it et al.}  [AMS Collaboration],
  Phys.\ Rev.\ Lett.\  {\bf 113}, 121101 (2014).

\bibitem{AMSAntiproton}
  AMS-02 Collaboration, 
  presentations at AMS Days at CERN, April 15-17, 2015.

\bibitem{peter}
  A.~H.~G.~Peter, C.~E.~Moody and M.~Kamionkowski,
  Phys.\ Rev.\ D {\bf 81}, 103501 (2010)
  [arXiv:1003.0419 [astro-ph.CO]].


\bibitem{MeiYu} 
  M.~Y.~Wang, A.~H.~G.~Peter, L.~E.~Strigari, A.~R.~Zentner, B.~Arant, S.~Garrison-Kimmel and M.~Rocha,
  Mon.\ Not.\ Roy.\ Astron.\ Soc.\  {\bf 445}, no. 1, 614 (2014)
  [arXiv:1406.0527 [astro-ph.CO]].

\bibitem{LensingConstraint} 
  M.~Y.~Wang and A.~R.~Zentner,
  Phys.\ Rev.\ D {\bf 82}, 123507 (2010)
  [arXiv:1011.2774 [astro-ph.CO]].

\bibitem{Gong} 
  Y.~Gong and X.~Chen,
  Phys.\ Rev.\ D {\bf 77}, 103511 (2008)
  [arXiv:0802.2296 [astro-ph]].

\bibitem{Savvas} 
  G.~Blackadder and S.~M.~Koushiappas,
  Phys.\ Rev.\ D {\bf 90}, no. 10, 103527 (2014)
  [arXiv:1410.0683 [astro-ph.CO]].



\bibitem{DeLopeAmigo}
     S.~De Lope Amigo, W.~M.~Y.~Cheung, Z.~Huang and S.~P.~Ng,
     JCAP {\bf 0906}, 005 (2009)      
     [arXiv:0812.4016 [hep-ph]].



\bibitem{Audren}
     B.~Audren, J.~Lesgourgues, G.~Mangano, P.~D.~Serpico and T.~Tram,
     JCAP {\bf 1412}, no. 12, 028 (2014)
     [arXiv:1407.2418 [astro-ph.CO]].

\bibitem{Aubourg}
     E.~Aubourg {\it et al.},
     Phys.\ Rev.\ D {\bf 92}, no. 12, 123516 (2015)
     [arXiv:1411.1074 [astro-ph.CO]]. 

\bibitem{Blackadder}
     G.~Blackadder and S.~M.~Koushiappas,
     arXiv:1510.06026 [astro-ph.CO].







\bibitem{NussinovTechnibaryon}
  S.~Nussinov,
  Phys.\ Lett.\ B {\bf 165}, 55 (1985).


\bibitem{BarrTechnibaryon}
  S.~M.~Barr, R.~S.~Chivukula and E.~Farhi,
  Phys.\ Lett.\ B {\bf 241}, 387 (1990).


\bibitem{GudnasonTechnibaryon}
  S.~B.~Gudnason, C.~Kouvaris and F.~Sannino,
  Phys.\ Rev.\ D {\bf 73}, 115003 (2006)
  [hep-ph/0603014].

\bibitem{RyttovTIMP} 
  T.~A.~Ryttov and F.~Sannino,
  Phys.\ Rev.\ D {\bf 78}, 115010 (2008)
  [arXiv:0809.0713 [hep-ph]].

\bibitem{GUTzilla} 
  K.~Harigaya, T.~Lin and H.~K.~Lou,
  JHEP {\bf 1609}, 014 (2016)
  [arXiv:1606.00923 [hep-ph]].




\bibitem{AppelquistLattice1}
  T.~Appelquist {\it et al.} [Lattice Strong Dynamics (LSD) Collaboration],
  Phys.\ Rev.\ D {\bf 89}, no. 9, 094508 (2014)
  [arXiv:1402.6656 [hep-lat]].


\bibitem{AppelquistLattice2}
  T.~Appelquist {\it et al.},
  Phys.\ Rev.\ D {\bf 92}, no. 7, 075030 (2015)
  [arXiv:1503.04203 [hep-ph]].


\bibitem{AppelquistLattice3} 
  T.~Appelquist {\it et al.},
  Phys.\ Rev.\ Lett.\  {\bf 115}, no. 17, 171803 (2015)
  [arXiv:1503.04205 [hep-ph]].



\bibitem{KribsReview} 
  G.~D.~Kribs and E.~T.~Neil,
  Int.\ J.\ Mod.\ Phys.\ A {\bf 31}, no. 22, 1643004 (2016)
  [arXiv:1604.04627 [hep-ph]].




\bibitem{KilicDarkPion} 
  C.~Kilic, T.~Okui and R.~Sundrum,
  JHEP {\bf 1002}, 018 (2010)
  [arXiv:0906.0577 [hep-ph]].


\bibitem{HurKoPion} 
  T.~Hur and P.~Ko,
  Phys.\ Rev.\ Lett.\  {\bf 106}, 141802 (2011)
  [arXiv:1103.2571 [hep-ph]].


\bibitem{HolthausenPion} 
  M.~Holthausen, J.~Kubo, K.~S.~Lim and M.~Lindner,
  JHEP {\bf 1312}, 076 (2013)
  [arXiv:1310.4423 [hep-ph]].

\bibitem{HatanakaPion} 
  H.~Hatanaka, D.~W.~Jung and P.~Ko,
  JHEP {\bf 1608}, 094 (2016)
  [arXiv:1606.02969 [hep-ph]].

\bibitem{AmetaniPNGB} 
  Y.~Ametani, M.~Aoki, H.~Goto and J.~Kubo,
  Phys.\ Rev.\ D {\bf 91}, no. 11, 115007 (2015)
  [arXiv:1505.00128 [hep-ph]].

\bibitem{BaiPion} 
  Y.~Bai and R.~J.~Hill,
  Phys.\ Rev.\ D {\bf 82}, 111701 (2010)
  [arXiv:1005.0008 [hep-ph]].

\bibitem{Ectocolor} 
  M.~R.~Buckley and E.~T.~Neil,
  Phys.\ Rev.\ D {\bf 87}, no. 4, 043510 (2013)
  [arXiv:1209.6054 [hep-ph]].

\bibitem{BhattacharyaPion} 
  S.~Bhattacharya, B.~Meli\'{c} and J.~Wudka,
  JHEP {\bf 1402}, 115 (2014)
  [arXiv:1307.2647 [hep-ph]].


\bibitem{CarmonaPNGB}
  A.~Carmona and M.~Chala,
  JHEP {\bf 1506}, 105 (2015)
  [arXiv:1504.00332 [hep-ph]].
  
\bibitem{FrigerioHeavyEta} 
  M.~Frigerio, A.~Pomarol, F.~Riva and A.~Urbano,
  JHEP {\bf 1207}, 015 (2012)
  [arXiv:1204.2808 [hep-ph]].
  



\bibitem{LewisLattice} 
  R.~Lewis, C.~Pica and F.~Sannino,
  Phys.\ Rev.\ D {\bf 85}, 014504 (2012)
  [arXiv:1109.3513 [hep-ph]].


\bibitem{HietanenLattice} 
  A.~Hietanen, C.~Pica, F.~Sannino and U.~I.~Sondergaard,
  Phys.\ Rev.\ D {\bf 87}, no. 3, 034508 (2013)
  [arXiv:1211.5021 [hep-lat]].





\bibitem{CiDM} 
  D.~S.~M.~Alves, S.~R.~Behbahani, P.~Schuster and J.~G.~Wacker,
  Phys.\ Lett.\ B {\bf 692}, 323 (2010)
  [arXiv:0903.3945 [hep-ph]].


\bibitem{CiDMParity} 
  M.~Lisanti and J.~G.~Wacker,
  Phys.\ Rev.\ D {\bf 82}, 055023 (2010)
  [arXiv:0911.4483 [hep-ph]].


\bibitem{CiDMCosmology} 
  D.~Spier Moreira Alves, S.~R.~Behbahani, P.~Schuster and J.~G.~Wacker,
  JHEP {\bf 1006}, 113 (2010)
  [arXiv:1003.4729 [hep-ph]].


\bibitem{KribsQuirks}
  G.~D.~Kribs, T.~S.~Roy, J.~Terning and K.~M.~Zurek,
  Phys.\ Rev.\ D {\bf 81}, 095001 (2010)
  [arXiv:0909.2034 [hep-ph]].




\bibitem{AntipinCompositeHS1}
  O.~Antipin, M.~Redi and A.~Strumia,
  JHEP {\bf 1501}, 157 (2015)
  [arXiv:1410.1817 [hep-ph]].


\bibitem{AntipinCompositeHS2}
  O.~Antipin, M.~Redi, A.~Strumia and E.~Vigiani,
  JHEP {\bf 1507}, 039 (2015)
  [arXiv:1503.08749 [hep-ph]].




\bibitem{OkunThetons1} 
  L.~B.~Okun,
  JETP Lett.\  {\bf 31}, 144 (1980)
  [Pisma Zh.\ Eksp.\ Teor.\ Fiz.\  {\bf 31}, 156 (1979)].


\bibitem{OkunThetons2} 
  L.~B.~Okun,
  Nucl.\ Phys.\ B {\bf 173}, 1 (1980).


\bibitem{SoniGlueballs} 
  A.~Soni and Y.~Zhang,
  Phys.\ Rev.\ D {\bf 93}, no. 11, 115025 (2016)
  [arXiv:1602.00714 [hep-ph]].


\bibitem{ForestellGlueballs}
  L.~Forestell, D.~E.~Morrissey and K.~Sigurdson,
  arXiv:1605.08048 [hep-ph].




\bibitem{FaraggiGlueballs}
  A.~E.~Faraggi and M.~Pospelov,
  Astropart.\ Phys.\  {\bf 16}, 451 (2002)
  [hep-ph/0008223].


\bibitem{HalversonGlueballs} 
  J.~Halverson, B.~D.~Nelson and F.~Ruehle,
  arXiv:1609.02151 [hep-ph].




\bibitem{FengShadmiWIMPless}
  J.~L.~Feng and Y.~Shadmi,
  Phys.\ Rev.\ D {\bf 83}, 095011 (2011)
  [arXiv:1102.0282 [hep-ph]].




\bibitem{DetmoldDarkNuclei1}
  W.~Detmold, M.~McCullough and A.~Pochinsky,
  Phys.\ Rev.\ D {\bf 90}, no. 11, 115013 (2014)
  [arXiv:1406.2276 [hep-ph]].


\bibitem{KrnjaicDarkNuclei}
  G.~Krnjaic and K.~Sigurdson,
  Phys.\ Lett.\ B {\bf 751}, 464 (2015)
  [arXiv:1406.1171 [hep-ph]].


\bibitem{ClineDarkMolecules}
  J.~M.~Cline, Z.~Liu, G.~Moore and W.~Xue,
  Phys.\ Rev.\ D {\bf 90}, no. 1, 015023 (2014)
  [arXiv:1312.3325 [hep-ph]].
  
\bibitem{BoddyDarkHydrogen} 
  K.~K.~Boddy, M.~Kaplinghat, A.~Kwa and A.~H.~G.~Peter,
  arXiv:1609.03592 [hep-ph].

  
\bibitem{DetmoldDarkNuclei2} 
  W.~Detmold, M.~McCullough and A.~Pochinsky,
  Phys.\ Rev.\ D {\bf 90}, no. 11, 114506 (2014)
  [arXiv:1406.4116 [hep-lat]].

\bibitem{LargeDarkNuclei1} 
  E.~Hardy, R.~Lasenby, J.~March-Russell and S.~M.~West,
  JHEP {\bf 1506}, 011 (2015)
  [arXiv:1411.3739 [hep-ph]].
  
\bibitem{LargeDarkNuclei2} 
  E.~Hardy, R.~Lasenby, J.~March-Russell and S.~M.~West,
  JHEP {\bf 1507}, 133 (2015)
  [arXiv:1504.05419 [hep-ph]].




\bibitem{SIMPParadigmWacker} 
  Y.~Hochberg, E.~Kuflik, T.~Volansky and J.~G.~Wacker,
  Phys.\ Rev.\ Lett.\  {\bf 113}, 171301 (2014)
  [arXiv:1402.5143 [hep-ph]].


\bibitem{SIMPParadigmCarlson} 
  E.~D.~Carlson, M.~E.~Machacek and L.~J.~Hall,
  Astrophys.\ J.\  {\bf 398}, 43 (1992).


\bibitem{SIMPParadigmdeLaix} 
  A.~A.~de Laix, R.~J.~Scherrer and R.~K.~Schaefer,
  Astrophys.\ J.\  {\bf 452}, 495 (1995)
  [astro-ph/9502087].


\bibitem{SIMPModelHochberg} 
  Y.~Hochberg, E.~Kuflik, H.~Murayama, T.~Volansky and J.~G.~Wacker,
  Phys.\ Rev.\ Lett.\  {\bf 115}, no. 2, 021301 (2015)
  [arXiv:1411.3727 [hep-ph]].


\bibitem{SIMPModelHansen} 
  M.~Hansen, K.~Lang{\ae}ble and F.~Sannino,
  Phys.\ Rev.\ D {\bf 92}, no. 7, 075036 (2015)
  [arXiv:1507.01590 [hep-ph]].


\bibitem{SIMPModelBernal} 
  N.~Bernal and X.~Chu,
  JCAP {\bf 1601}, 006 (2016)
  [arXiv:1510.08527 [hep-ph]].


\bibitem{SIMPModelKamada} 
  A.~Kamada, M.~Yamada, T.~T.~Yanagida and K.~Yonekura,
  arXiv:1606.01628 [hep-ph].


\bibitem{BoddySIDM1}
  K.~K.~Boddy, J.~L.~Feng, M.~Kaplinghat and T.~M.~P.~Tait,
  Phys.\ Rev.\ D {\bf 89}, no. 11, 115017 (2014)
  [arXiv:1402.3629 [hep-ph]].


\bibitem{BoddySIDM2}
  K.~K.~Boddy, J.~L.~Feng, M.~Kaplinghat, Y.~Shadmi and T.~M.~P.~Tait,
  Phys.\ Rev.\ D {\bf 90}, no. 9, 095016 (2014)
  [arXiv:1408.6532 [hep-ph]].

\bibitem{DarkShower} 
  M.~Freytsis, D.~J.~Robinson and Y.~Tsai,
  Phys.\ Rev.\ D {\bf 91}, no. 3, 035028 (2015)
  [arXiv:1410.3818 [hep-ph]].


\bibitem{SIMPPhenoLee} 
  H.~M.~Lee and M.~S.~Seo,
  Phys.\ Lett.\ B {\bf 748}, 316 (2015)
  [arXiv:1504.00745 [hep-ph]].


\bibitem{SIMPPhenoHochberg} 
  Y.~Hochberg, E.~Kuflik and H.~Murayama,
  JHEP {\bf 1605}, 090 (2016)
  [arXiv:1512.07917 [hep-ph]].

  
\bibitem{EnglertSIMPCollder} 
  C.~Englert, K.~Nordstrom and M.~Spannowsky,
  arXiv:1606.05359 [hep-ph].


\bibitem{BruggisserSIMPCollider} 
  S.~Bruggisser, F.~Riva and A.~Urbano,
  arXiv:1607.02474 [hep-ph].



\bibitem{MultipleThetaAngles} 
  P.~Di Vecchia and F.~Sannino,
  Eur.\ Phys.\ J.\ Plus {\bf 129}, 262 (2014)
  [arXiv:1310.0954 [hep-ph]].



\bibitem{AudrenConstraintsOnDarkSectorDMDecays} 
  B.~Audren, J.~Lesgourgues, G.~Mangano, P.~D.~Serpico and T.~Tram,
  JCAP {\bf 1412}, no. 12, 028 (2014)
  [arXiv:1407.2418 [astro-ph.CO]].

\bibitem{PoulinConstraintsOnDarkSectorDMDecays} 
  V.~Poulin, P.~D.~Serpico and J.~Lesgourgues,
  JCAP {\bf 1608}, no. 08, 036 (2016)
  [arXiv:1606.02073 [astro-ph.CO]].


\bibitem{toappear}
  K.~R. Dienes, F.~Huang, J.~Kost, S.~Su, and B.~Thomas, to appear.


\end{references}
\end{document}